\pdfoutput=1  

\documentclass[preprints,article,accept,moreauthors,pdftex]{Definitions/mdpi}

\usepackage{bm}
\firstpage{1} 
\pubvolume{0}
\issuenum{1}
\articlenumber{5}
\pubyear{2020}
\copyrightyear{2020}
\history{}

\usepackage{xfrac}  

\newcommand*\nc{\newcommand*}  \nc\longnc{\newcommand}
\longnc\VOMIT[1]{#1}  \longnc\OMIT[1]{}  

\nc\Par{}  
\nc\qhl[1]{#1}
\nc\qred{black}

\nc\qwidth{0.495\columnwidth}
\nc\qqwidth{0.99\columnwidth}
\nc\Qwidth{0.393\columnwidth}
\nc\QWidth{0.43\columnwidth}
\nc\qfigx{\hspace{.2em}}
\nc\qfigw{0.535\columnwidth}
\nc\qfigW{0.484\columnwidth}
\nc\qfiga[2]{\hspace{0.29\columnwidth}#2\hspace
{-1.04\columnwidth}#1\hspace{0.01\columnwidth}}

\nc\quot[1]{`#1'}      \nc\quott[1]{``#1''}
\nc\lat\textit  
\nc\ie{\lat{i.e.,\ }}  \nc\etal{\lat{et al.}}    \nc\etc{\lat{etc.\ }}
\nc\eg{\lat{e.g.,\ }}  \nc\ergo{\lat{ergo}}
\nc\cf{cf.\ }
\nc\lhs{lhs}      \nc\rhs{rhs}
\nc\PTZ{Poynting--Thomson--Zener}

\nc\m[1]{$ #1 $}                     
\nc\re[1]{(\ref{#1})}

\nc\0[2]{\mathopen{}\mathclose{\9#1{#2}}}  
\nc\1[2]{\Delim#1#2\deliM#1}               
\nc\9[2]{\left\DElim#1#2\right\delIM#1}    
\nc\Delim[1]{\ifcase #1\or(\or[\or\{\or\mathord<\or\langle\or|\or\|\fi}
\nc\deliM[1]{\ifcase #1\or)\or]\or\}\or\mathord>\or\rangle\or|\or\|\fi}
\nc\DElim[1]{\ifcase #1.\or(\or[\or\{\or<\or\langle\or|\or\|\fi}
\nc\delIM[1]{\ifcase #1.\or)\or]\or\}\or>\or\rangle\or|\or\|\fi}

\nc\f\frac   \nc\ff[2]{{\textstyle \f{#1}{#2}}}
\nc\F[5]{\1#3{\1#1{#4}/\1#2{#5}}} 
\nc\tr{\operatorname{tr}}

\nc\e{\mathrm{e}} 
\nc\ii{\mathrm{i}} 

\nc\dd{\mathrm{d}} 
\nc\pd\partial
\nc\var\delta
\nc\lta[1]{{\overset{{\scriptscriptstyle \leftarrow}}{#1}}}
\nc\rta[1]{{\overset{{\scriptscriptstyle \rightarrow}}{#1}}}
\nc\doubledot{:}
\nc\sd{\cdot}
\nc\dyad{\otimes}
\nc\Trans{\text{T}}
\nc\tensor\mathbf  \nc\Tensor\boldsymbol  \nc\tensoR\mathsf
\nc\dev[1]{{#1}^\text{dev}}            \nc\sph[1]{{#1}^\text{sph}}

\nc\devv[1]{{#1}\bit^\text{d}}         \nc\sphh[1]{{#1}\bit^\text{s}}
\nc\tot[1]{{#1}_\text{total}}            \nc\kin[1]{{#1}_\text{kinetic}}
\nc\ther[1]{{#1}_\text{thermal}}            \nc\ela[1]{{#1}_\text{elastic}}
\nc\rheol[1]{{#1}_\text{rheological}}
\nc\dyn[1]{{#1}_\text{dyn}}
\nc\aux[1]{{#1}_\text{ref}}
\nc\nd[1]{\tilde{#1}}  
\nc\qinner{\Delta}
\nc\Quad{\:\:}
\nc\qdot[1]{\f {\pd #1}{\pd \qt}}  \nc\Pri[1]{\f {\pd #1}{\pd \qx}}
\nc\RE{\operatorname{Re}}          \nc\IM{\operatorname{Im}}
\nc\Ordo{\mathcal{O}}
\nc\qqh[1]{#1}  \nc\qqp[1]{\hat{#1}}
\nc\qldsym[1]{l^\text{d}_\text{S}}     \nc\qlssym[1]{l^\text{s}_\text{S}}
\nc\qldskw[1]{l^\text{d}_\text{A}}     \nc\qlsskw[1]{l^\text{s}_\text{A}}
\nc\qlsym{l_\text{S}}                  \nc\qlskw{l_\text{A}}
\nc\qdx[1]{\f {\pd #1}{\pd \qx}}
\nc\qdy[1]{\f {\pd #1}{\pd \qy}}
\nc\qdz[1]{\f {\pd #1}{\pd \qz}}
\nc\qsph{\text{S}}  \nc\qdev{\text{D}}
\nc\qvm{\text{vm}}
\nc\iOii{\sfrac{1}{2}}  
\nc\qxx{xx}  \nc\qyy{yy}  \nc\qzz{zz}  \nc\qxy{xy}  \nc\qyx{yx}
\nc\qxz{xz}  \nc\qzx{zx}  \nc\qyz{yz}  \nc\qzy{zy}
\nc\qpp{pp}  \nc\qpq{pq}
\nc\qlmn{\ql',\qm',\qn'}

\nc\Qa{a}              \nc\qai[1]{\Qa_{#1}}
\nc\qb{b}              \nc\qa[1]{\qb_{#1}}
\nc\qc{c}              \nc\qcc{\qqp{\qc}}
\nc\qcp{c_{\Qsig}}
\nc\qcC{\f {\qEY}{\qrho}}  \nc\qcCF{\ff {\qEY}{\qrho}}
\nc\qcl{\qc_l}  \nc\qCl{\qC_l}  \nc\qCcl{\hat{\qC_l}}
\nc\qct{\qc_t}  \nc\qCt{\qC_t}  \nc\qCct{\hat{\qC_t}}
\nc\qe{e}
\nc\qi{i}
\nc\qj{j}  \nc\qjm{\qj - \iOii}    \nc\qjp{\qj + \iOii}
\nc\qk{k}
\nc\ql{l}  \nc\qlm{\ql - \iOii}    \nc\qlp{\ql + \iOii}
\nc\qm{m}  \nc\qmm{\qm - \iOii}    \nc\qmp{\qm + \iOii}
\nc\qn{n}  \nc\qnm{\qn - \iOii}    \nc\qnp{\qn + \iOii}
\nc\qnn{n}
\nc\qp{p}
\nc\qq{q}
\nc\qs{s}
\nc\qt{t}  \nc\qDt{\Delta \qt}  \nc\qDtt{\Delta \nd{\qt}}
\nc\qu{u}
\nc\qv{v}  \nc\Qv{\tensor{\qv}}
\nc\qx{x}  \nc\qDx{\Delta \qx}
\nc\qy{y}  \nc\qDy{\Delta \qy}
\nc\qz{z}

\nc\qAv{A_{\qv}}       \nc\qAvv[1]{\qAv\01{#1}}
\nc\qAeps{A_{\qeps}}   \nc\qAepss[1]{\qAeps\01{#1}}
\nc\qAsig{A_{\qsig}}   \nc\qAsigg[1]{\qAsig\01{#1}}
\nc\qAAsig[1]{A_{\qsig,#1}}
\nc\qC{C}             \nc\qCC{\qqp{\qC}}
\nc\qCc{\hat{\qC}}
\nc\qEY{E}              \nc\qEE{\hat{\qEY}}     \nc\qEEE{\hat{\hat{\qEY}}}
\nc\qEsph{\qEY^S}  \nc\qEdev{\qEY^D}  
\nc\qEinf{\qEY_\infty}
\nc\qH{H}
\nc\qII{\hat{I}}
\nc\qJ{J}
\nc\qJac{\text{J}}
\nc\qM{M}
\nc\qN{N}
\nc\qP{P}              \nc\qPP{\qqp{\qP}}
\nc\qQ{Q}              \nc\qQQ{\qqp{\qQ}}
\nc\qS{S}
\nc\qSS{\sin{\f{\qk \qDx}{2}}}
\nc\qT{T}              \nc\qqT{\tensoR T}     \nc\qqTT{\qqp{\qqT}}
\nc\qWb{W_\text{b}}
\nc\qX{X}  \nc\qY{Y}

\nc\qalp{\alpha}
\nc\qbet{\beta}
\nc\qeps{\varepsilon}  \nc\Qeps{\Tensor{\qeps}}
\nc\qpi{\pi}
\nc\qrho{\varrho}
\nc\qsig{\sigma}       \nc\Qsig{\Tensor{\qsig}}
     \nc\qsigb{\qsig_\text{b}}
\nc\Qsighat{\hat\Qsig}  \nc\Qsighatdev{\dev{\Qsighat}}
                        \nc\Qsighatsph{\sph{\Qsighat}}
\nc\Qsighatdevhack{\Qsighatdev{\vphantom{\dev\Qsig}}}  
\nc\Qsighatsphhack{\Qsighatsph{\vphantom{\sph\Qsig}}}  

\nc\qtau{\tau}         \nc\qtauu{\hat{\qtau}}
\nc\qtaub{\tau_\text{b}}  \nc\qndtaub{{\nd{\tau}}_\text{b}}
\nc\qxi{\xi}  \nc\qxii{\qxi_+}  \nc\qxiii{\qxi_-}  \nc\qxiiii{\qxi_\pm}
\nc\qeta{\eta}
\nc\qome{\omega}

\Title{Four Spacetime Dimensional Simulation of Rheological Waves in Solids
and the Merits of~Thermodynamics}

\Author{\qhl{\'Aron Pozs\'ar} 
 $^{1
}$\orcidA{},
M\'aty\'as Sz\"ucs $^{1,2
}$\orcidB{},
R\'obert Kov\'acs $^{1,2,3,
}$*\orcidC{}
and Tam\'as F\"ul\"op $^{1,2}$\orcidD{}}

\AuthorNames{\'Aron Pozs\'ar, M\'aty\'as Sz\"ucs, R\'obert Kov\'acs and
Tam\'as F\"ul\"op}

\address{%
$^{1}$ \quad Department of Energy Engineering, Faculty of Mechanical
Engineering, BME, 1521 Budapest, Hungary; paron05@gmail.com (\'A.P.); szucsmatyas@energia.bme.hu (M.S.); fulop@energia.bme.hu (T.F.)
 \\
$^{2}$ \quad Montavid Thermodynamic Research Group, 1112 Budapest, Hungary
 \\
$^{3}$ \quad Department of Theoretical Physics, Wigner Research Centre for
Physics, Institute for Particle and Nuclear Physics, 1525 Budapest, Hungary
}

\corres{Correspondence: kovacsrobert@energia.bme.hu}

\abstract{%
The recent results attained from a \textcolor{\qred}{thermodynamically}
conceived numerical scheme applied on wave propagation in viscoelastic/rheological solids are generalized here, both in the sense that the scheme is extended to four spacetime dimensions and in the aspect of the virtues of a thermodynamical approach. Regarding the scheme, the arrangement of which quantity is
represented where in discretized spacetime, including the question of
appropriately realizing the boundary conditions, is nontrivial. In parallel, placing the
problem in the thermodynamical framework proves to be beneficial in regards to
monitoring and controlling numerical artefacts---instability, dissipation
error, and dispersion error. This, in addition to the observed preciseness,
speed, and resource-friendliness, makes the thermodynamically extended
symplectic approach that is presented here advantageous above commercial finite element software solutions.
 }

\keyword{symplectic numerical methods; rheology; solids; waves; spacetime;
thermodynamics}

\begin{document}

\section{Introduction}

Solids may be less \quott{solid} than expected. Beyond elastic behaviour,
they may exhibit damped and delayed response. This viscoelastic/rheological
reaction may not be simply explained by a viscosity-related additional stress
\1 1 {the Kelvin--Voigt model of rheology}, but the time derivative of stress
may also be needed in the description, with the simplest such model being the
so-called standard or Poynting--Thomson--Zener (PTZ) one \1 2 {see its
details below}. Namely, the PTZ model is the simplest model that enables
describing both creep \1 1 {declining increase of strain during constant
stress} and relaxation \1 1 {declining decrease of stress during constant
strain}, as well as the simplest one, via which it is possible to interpret that
the dynamic elasticity coefficients of rocks are different from, and larger than,
their static counterpart \cite{Barnafoldi17,Van19,Davarpanah20,Fulop20}.
Related to the latter aspect, high-frequency waves have a larger propagation
speed in PTZ media than low-frequency ones \cite{Fulop20}, which makes this
model relevant for, e.g., seismic phenomena and acoustic rock mechanical
measurement methods.

Analytical solutions for problems in PTZ and more complex rheological solid
media exist \1 1 {see, e.g., \cite{VolterraEUROCK,Volterraarxiv}}, but mostly in
the force-equilibrial/quasistatic approximation, which cannot give account of
transients and waves. Incorporating such \quot{fast} effects is expected to
only be realizable by means of numerical calculations in most practical
situations.

In many of the practical applications, such a numerical calculation must be
performed many times with different material coefficients, for example, as
part of a fitting procedure where the experimental data are to be fitted. Hence,
the numerical scheme should be fast, resource-friendly, yet reliable and
precise enough.

In addition, numerical calculations face three frequent challenges:
instability \1 1 {exponential blow-up of the solution}, dissipation error \1
1 {artificial decrease of amplitudes and energies}, and dispersion error \1 1
{artificial oscillations near fast changes}. A good scheme keeps these
artefacts under control.

Being driven by \1 1 {primarily rock mechanical} applications in scope, we have
tried to use commercial finite element softwares for wave phenomena in PTZ
models. What we found---already for the Hookean case \1 1 {but also for
non-Fourier heat conduction \cite{Rieth18}}---was disappointing: the
solutions ran very slowly, with large memory and CPU demand, and they were
burdened by considerable numerical artefacts of the mentioned kinds.

Now, if a numerical scheme exhibits dissipation error for conservative
systems, then it is expected to behave similarly for nonconservative ones, so
that one cannot separate the real dissipation of mechanical energy from the
dissipation artefact of the scheme. Additionally, similarly, a real wavy behaviour
cannot be distinguished from the dispersion/wavy artefact.

These have motivated us to develop our own numerical scheme, which performs
better \cite{1Dcikk}. \textcolor{\qred}{Similarly to that the PTZ model 
can be obtained in a thermodynamical approach as an internal variable extension
of Hooke elasticity \cite{asszonyi-fulop-van:2015}, our starting point was a
symplectic scheme for Hooke elasticity.}
Symplectic numerical schemes \1 1
{see, e.g., \cite{hairer}} provide much better large-time approximations,
thanks to the fact that a symplectic numerical integrator of a conservative
system is actually the exact integrator of a \quot{nearby} \1 1 {coinciding
in the zeroth order of the time step} conservative system.

In recent years, numerous works have been born in order to develop extensions of
symplectic schemes to nonconservative systems
\cite{zinner-ottinger:2019,shang-ottinger:2018,portillo-etal:2019,
vermeeren-etal:2019,gayos18,couga20,romero1:2010,romero2:2010,BerVan17b,czech}.
We also took this path, and devised such an extension, on the example of the
PTZ model, in one space dimension \cite{1Dcikk}, with the novelties that some
of the discretized field values reside with half space and time steps with respect
to some other field values. This made the symplectic Euler method---an
originally order-one accurate scheme---accurate to a second order, and the spatial accuracy was also second order.

\begin{figure}[H]
\qfiga{\includegraphics[width=\qfigw]{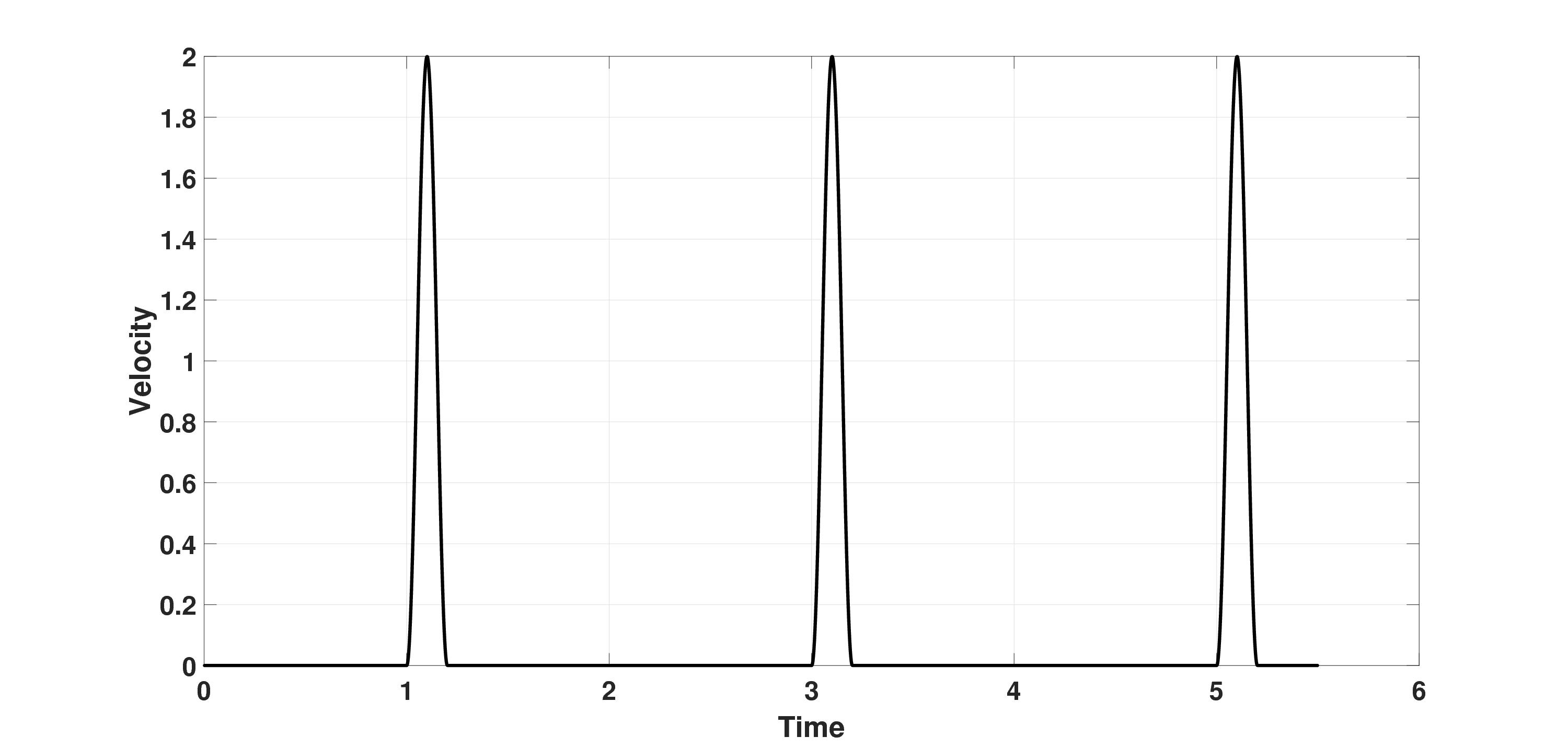}}
{\includegraphics[width=\qfigw]{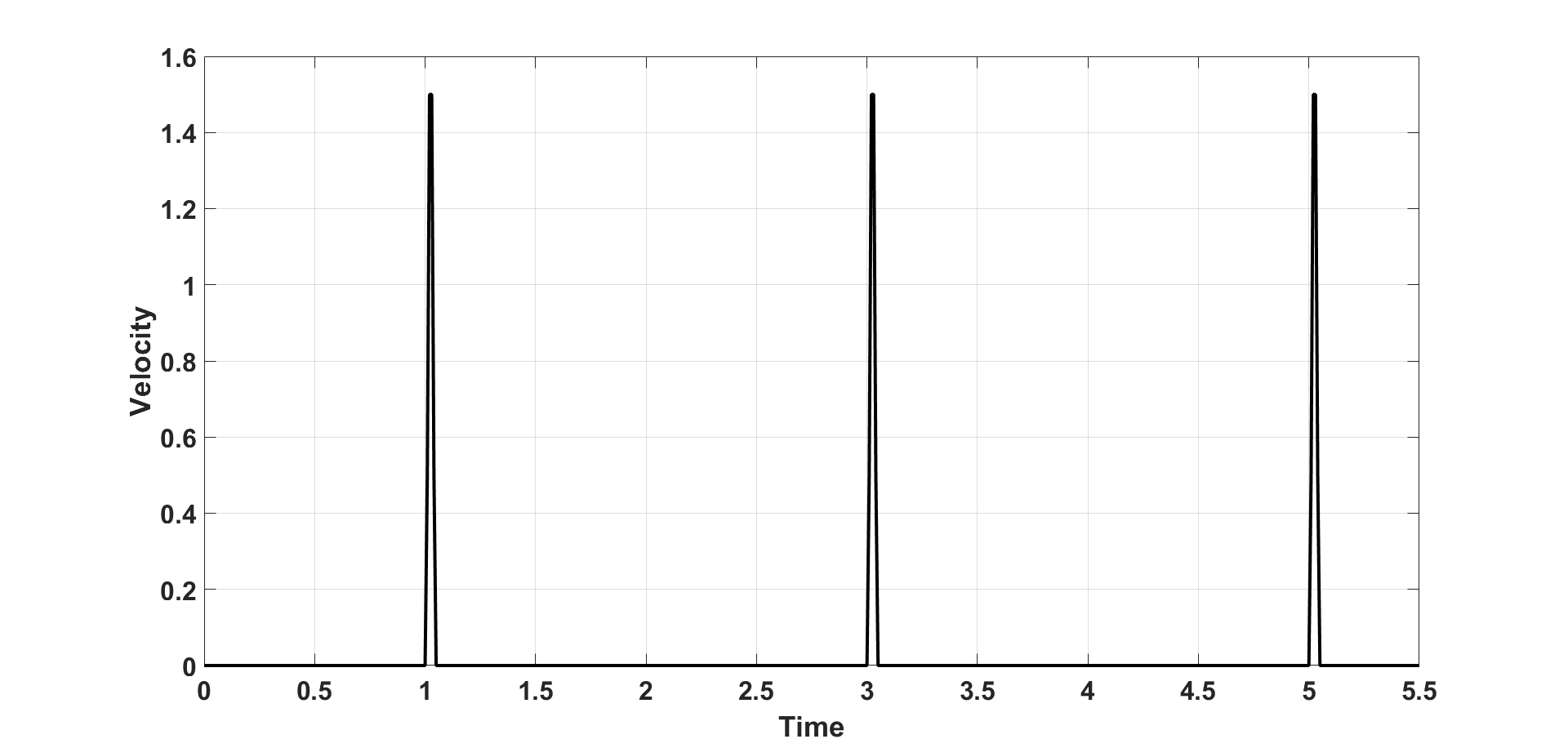}}
 \\
 \centering
\includegraphics[width=\qfigW]{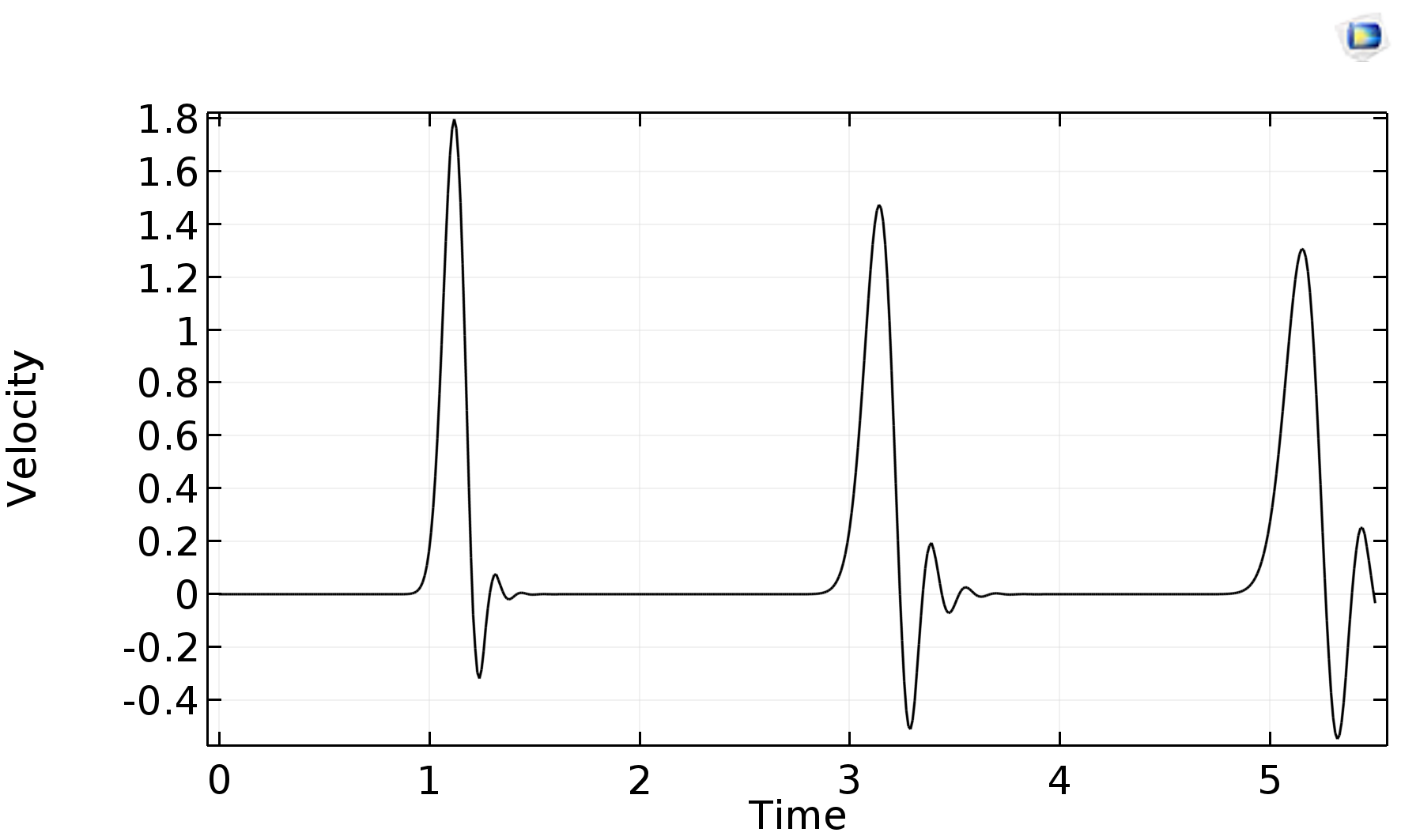}
\hfill\includegraphics[width=\qfigW]{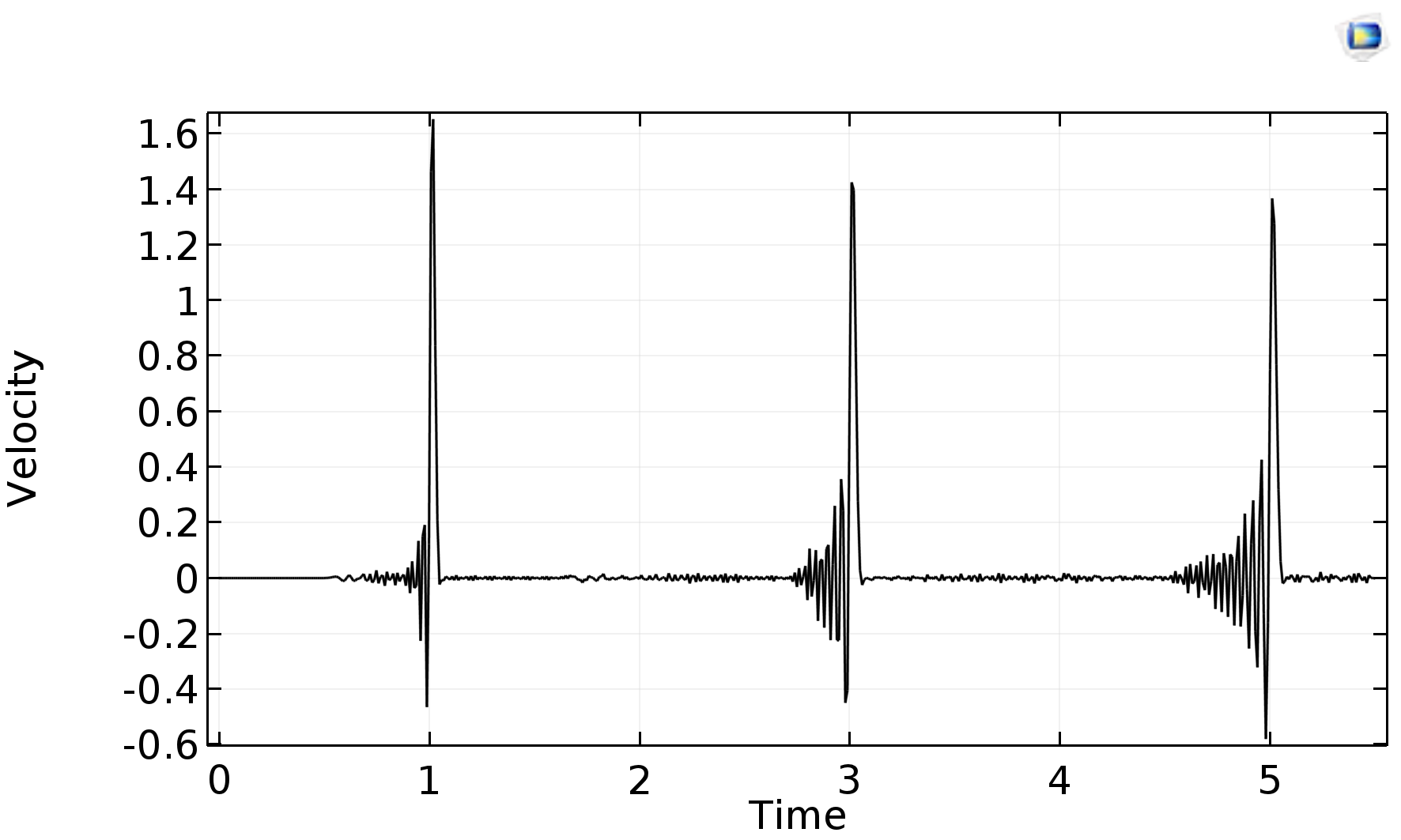}
 \\
\includegraphics[width=\qfigW]{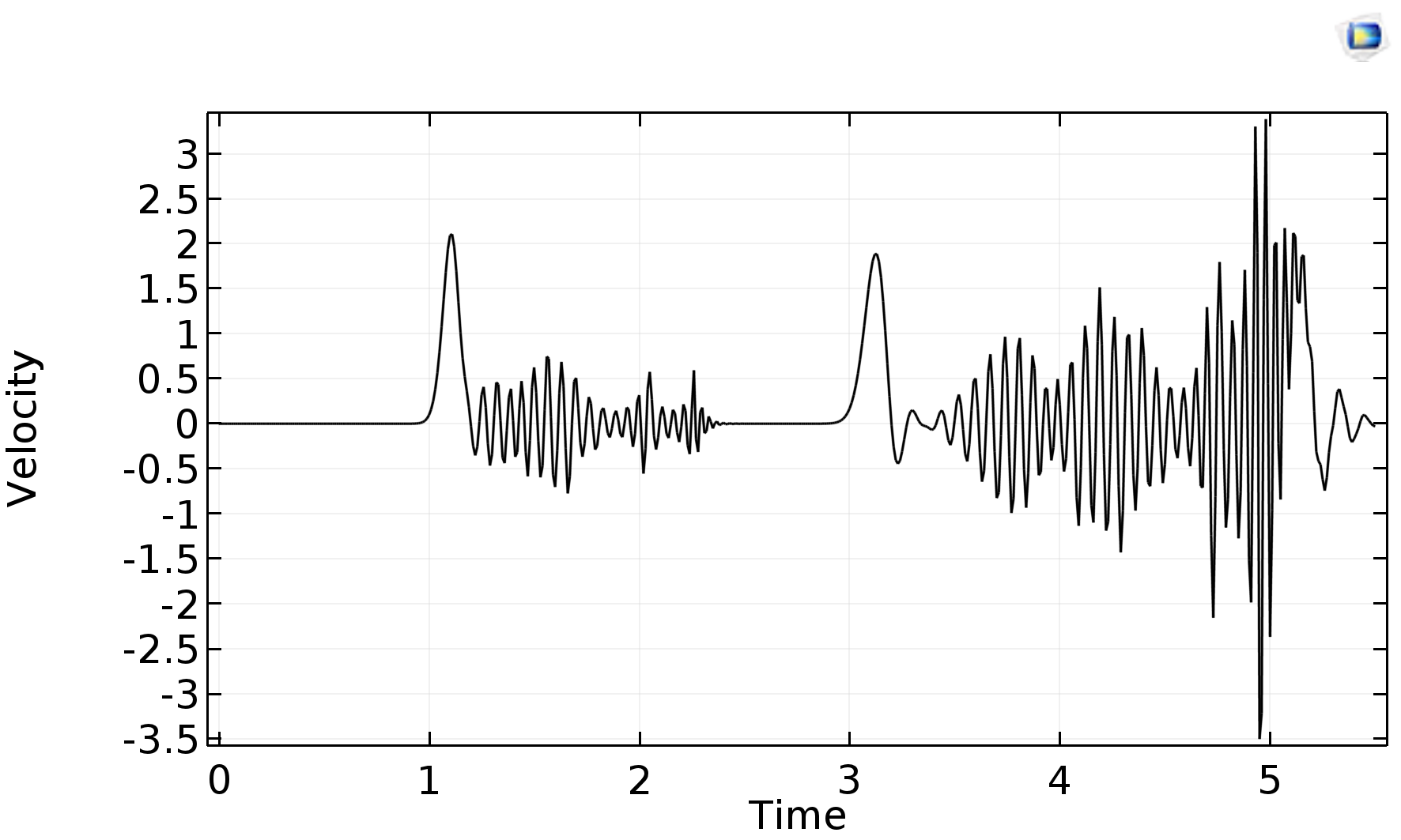}
 \hfill\includegraphics[width=\qfigW]{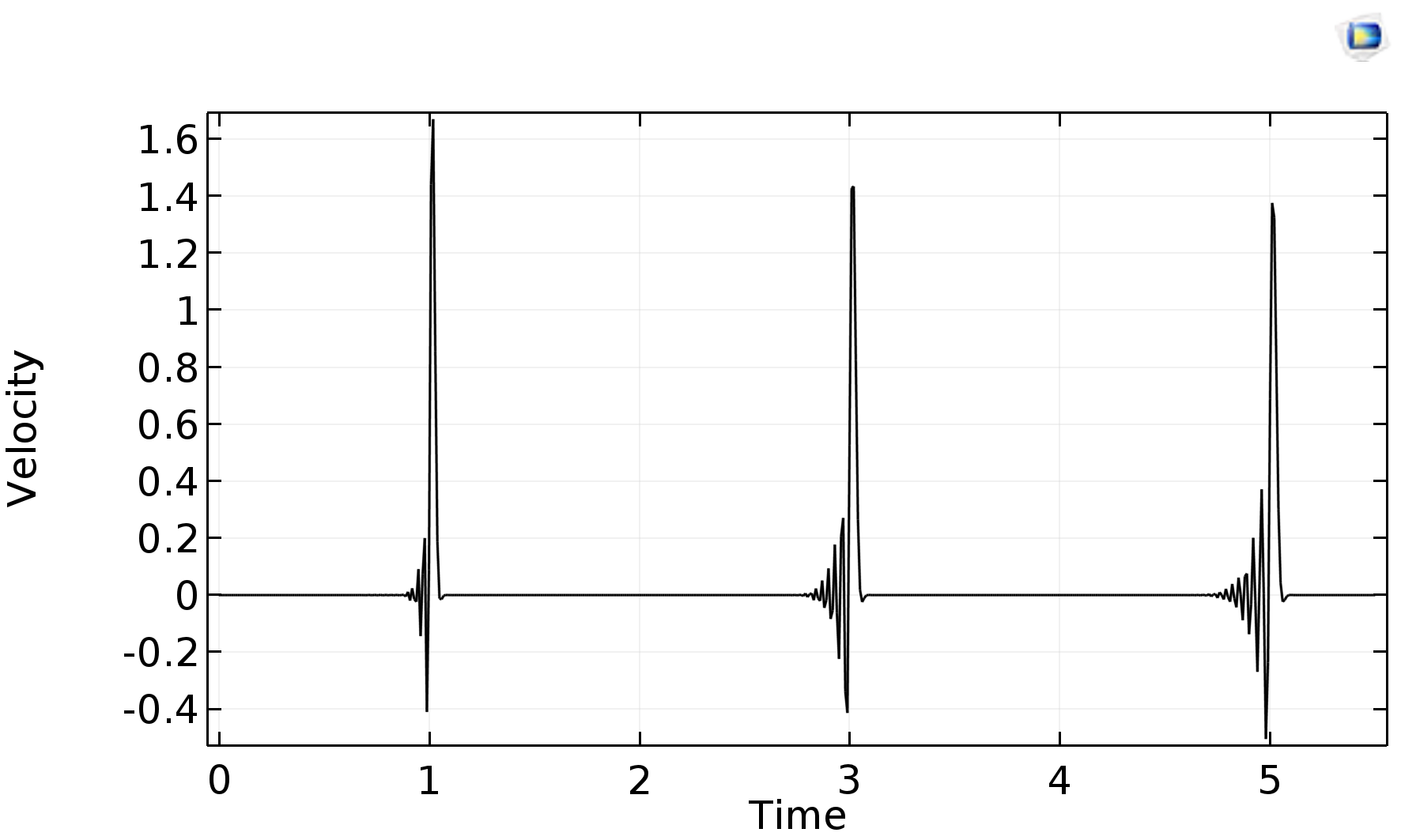}
\caption{An excitation pulse, generated at the left endpoint of a finite-size
one space dimensional Hookean sample, regularly arrives at the right endpoint. First row: the results from the scheme that was introduced in \cite{1Dcikk} for
two different pulse lengths. Second and third rows: the corresponding results
obtained by the finite element software COMSOL \cite{1Dcikk}, at various
settings: left column second row: Backward Differentiation Formula (BDF)
Maximum order 2, third row: BDF Maximum order~5; right~column second row:
Dormand--Prince \1 1 {DP} 5, third row: Runge--Kutta \1 1 {RK} 34. In each
finite element solution, dissipation error \1 1 {decrease of the
amplitude} and dispersion error \1 1 {artificial oscillations} are both observable,
even during the first three bounces. Meanwhile, the pulses in the first row
keep their shape even after many bounces \cite{1Dcikk}.}
  \label{com}
 \end{figure}

Indeed, our scheme has performed well: produced, in a much faster and
resource-friendlier way, much more artefact-free solutions, as demonstrated
in Figure \ref{com}. We note that, although the PTZ model allows for an exact
integrator in the nonconservative part of the model, along analogous lines, as~
in~\cite{shang-ottinger:2018}, we have refrained from using it, since, in the
future, we also wish to use the same scheme for more general nonconservative
systems, so we intended to test robustness in the dissipative aspect.

In addition to comparison to finite-element solutions, in \cite{1Dcikk}, we
analytically derived the criteria for stability, and showed how dissipation
and dispersion error can be kept small.

The first study done while using our scheme was numerically measuring wave
propagation speed in one space dimensional samples, and finding good
agreement with the corresponding analytical result \cite{Fulop20}.

The next step is reported here, with two novelties. The first is
the generalization of the scheme to three space dimensions (3D), and the other is
exploiting the whole thermodynamical theory around the PTZ model for
diagnostics regarding the credibility of the numerical solution, since
stability, dissipation, and dispersion error are much harder to investigate in
the case of a 3D model, with its numerous vectorial and tensorial degrees of
freedom.

The 3D scheme is designed in order to keep the nice, second-order, behaviour of the
discretization in both the spatial and in temporal direction. Achieving this is
not so trivial---different components of vectors and tensors are placed
at different discretized positions in order to fulfil the aim. In parallel, the
boundaries also pose a challenge: quantities must be placed in such a way
that the set of equations becomes closed. We succeed in finding a rule for
this that is general enough to hold for both stress and displacment boundary
conditions, where these two may differ at different sides of the 3D sample.

In finding the arrangement of discretized quantities suggested here, the
spacetime perspective has helped us a lot. Specifically, on one side,
thermodynamical balances in their differential form are four-divergences from
the spacetime aspect, and they have an integral counterpart, which, via~Gauss'
theorem, helps one to find where to represent which flux-type quantity.
Additionally, knowing that, from the spacetime point of view, velocity is a timelike
four-vector, \cite{fulop-van:2012,Van17} gives the information that velocity
should be shifted not only spatially, but also temporally. Oppositely,
stress is a spacelike tensor, so no temporal shift is needed.

In parallel, thermodynamics is not only important from the aspects of
balances.
\textcolor{\qred}{Namely, commercial finite element softwares only focus on the set
of equations to solve, i.e., on the minimally necessary equations to
follow the minimally necessary quantities.}
However, knowing, from~the
spacetime perspective, that momentum and energy form a four-quantity \1 1
{also in continuum theory on Galilean spacetime} \cite{Van17}, in addition to
the customarily taken balance of momentum, the balance of energy is also
present. This enables one to follow, in addition to the mechanically
considered quantities, internal energy---or, if practice favours so,
\textcolor{\qred}{temperature}.
The point in doing so \1 1 {even in situations where
thermomechanical coupling and heat conduction are neglected} is that, if,
say, temperature is followed via a separate discretized evolution equation,
then the conservation of total energy---at the discretized level---is not
built-in, but rather is a property that will hold only approximately. Subsequently, checking how well this conservation holds along the numerical solution can provide a
diagnostic tool. Thus, one may check the degree of dissipation error \1 1 {i.e.,
the degree of violation of total energy conservation} and of dispersion error
\1 1 {spurious oscillations on total energy that should be a constant}. This
idea is demonstrated below, on the example of the PTZ model \1 1 {also~equipped
with the thermodynamical constituents}.

Furthermore, thermodynamics also
\textcolor{\qred}{provides}
entropy, which is known to serve
as a Lyapunov function, ensuring asymptotic stability \1 1 {see, e.g.,
\cite{MT-kek-konyve}}. Now, stability also becomes challenged at the numerical
level. Accordingly, entropy, and the related entropy production, may serve as
an aid for reliable numerical calculations. Certain efforts in this direction
have already been made \cite{zinner-ottinger:2019,shang-ottinger:2018}. Here,
we introduce another way of utilizing this general idea.

Specifically, we focus on entropy production. At the continuum level, it must
be positive definite according to the second law of thermodynamics. However,
when discretized, this property may also become challenged. Naturally, if an
explicitly positive definite expression is discretized, then it remains
positive definite. However, alternative forms---which only turn out to be
positive definite when the further thermodynamical equations also hold---are not \lat{ab ovo} positive definite and, correspondingly, may fail in
being/remaining so along a numerical solution. Such forms are provided in a
natural way, for instance, when the balance of entropy is connected to the
balance of internal energy, such as when rheological models, like the PTZ one, are derived from the internal variable approach \cite{asszonyi-fulop-van:2015}. Here, we~discretize such an expression of entropy production and show that its
value becoming negative can forecast a loss of stability and blowing-up of the
solution.

\OMIT{%
The outline of the paper is as follows. We start with describing the
continuum PTZ model together with its thermodynamical embedding. Subsequently, we
present our discretization scheme for this model. Two examples follow, with the
first being a long rectangular beam excited by a stress pulse on one of its
sides---here, we show wave propagation via snapshots for various quantities
for both the simpler Hooke model and PTZ one, and demonstrate the
informative role of the various energy terms and total energy. In the
second example, a cubic sample is treated, where, after similar snapshots,
not only energies, but entropy production rate, is also displayed, in four
discretized forms, and a comparison of a stable setting and a corresponding
unstable one shows the possible role of entropy production rate in monitoring
the loss of stability. Finally, finite element calculations for a similar, but
much simpler, wave propagation problem are presented while using the built-in
facilities of a commercial finite element software, which illustrate both

 \begin{itemize}
 \item
the need for new powerful numerical schemes \1 1 {with symplectic roots,
spacetime friendliness and other features}, and
 \item
the need for means of validating, testing, and monitoring the solutions, for
which thermodynamics prove to provide excellent tools.
 \end{itemize}
}

\section{The Continuum PTZ Model and the Thermodynamics Behind}

We consider a homogeneous and isotropic solid, in the small-deformation
approximation \1 1 {with~respect to an inertial reference system}, due to
which we do not have to differentiate between Eulerian and Lagrangian
position or make a distinction between spatial spacetime vectors,
covectors, tensors, etc, as well as material manifold related ones, mass density
\m { \qrho } can also be treated as constant, and the relationship between
the symmetric strain tensor \m { \Qeps }
to the velocity field
\m { \Qv }
is
  \Par
 \begin{align}  \label{B}
\qdot{\Tensor{\qeps}} = \f{1}{2} \9 1 { \rta{\nabla} \otimes \Qv + \Qv
\otimes \lta{\nabla} } ,
 \end{align}
  \Par
with \m { \rta{\nabla} } and \m { \lta{\nabla} } denoting the spatial
derivative operation acting to the right and left, respectively.

The stress tensor \m { \Qsig } is also assumed to be symmetric, and it governs
the time evolution of \m { \Qv } according to
  \Par
 \begin{align}  \label{A}
\qrho \qdot{\Qv} = \Qsig \cdot \lta{\nabla}
 \end{align}

With the deviatoric and spherical parts of tensors,
  \Par
 \begin{align} \label{devdecomp}
\sph\Qsig = \f{1}{3} \9 1 {\tr \Qsig} \tensor{1} ,
 \quad
\dev\Qsig = \Qsig - \sph\Qsig ,
 \qquad\quad
\sph\Qeps = \f{1}{3} \9 1 {\tr \Qeps} \tensor{1} ,
 \quad
\dev\Qeps = \Qeps - \sph\Qeps
 \end{align}
  \Par
\1 1 {\m { \tensor{1} } denoting the unit tensor}, the Hooke elasticity can be
expressed as
  \Par
 \begin{align}  \label{Ho}
\dev\Qsig = \dev \qEY \dev\Qeps ,  \quad  \sph\Qsig = \sph \qEY \sph\Qeps ,
 \qquad\quad
\dev \qEY = 2G ,  \quad  \sph \qEY = 3K ,
 \end{align}
  \Par
and its PTZ generalization \1 3 {among its vast literature, see
\cite{asszonyi-fulop-van:2015,Godollo,Fulop20} for its treatment in the
irreversible thermodynamical internal variable approach and \cite{GENERIC},
for its presentation in the GENERIC \1 1 {General~Equation for the Non-
Equilibrium Reversible--Irreversible Coupling} framework}, is
  \Par
 \begin{align}  \label{PTZ}
\dev\Qsig + \dev\qtau \qdot{\dev\Qsig} = \dev\qEY \dev\Qeps + \dev{\qEE}
\qdot{\dev\Qeps} ,
 \qquad
\sph\Qsig + \sph\qtau \qdot{\sph\Qsig} = \sph\qEY \sph\Qeps + \sph{\qEE}
\qdot{\sph\Qeps} ,
 \end{align}
  \Par
the coefficients will be treated as constants hereafter.

In order to make the subsequent formulae more intelligible, we introduce
  \Par
 \begin{align}  \label{sighat}
\Qsighatdev = \dev\Qsig - \dev\qEY \dev\Qeps ,
 \qquad
\Qsighatsph = \sph\Qsig - \sph\qEY \sph\Qeps
 \end{align}
  \Par
and the coefficient combinations
  \Par
  \begin{align}  \label{damping}
\dev\qII = \dev{\qEE} -  \dev\qtau \dev\qEY ,
 \qquad
\sph\qII = \sph{\qEE} -  \sph\qtau \sph\qEY ,
 \end{align}
  \Par
with the aid of which \re{PTZ} gets simplified to
  \Par
 \begin{align}  \label{ptzsighat}
\Qsighatdev + \dev\qtau \qdot{\Qsighatdev} = \dev\qII\qdot{\dev\Qeps} ,
 \qquad
\Qsighatsph + \sph\qtau \qdot{\Qsighatsph} = \sph\qII\qdot{\sph\Qeps} .
 \end{align}

Taking \1 1 {also for simplicity} a constant \quot{isobaric} specific heat \m
{ \qcp } as well as neglected thermal expansion and heat conduction, the
internal variable approach puts the following thermodynamical background
behind the PTZ model: after eliminating the internal variable, its specific
total energy
  \Par
 \begin{align}  \label{energy}
\tot\qe & = \kin\qe + \ther\qe + \ela\qe + \rheol\qe ,
 \hspace{-30em}  &&  
 \\
\kin\qe & = \f {1}{2} \Qv^2 ,
 &
\ela\qe & = \f {\dev\qEY}{2 \qrho} \tr \0 1 {{\dev\Qeps}^2}
 + \f {\sph\qEY}{2 \qrho} \tr \0 1 {{\sph\Qeps}^2} ,
 \notag \\ \notag
\ther\qe & = \qcp \qT ,
 &
\rheol\qe & = \f{\dev\qtau}{2 \qrho \dev\qII} \tr \9 1 { \Qsighatdevhack^2 }
+ \f{\sph\qtau}{2 \qrho \sph\qII} \tr \9 1 { \Qsighatsphhack^2 }
 \end{align}
  \Par
with absolute temperature \m { \qT }, accompanied with specific entropy
\m { \qs } and entropy production rate density~\m { \qpi_{\qs} }
  \Par
\begingroup\makeatletter\def\f@size{9}\check@mathfonts
\def\maketag@@@#1{\hbox{\m@th\normalsize\normalfont#1}}%
 \begin{align}  \label{entr}
\qs & = \qcp \ln \f {\qT}{\aux\qT} ,
 \\ \label{entprodA}
\qpi_{\qs} & = \f {1}{\qT} \9 3 { \f {1}{\dev\qII} \tr \9 2 { \Qsighatdev \0
1 { \dev{\qEE} \dev{\qdot\Qeps} - \dev\qtau \dev{\qdot\Qsig} } } + \f
{1}{\sph\qII} \tr \9 2 { \Qsighatsph \0 1 { \sph{\qEE} \sph{\qdot\Qeps} -
\sph\qtau \dev{\qdot\Qsig}} }  }
 \\ \label{entprodB}
& = \f {1}{\qT} \9 3 { \f {1}{\dev\qII} \tr \9 1 { \Qsighatdevhack^2 } +
\f{1}{\sph\qII} \tr \9 1 { \Qsighatsphhack^2 } } ,
 \end{align}
\endgroup
  \Par
for which the specific internal energy part \m { \tot\qe - \kin\qe } fulfils
the balance
  \Par
 \begin{align}  \label{benmerleg}
\qrho \qdot {\0 1 {\tot\qe - \kin\qe} } = \tr \0 1 { \Qsig \qdot\Qeps }
 \end{align}
  \Par
and specific entropy the balance
  \Par
 \begin{align}  \label{entmerleg}
\qrho \qdot\qs = \qpi_{\qs} ,
 \end{align}
  \Par
as can be found along the lines of \cite{asszonyi-fulop-van:2015} \1
1{including its \qhl{Appendix B} 
}, and it is straightforward to check.
Because of the second law of thermodynamics,
  \Par
 \begin{align}  \label{ineqA}
\dev\qII > 0 ,  \qquad  \sph\qII > 0
 \end{align}
  \Par
and
  \Par
 \begin{align}  \label{ineqB}
\qpi_{\qs} \ge 0
 \end{align}
  \Par
follow for the PTZ model \cite{asszonyi-fulop-van:2015}, where \re{ineqB} is
already apparent from the form \re{entprodB}. \1 1 {Recall that heat
conduction is neglected, so there are no heat and entropy flux terms in the
balances. In parallel, there is no term in \m { \tot\qe } that couples \m
{\qT} and \m { \Qeps }, and---correspondingly---there is no \m { \Qeps }
dependent term in \m { \qs }, due~to neglected thermal expansion.}

Superficially, it seems redundant to also provide \m { \qpi_{\qs} } in the equivalent
form \re{entprodA}. However, it is just this
not-automatically-positive-definite form that will prove to be beneficial in the
diagnostics of the numerical solution.

From either balance \re{benmerleg} or \re{entmerleg}, the time derivative
of temperature can also be expressed:
  \Par
 \begin{align}  \label{Tdot}
\qdot\qT = \f {\qT}{\qrho \qcp} \qpi_{\qs} .
 \end{align}

As a simple analysis of the PTZ model, for \quot{slow} processes, which is to
be understood with respect to the time scales
  \Par
 \begin{align}  \label{timesc}
\dev\qtau,  \qquad  \dev{\qtauu} = \F000{\dev\qEE}{\dev\qEY} ,  \qquad
\sph\qtau,  \qquad  \sph{\qtauu} = \F000{\sph\qEE}{\sph\qEY} ,
 \end{align}
  \Par
a rule-of-thumb approximation is to neglect the time derivative terms \1 1
{to only keep the lowest time derivative term for each quantity} in
\re{PTZ}. The result is nothing but the Hooke model \re{Ho}, for which the
longitudinal and transversal wave propagation speeds are
  \Par
 \begin{align}  \label{lowspeed}
\qc_\text{longitudinal} = \sqrt{ \f {2 \dev\qEY + \sph\qEY}{3 \qrho} } ,
 \qquad
\qc_\text{transversal} = \sqrt{ \f {\dev\qEY}{2 \qrho} } .
 \end{align}

Now, as opposed to this \quot{static} limit, let us consider the limit of
\quot{fast} processes: then, it is the time derivative terms \1 1 {the highest
time derivative term for each quantity} that we keep. The result is the time
derivative of an effective/\quot{dynamic} Hooke model:
  \Par
\begingroup\makeatletter\def\f@size{9}\check@mathfonts
\def\maketag@@@#1{\hbox{\m@th\normalsize\normalfont#1}}%
 \begin{align}  \label{Hohoho}
\dev\Qsig = \dev \qEY_\infty \dev\Qeps ,  \quad
\sph\Qsig = \sph \qEY_\infty \sph\Qeps ,
 \qquad\quad
\dev{\qEY_\infty} = \F000{\dev\qEE}{\dev\qtau} > \dev\qEY ,  \quad
\sph{\qEY_\infty} = \F000{\sph\qEE}{\sph\qtau} > \sph\qEY ,
 \end{align}
\endgroup
  \Par
where the inequalities follow from \re{ineqA}. Accordingly, the wave
propagation speeds
  \Par
 \begin{align}  \label{highspeed}
\hat\qc_\text{longitudinal} = \sqrt{ \f {2 \dev\qEE_\infty +
\sph\qEE_\infty}{3 \qrho} } > \qc_\text{longitudinal} ,
 \qquad
\hat\qc_\text{transversal} = \sqrt{ \f {\dev\qEE_\infty}{2 \qrho} } >
\qc_\text{transversal}
 \end{align}
  \Par
follow. This, on one side, illustrates how the PTZ model can interpret that
the dynamic elasticity coefficients of rocks are larger than their static
counterpart \cite{Barnafoldi17,Van19,Davarpanah20,Fulop20}. On the other
side, the nontrivial---frequency dependent, therefore, dispersive---wave
propagation indicates that the numerical solution of PTZ wave propagation
problems should contain the minimal possible amount of dispersion error, in order to give account of the dispersive property of the continuum model itself. In~
parallel, the dissipative nature of the PTZ model requires the minimal
possible amount of dissipation error to reliably describe the decrease of
wave amplitudes.

\section{The Numerical Scheme}

\begin{figure}[H]
\centering
\includegraphics[width=0.35\columnwidth]{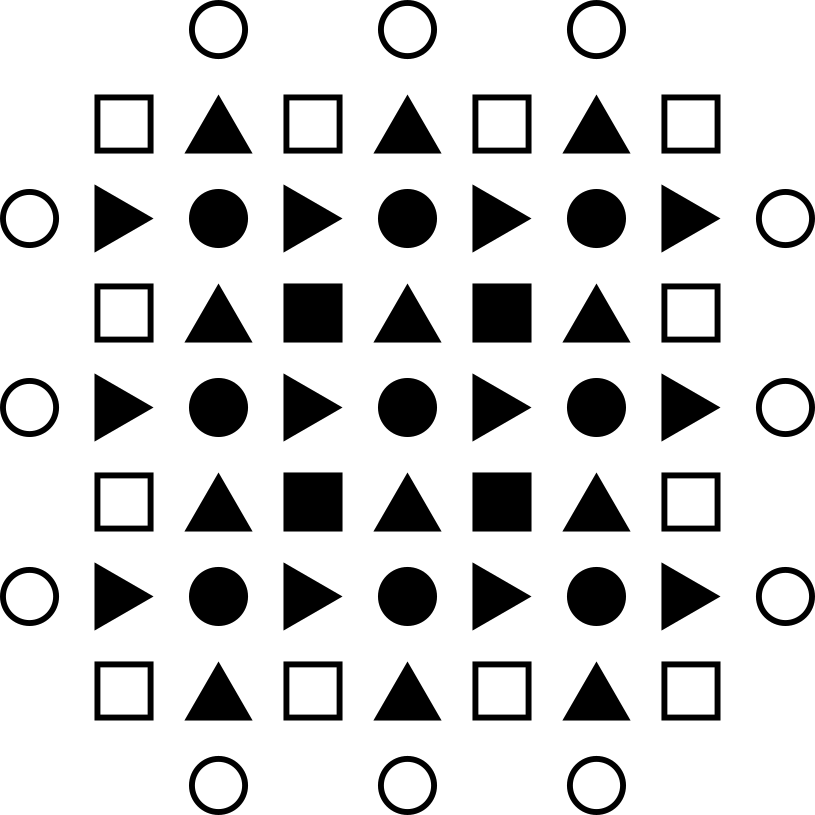}
 \caption{\qhl{Spatial} 
 arrangement of the discretized quantities \1 1
{two-dimensional projection}. The~circles stand for diagonal tensor components,
squares for offdiagonal ones, and triangles for vector components, different
components with differently oriented triangles. Void quantities are
prescribed by boundary condition \1 1 {in the case stress boundary conditions are
considered, like here.}} \label{FIG2}
 \end{figure}

We take a Cartesian grid with spacings \m{\Delta x}, \m{\Delta y}, \m{\Delta
z}, and time step \m { \Delta t }. Corresponding to the continuum formula
\re{A}, we introduce the finite difference discretization
  \Par
 \begin{equation}  \label{fdvx}
\begin{array}{ll}
\qrho \f { \9 1 {\qv_{\qx}}^{\qjp}_{\qlp,\qm,\qn} -
\9 1 {\qv_{\qx}}^{\qjm}_{\qlp,\qm,\qn} }{\Delta t}
= \f { \9 1 { \qsig_{\qxx} }^{\qj}_{\ql+1,\qm,\qn} - \9 1 { \qsig_{\qxx}
}^{\qj}_{\ql,\qm,\qn} }{\Delta x}
& + \f { \9 1 { \qsig_{\qxy} }^{\qj}_{\qlp,\qmp,\qn} - \9 1 { \qsig_{\qxy}
}^{\qj}_{\qlp,\qmm,\qn} }{\Delta y}
 \\
& + \f { \9 1 { \qsig_{\qxz} }^{\qj}_{\qlp,\qm,\qnp} - \9 1 { \qsig_{\qxz}
}^{\qj}_{\qlp,\qm,\qnm} }{\Delta z} ,
\end{array}
 \end{equation}
  \Par
\begingroup\makeatletter\def\f@size{9}\check@mathfonts
\def\maketag@@@#1{\hbox{\m@th\normalsize\normalfont#1}}%
 \begin{equation}  \label{fdvy}
\begin{array}{ll}
\qrho \f { \9 1 {\qv_{\qy}}^{\qjp}_{\ql,\qmp,\qn} -
\9 1 {\qv_{\qy}}^{\qjm}_{\ql,\qmp,\qn} }{\Delta t}
& = \f { \9 1 { \qsig_{\qyx} }^{\qj}_{\qlp,\qmp,\qn} - \9 1 { \qsig_{\qyx}
}^{\qj}_{\qlm,\qmp,\qn} }{\Delta x}
+ \f { \9 1 { \qsig_{\qyy} }^{\qj}_{\ql,\qm+1,\qn} - \9 1 { \qsig_{\qyy}
}^{\qj}_{\ql,\qm,\qn} }{\Delta y}
  \\
& + \f { \9 1 { \qsig_{\qyz} }^{\qj}_{\ql,\qmp,\qnp} - \9 1 { \qsig_{\qyz}
}^{\qj}_{\ql,\qmp,\qnm} }{\Delta z} ,
\end{array}
 \end{equation}
\endgroup
  \Par
 \begin{equation}  \label{fdvz}
\begin{array}{ll}
\qrho \f { \9 1 {\qv_{\qz}}^{\qjp}_{\ql,\qm,\qnp} -
\9 1 {\qv_{\qz}}^{\qjm}_{\ql,\qm,\qnp} }{\Delta t}
& = \f { \9 1 { \qsig_{\qzx} }^{\qj}_{\qlp,\qm,\qnp} - \9 1 { \qsig_{\qzx}
}^{\qj}_{\qlm,\qm,\qnp} }{\Delta x}
  \\
& + \f { \9 1 { \qsig_{\qzy} }^{\qj}_{\ql,\qmp,\qnp} - \9 1 { \qsig_{\qzy}
}^{\qj}_{\ql,\qmm,\qnp} }{\Delta y}
+ \f { \9 1 { \qsig_{\qzz} }^{\qj}_{\ql,\qm,\qn+1} - \9 1 { \qsig_{\qzz}
}^{\qj}_{\ql,\qm,\qn} }{\Delta z} ,
\end{array}
 \end{equation}
  \Par
where the time index \m { \qj } refers to a value at \m { \qt^{\qj} = \qj
\cdot \Delta t }, \m { \qjp } to a value at \m { \qt^{\qjp} = \9 1 {\qjp}
\cdot \Delta t }, the~space index \m { \ql } refers to a value at \m {
\qx_{\ql} = \ql \cdot \Delta x }, \m { \qm } is the space index in the \m
{\qy} direction, and \m { \qn } in the \m { \qz } direction. Accordingly,
stress \1 1 {and strain} values reside at integer time instants, while
the velocity ones are shifted in time by half; diagonal stress \1 1 {and strain}
components reside at integer positions, off-diagonal ones are shifted in the
two directions that match with the two Cartesian indices; and, velocity
components are only shifted in the direction matching with their Cartesian
index
\textcolor{\qred}{(see Figure~\ref{FIG2})}.
From these formulae, the \m { \qjp } indexed velocities can be
expressed explicitly \1 1 {as functions of earlier quantities}.

This same pattern---distribution of quantities---is used for the
discretization of \re{A}:
  \Par
 \begin{align}  \label{fdepsxx}
\f { \9 1 {\qeps_{\qxx}}^{\qj+1}_{\ql,\qm,\qn} - \9 1
{\qeps_{\qxx}}^{\qj}_{\ql,\qm,\qn} }{\Delta t}
 & =
\f { \9 1 {\qv_{\qx}}^{\qjp}_{\qlp,\qm,\qn} - \9 1
{\qv_{\qx}}^{\qjp}_{\qlm,\qm,\qn} }{\Delta x} ,
 \\  \label{fdepsyy}
\f { \9 1 {\qeps_{\qyy}}^{\qj+1}_{\ql,\qm,\qn} - \9 1
{\qeps_{\qyy}}^{\qj}_{\ql,\qm,\qn} }{\Delta t}
 & =
\f { \9 1 {\qv_{\qy}}^{\qjp}_{\ql,\qmp,\qn} - \9 1
{\qv_{\qy}}^{\qjp}_{\ql,\qmm,\qn} }{\Delta y} ,
 \\  \label{fdepszz}
\f { \9 1 {\qeps_{\qzz}}^{\qj+1}_{\ql,\qm,\qn} - \9 1
{\qeps_{\qzz}}^{\qj}_{\ql,\qm,\qn} }{\Delta t}
 & =
\f { \9 1 {\qv_{\qz}}^{\qjp}_{\ql,\qm,\qnp} - \9 1
{\qv_{\qz}}^{\qjp}_{\ql,\qm,\qnm} }{\Delta z} ,
  \end{align}
  \Par
 \begin{equation}  \label{fdepsxy}
\begin{array}{ll}
& \f { \9 1 {\qeps_{\qxy}}^{\qj+1}_{\qlp,\qmp,\qn} - \9 1
{\qeps_{\qxy}}^{\qj}_{\qlp,\qmp,\qn} }{\Delta t}
  \\
 & \hspace{7em} =
\f {1}{2} \9 3 { \f { \9 1 {\qv_{\qx}}^{\qjp}_{\qlp,\qm+1,\qn} - \9 1
{\qv_{\qx}}^{\qjp}_{\qlp,\qm,\qn} }{\Delta y} + \f { \9 1
{\qv_{\qy}}^{\qjp}_{\ql+1,\qmp,\qn} - \9 1 {\qv_{\qy}}^{\qjp}_{\ql,\qmp,\qn}
}{\Delta x} } ,
 \\ 
\end{array}
 \end{equation}
 \begin{equation}
\begin{array}{ll} \label{fdepsxz}
& \f { \9 1 {\qeps_{\qxz}}^{\qj+1}_{\qlp,\qm,\qnp} - \9 1
{\qeps_{\qxz}}^{\qj}_{\qlp,\qm,\qnp} }{\Delta t}
 \\
 & \hspace{7em} =
\f {1}{2} \9 3 { \f { \9 1 {\qv_{\qx}}^{\qjp}_{\qlp,\qm,\qn+1} - \9 1
{\qv_{\qx}}^{\qjp}_{\qlp,\qm,\qn} }{\Delta z} + \f { \9 1
{\qv_{\qz}}^{\qjp}_{\ql+1,\qm,\qnp} - \9 1 {\qv_{\qz}}^{\qjp}_{\ql,\qm,\qnp}
}{\Delta x} } ,
\\ 
 \end{array}
\end{equation}
 \begin{equation}
\begin{array}{ll} \label{fdepsyz}
& \f { \9 1 {\qeps_{\qyz}}^{\qj+1}_{\ql,\qmp,\qnp} - \9 1
{\qeps_{\qyz}}^{\qj}_{\ql,\qmp,\qnp} }{\Delta t}
  \\
 & \hspace{7em} =
\f {1}{2} \9 3 { \f { \9 1 {\qv_{\qy}}^{\qjp}_{\ql,\qmp,\qn+1} - \9 1
{\qv_{\qy}}^{\qjp}_{\ql,\qmp,\qn} }{\Delta z} + \f { \9 1
{\qv_{\qz}}^{\qjp}_{\ql,\qm+1,\qnp} - \9 1 {\qv_{\qz}}^{\qjp}_{\ql,\qm,\qnp}
}{\Delta y} } ,
\end{array}
 \end{equation}
  \Par
from which formulae the \m { \qj+1 } indexed strains can be explicitly expressed, 
and for the discretized version of \re{PTZ},
  \Par
\begingroup\makeatletter\def\f@size{9}\check@mathfonts
\def\maketag@@@#1{\hbox{\m@th\normalsize\normalfont#1}}%
\begin{equation}
 \begin{array}{ll}  \label{fdPTZdev}
\qalp \9 1 {\dev\qsig_{\qpq}}^{\qj}_{\qlmn} & \mathrel+ \1 1 {1 - \qalp} \9 1
{\dev\qsig_{\qpq}}^{\qj+1}_{\qlmn} + \dev\qtau \f { \9 1
{\dev\qsig_{\qpq}}^{\qj+1}_{\qlmn} - \9 1 {\dev\qsig_{\qpq}}^{\qj}_{\qlmn}
}{\Delta t}
  \\
& = \dev\qEY \9 2 { \qalp \9 1 {\dev\qeps_{\qpq}}^{\qj}_{\qlmn} + \1 1 {1 -
\qalp} \9 1 {\dev\qeps_{\qpq}}^{\qj+1}_{\qlmn} } + \dev{\qEE} \f { \9 1
{\dev\qeps_{\qpq}}^{\qj+1}_{\qlmn} - \9 1 {\dev\qeps_{\qpq}}^{\qj}_{\qlmn}
}{\Delta t} ,
 \\
\end{array}
 \end{equation}
\begin{equation}
 \begin{array}{ll}
 \label{fdPTZsph}
\qalp \9 1 {\sph\qsig_{\qpq}}^{\qj}_{\qlmn} & \mathrel+ \1 1 {1 - \qalp} \9 1
{\sph\qsig_{\qpq}}^{\qj+1}_{\qlmn} + \sph\qtau \f { \9 1
{\sph\qsig_{\qpq}}^{\qj+1}_{\qlmn} - \9 1 {\sph\qsig_{\qpq}}^{\qj}_{\qlmn}
}{\Delta t}
  \\
& = \sph\qEY \9 2 { \qalp \9 1 {\sph\qeps_{\qpq}}^{\qj}_{\qlmn} + \1 1 {1 -
\qalp} \9 1 {\sph\qeps_{\qpq}}^{\qj+1}_{\qlmn} } + \sph{\qEE} \f { \9 1
{\sph\qeps_{\qpq}}^{\qj+1}_{\qlmn} - \9 1 {\sph\qeps_{\qpq}}^{\qj}_{\qlmn}
}{\Delta t} ,
 \\ 
& \1 0 {\qp, \qq = \qx, \qy, \qz} , \quad \qlmn = \text{integers or
half-integers depending on } \qp, \qq,
 \end{array}
\end{equation}
\endgroup
  \Par
where \m { \qalp = 1/2 } ensures second-order accuracy of the whole scheme \1
1 {the proof is analogous to the one in~\cite{1Dcikk}}, from which---
together with \re{devdecomp}---the \m { \qj+1 } indexed stresses can be explicitly expressed, except~for stress boundary locations, where we express
strain \1 1 {and know stress from the boundary~condition}.

Actually, regarding boundary conditions, the rule we found for both stress
boundary condition and velocity \1 1 {or displacement} boundary condition is
that, if a quantity is missing for determining another boundary quantity, then
that missing quantity is to be added outside the boundary. This
also works for mixed boundary conditions, with different ones meeting at
edges of a rectangular sample, for example. In what follows, we present
stress boundary condition examples \1 1 {relevant, e.g., for a wide class of
rock mechanical applications}.

The pattern of which quantity to reside where---at integer or half-integer
space and time indices---could also be conjectured from the structure of the
equations, but, as said in the Introduction, the spacetime viewpoint helps a
lot to find this arrangement a geometrically---spacetime geometrically---
natural one.

In the reversible special case of the Hooke system, this scheme is
symplectic. It is actually the symplectic Euler method \1 1 {in words:
\quot{new\m{_1} from old\m{_1} and old\m{_2}, new\m{_2} from new\m{_1} and
old\m{_2}}}. The~interpretation is the improvement: here, new\m{_1} and
new\m{_2} are shifted in time with respect to each other, so second-order
accuracy is achieved, while the conventional interpretation of the symplectic
Euler method is first-order only. In parallel, since mechanical energy \1 1
{the Hamiltonian} is a velocity dependent term plus a strain dependent term
\1 1 {stress becomes a simple linear function of strain}, our~scheme is
explicit. This also remains true at the PTZ level, so one can expect---and
find, actually---a~fast-running program code.

For the aspects of thermodynamics, we also discretize \re{Tdot} \1 2 {here
using the form \re{entprodB}}, explicitly expressing the \m { \qjp } indexed
temperature values from
  \Par
 \begin{align}  \label{fdTdot}
  \f { \qT^{\qjp}_{\ql,\qm,\qn} - \qT^{\qjm}_{\ql,\qm,\qn} }{\Delta t} =
\f {1}{\qrho \qcp} \9 3 { \f {1}{\dev\qII} \tr \9 1 { \Qsighatdevhack^2
}^{\qj}_{\ql,\qm,\qn} +  \f{1}{\sph\qII} \tr \9 1 { \Qsighatsphhack^2
}^{\qj}_{\ql,\qm,\qn} } ,
 \end{align}
  \Par
where the notation \re{sighat} is utilized, and the traces are to be expanded
in Cartesian components and the terms containing offdiagonal components---that reside at half space-shifted locations in two indices---are averaged
around the location \m { \ql,\qm,\qn }, first neighbours only.

\section{Solution for a Rectangular Beam, and the Role of Total Energy}

\begin{figure}[H]
\centering
 \includegraphics[width=.5\columnwidth]{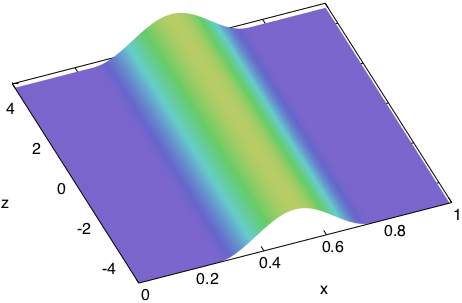}
\caption{The spatial distribution of normal stress as the boundary condition
on one side of a square cross-sectioned beam that is infinitely long in the
\m{z} direction. The excitation is a single cosine-shaped \quot{bump} in
time, too.}  \label{2dbump}
 \end{figure}

The first example on which we demonstrate the scheme is a square
cross-sectioned long beam, being thus treated as a plane-strain problem. Initially,
the beam is in relaxed/equilibrium state \1 1 {zero~stress, strain and
velocity, and homogeneous temperature}. Subsequently, on one of
its sides, a~single normal stress pulse is applied, with profile
  \Par
 \begin{align}  \label{cos}
\qsig_{\qyy} \0 1 { \qt, \qx, 0, \qz } =
 \begin{cases}
\qsigb \9 3 {\f {1}{2} \9 2 { 1 - \cos \0 1 { 2 \pi \f {\qt}{\qtaub} } }
 \cdot
\f {1}{2} \9 2 { 1 - \cos \0 1 {2 \pi \f {\qx - \qX/2 }{\qWb} } } }
& \text{if } \:
\1 0 { 0 \le \qt \le \qtaub \: \text{ and}
 \\ &
\f{\qX-\qWb}{2} \le \qx \le \f{\qX+\qWb}{2} } ,
 \\
0
& \1 0 {\text{otherwise}}
 \end{cases}
 \end{align}
  \Par
\1 1 {see Figure~\ref{2dbump}}, where \m { \qtaub } is the duration of the
pulse, \m { \qsigb } is its amplitude, \m { \qWb } is its spatial width, and
\m { \qX } is the width of the beam.
On the other sides, normal stress is constantly zero \1 1 {free surfaces}.

A \m { 50 \times 50 } grid is considered in the \m{x}--\m{y} plane, for 240
time steps, where the time step is the largest at which stability is
maintained. Notably, stability investigation is fairly involved for this
problem and
\textcolor{\qred}{it}
requires a separate whole study.
 Further settings \11{in appropriate units} are \m{\qrho=1}, \m{\dev\qEY=4},
\m{\sph\qEY=10}, \m{\qtaub=0.3}, \m{\qsigb=5}, \m { \qWb/\qX=0.6}.

\subsection{Hooke Case}

Figures \ref{Hs} and \ref{Hv} show snapshots of a stress component distribution and a velocity component.

 \begin{figure}[H]
\centering
\includegraphics[width=\qwidth]{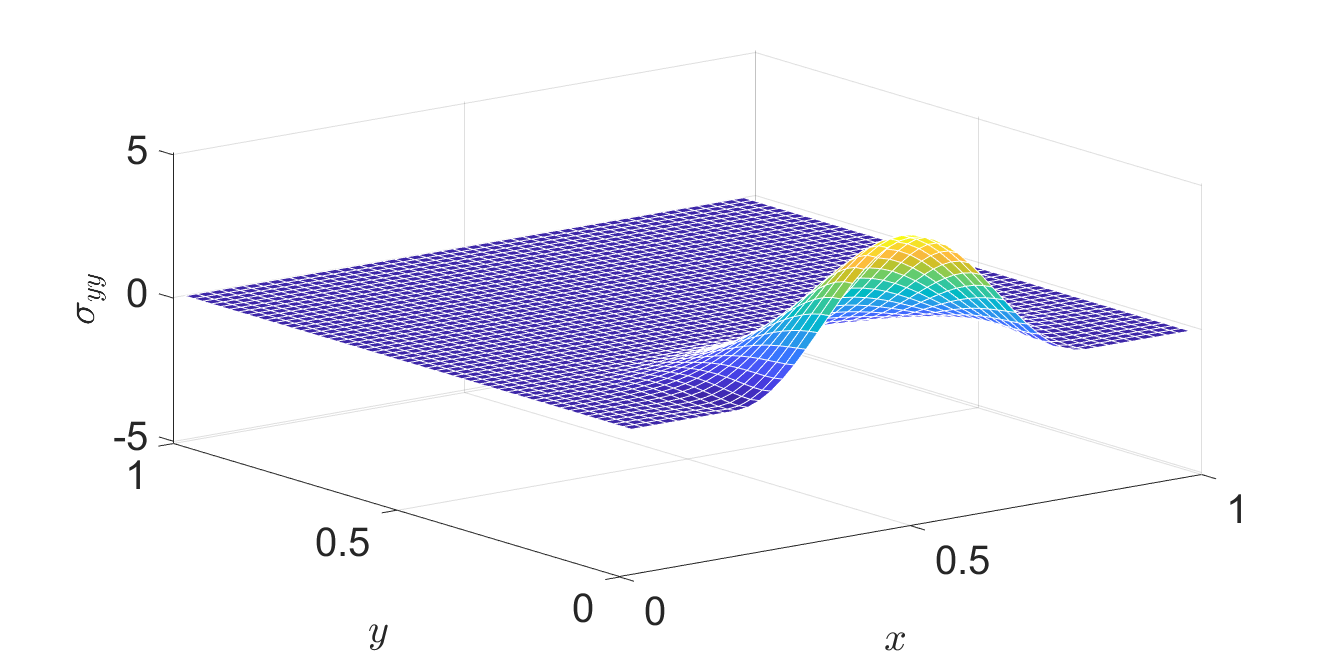}
 \hfill
\includegraphics[width=\qwidth]{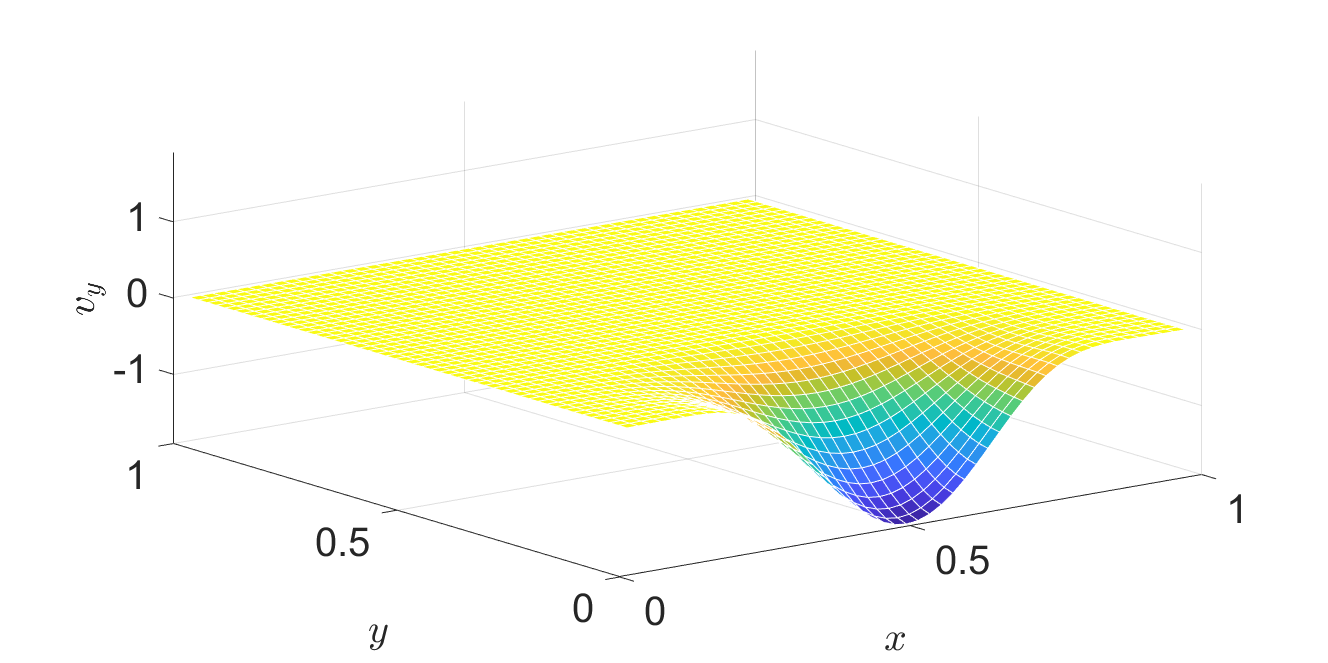}
 \\
\includegraphics[width=\qwidth]{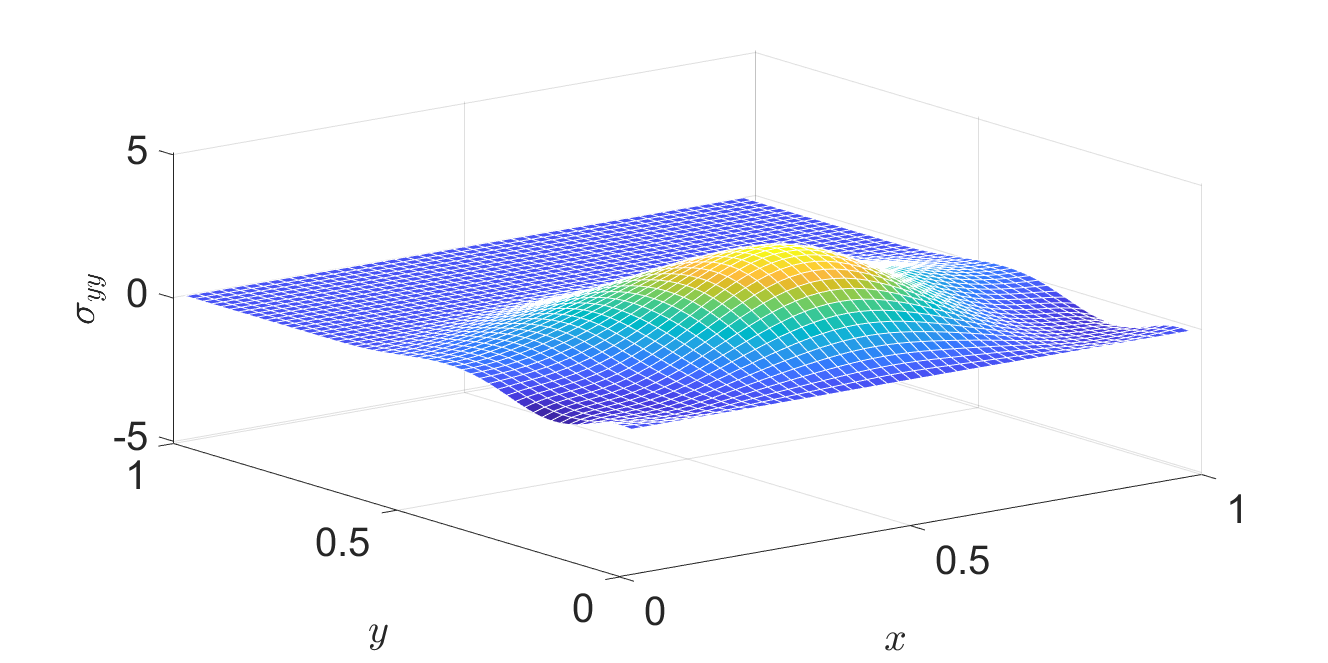}
 \hfill
\includegraphics[width=\qwidth]{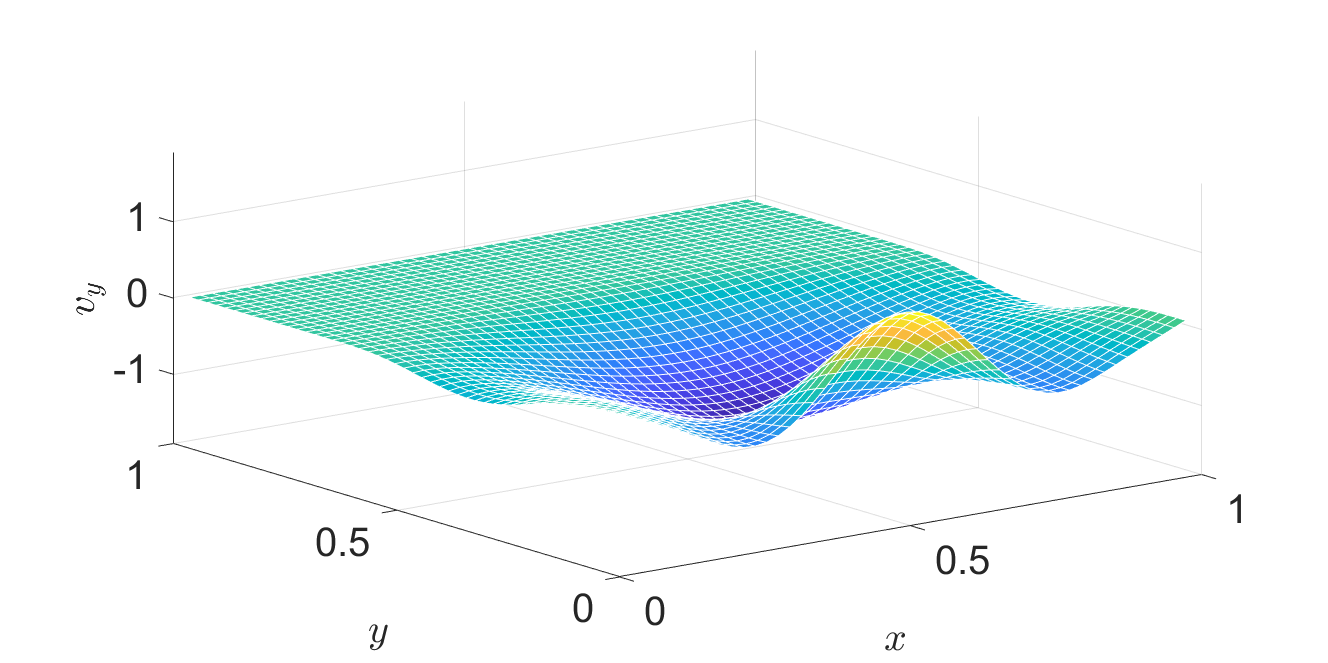}
 \\
\includegraphics[width=\qwidth]{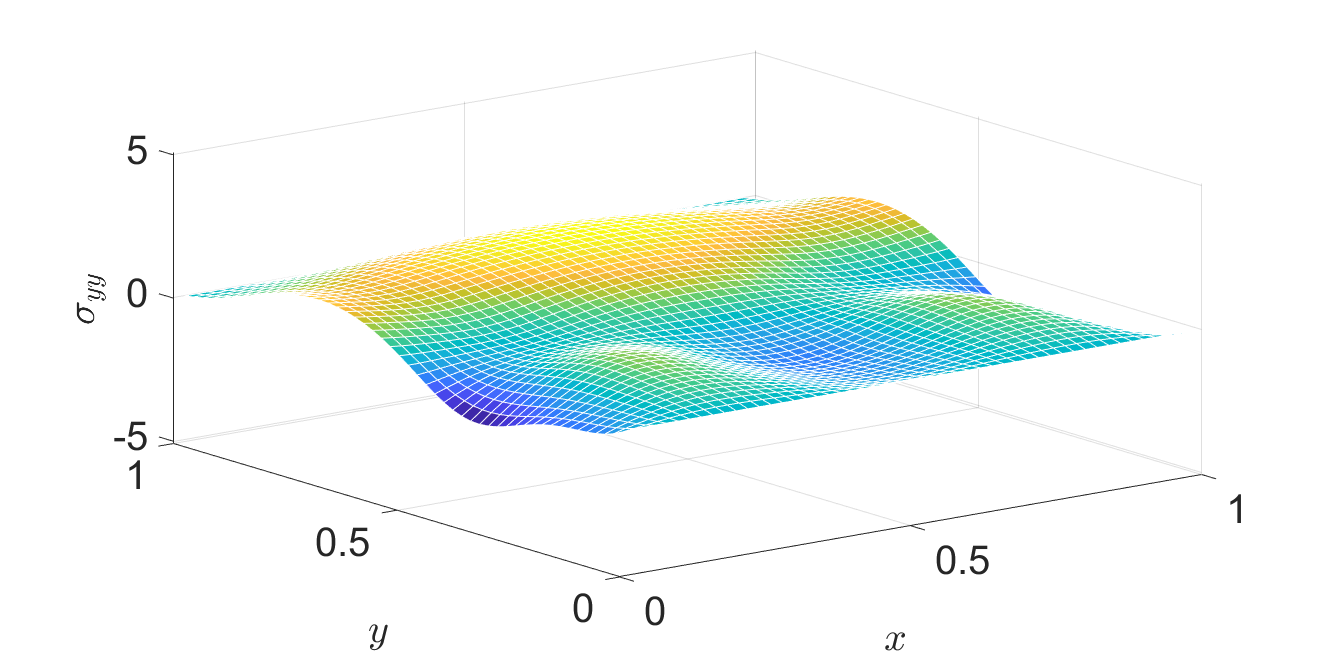}
 \hfill
\includegraphics[width=\qwidth]{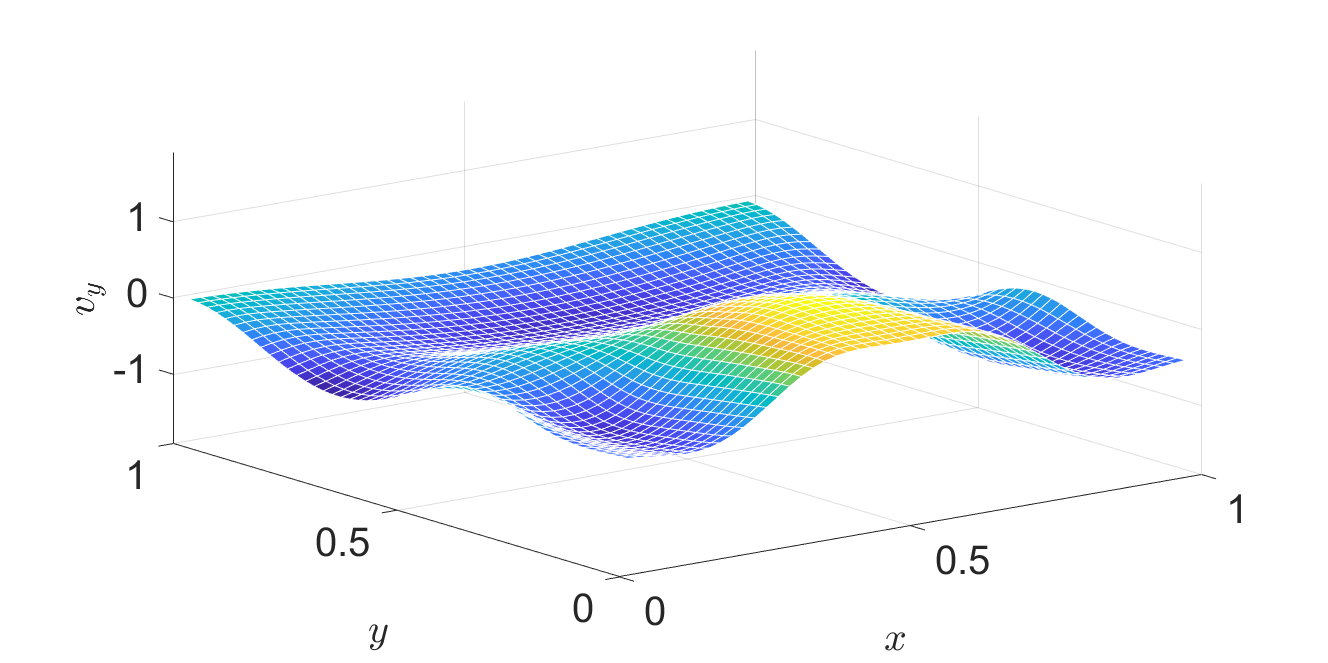}
 \\
\includegraphics[width=\qwidth]{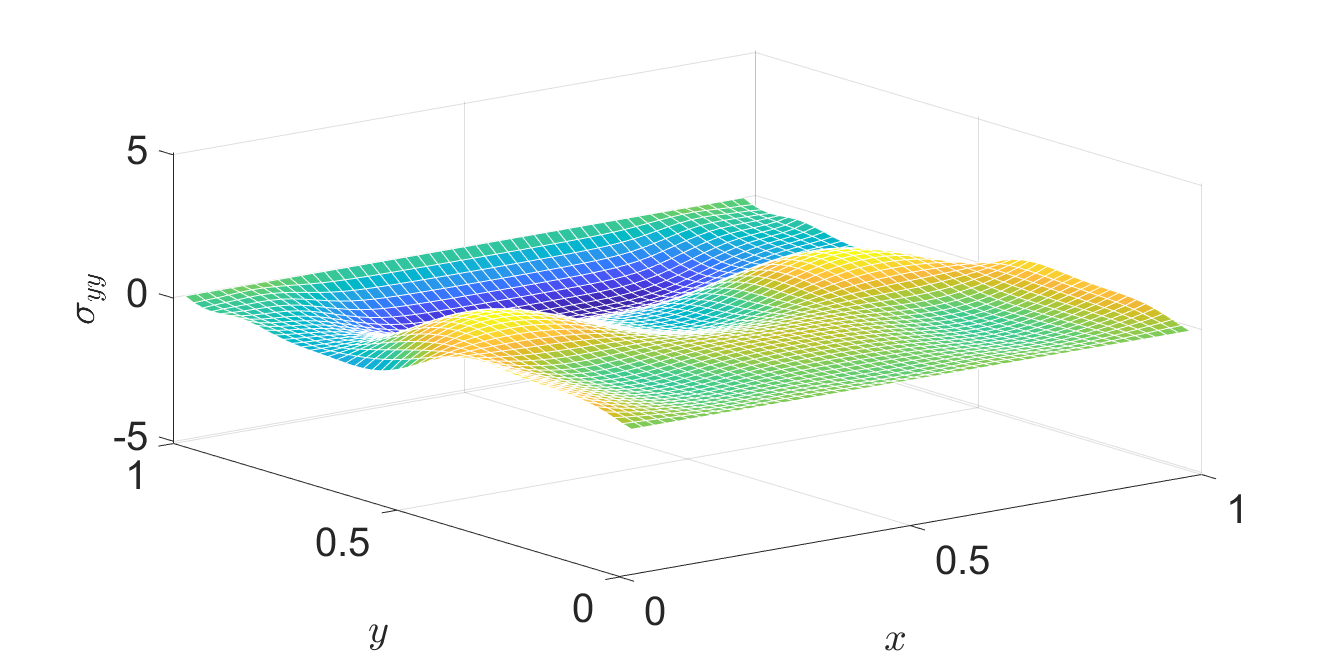}
 \hfill
\includegraphics[width=\qwidth]{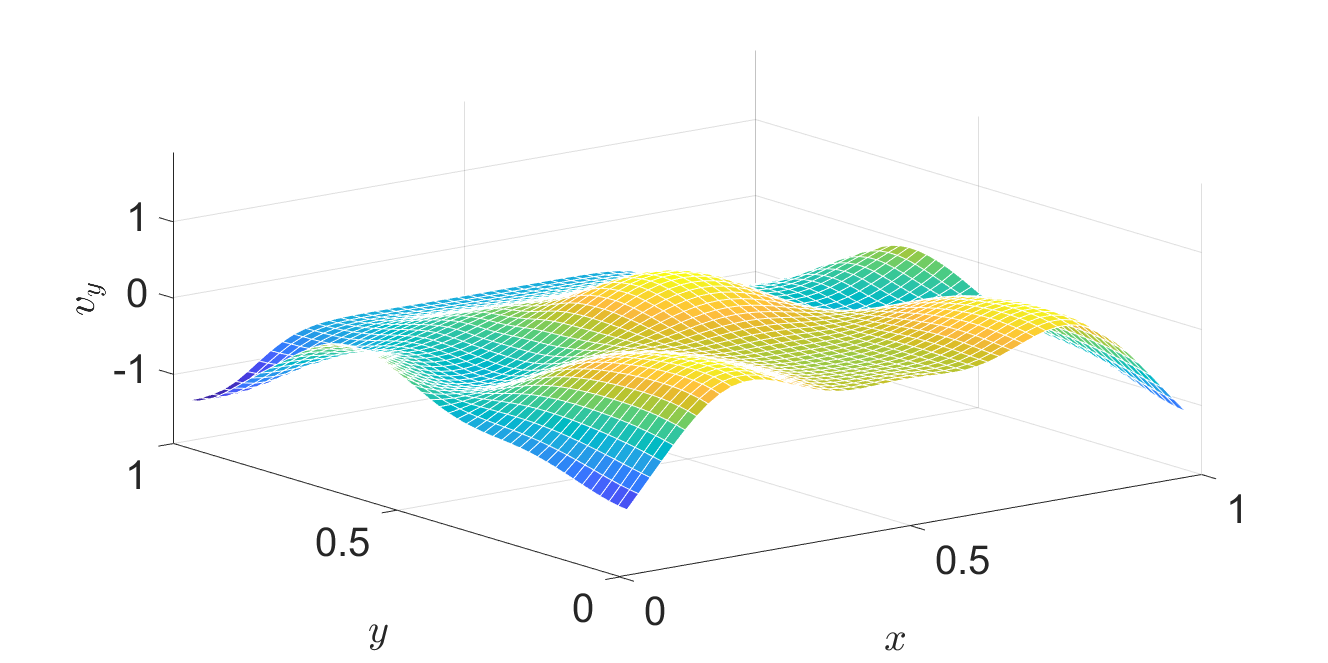}
 \caption{Distribution of a stress component \1 1 {left column} and of a
velocity component \1 1 {right column} at various instants, in the Hooke
case. From top to bottom: snapshots at instants \m { \F001{1}{2} \qtaub },
\m{\qtaub}, \m { \F001{3}{2} \qtaub }, \m { 2 \qtaub }, respectively.} 
\label{Hs}
 \end{figure}

In a movie format, it is more spectacular how reliably the simulation performs.

Furthermore, it is not only the eye that could judge the reliability: with
the help of thermodynamics, energy---in the Hooke case, mechanical energy---
proves to be a useful diagnostic~tool:

 \begin{itemize}[leftmargin=*,labelsep=5.8mm]
 \item if it explodes then there is instability;
 \item if it deviates from a constant then there is dissipation error; and,
 \item if it is wavy/oscillating then there is dispersion error.
 \end{itemize}

The scheme presented here also functions satisfactorily in this aspect, as
displayed in Figure~\ref{He}. For~energy, we perform summation, over the
integer centred discrete cells, of the energy terms discretized along the
above lines, including that averages, like for \re{fdTdot}, are taken wherever
necessary, also in the time direction \1 1 {for kinetic energy}.

 \begin{figure}[H]
\centering
\includegraphics[width=\qwidth]{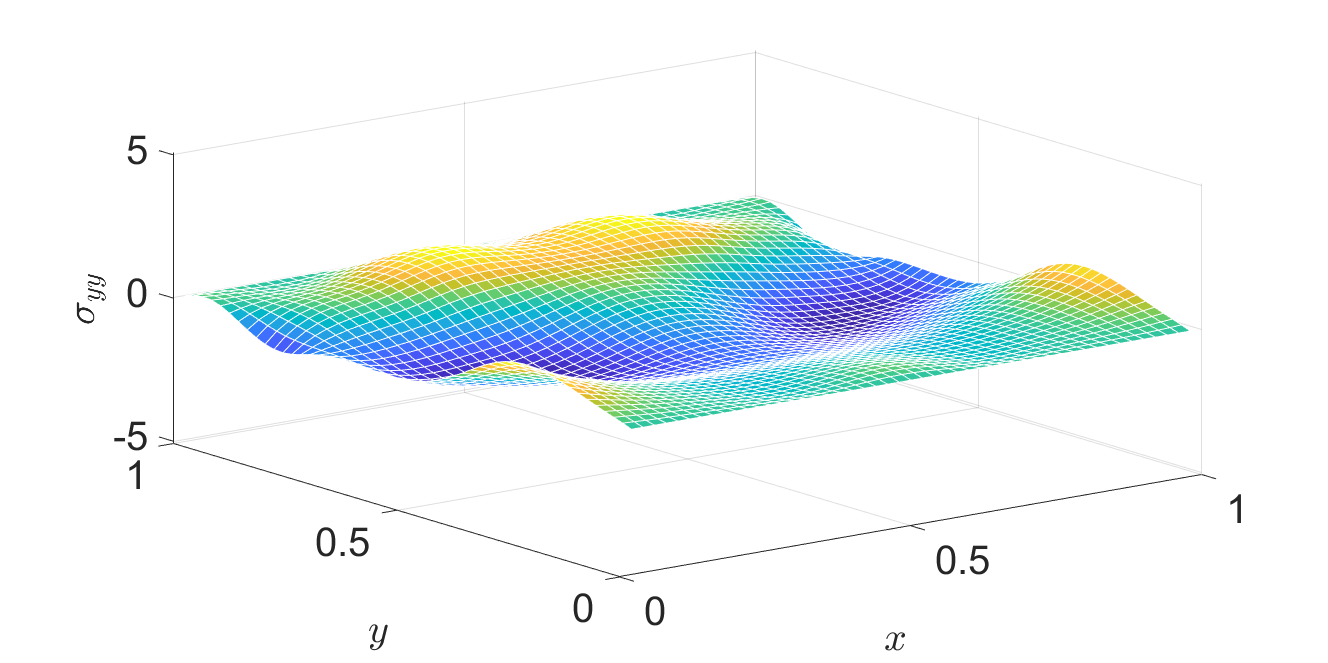}
 \hfill
\includegraphics[width=\qwidth]{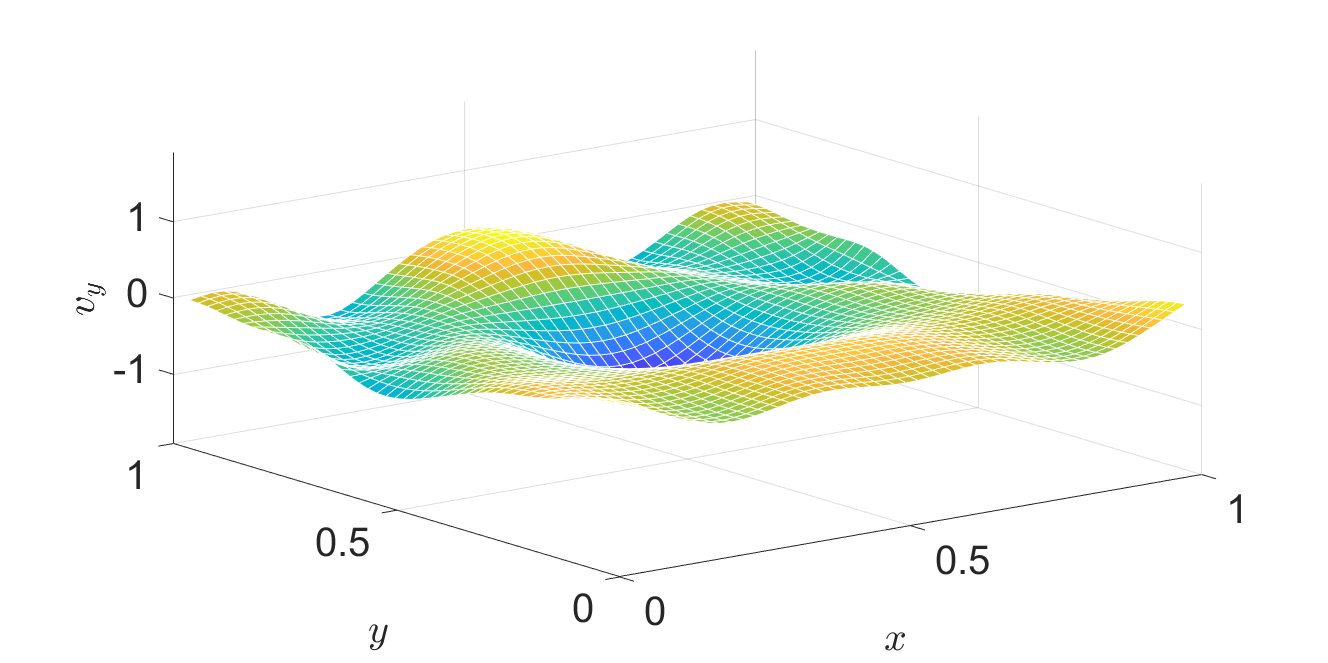}
 \\
\includegraphics[width=\qwidth]{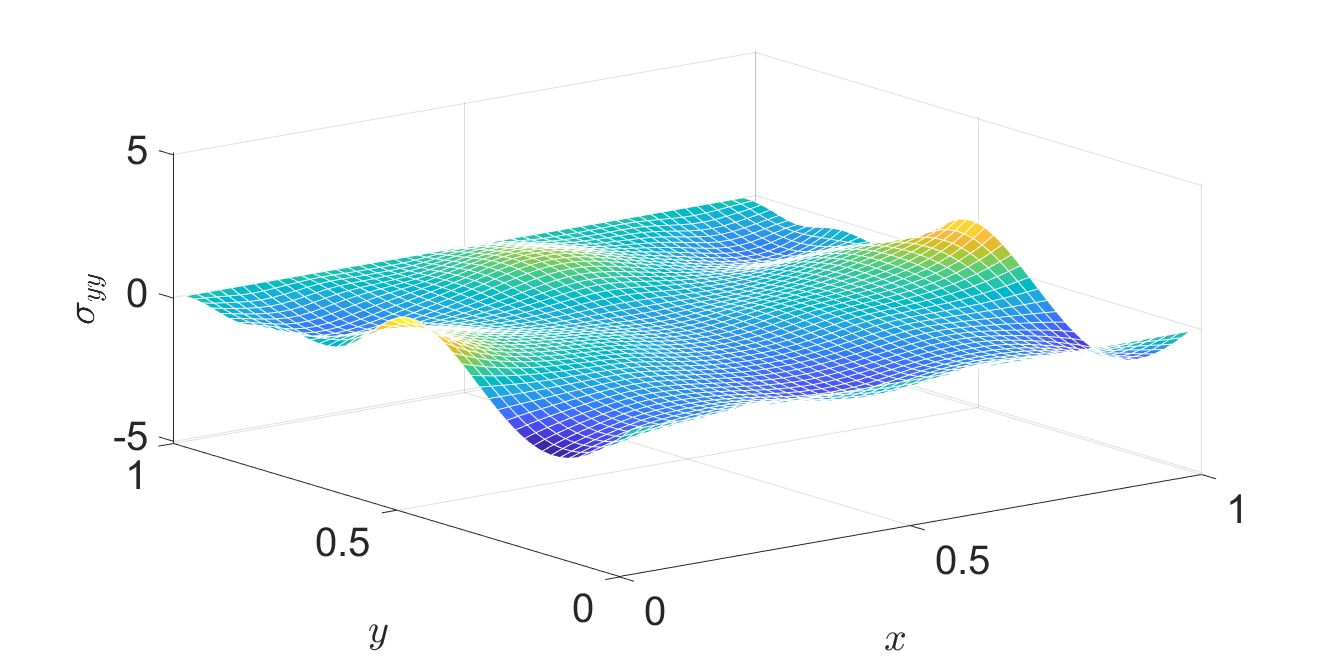}
 \hfill
\includegraphics[width=\qwidth]{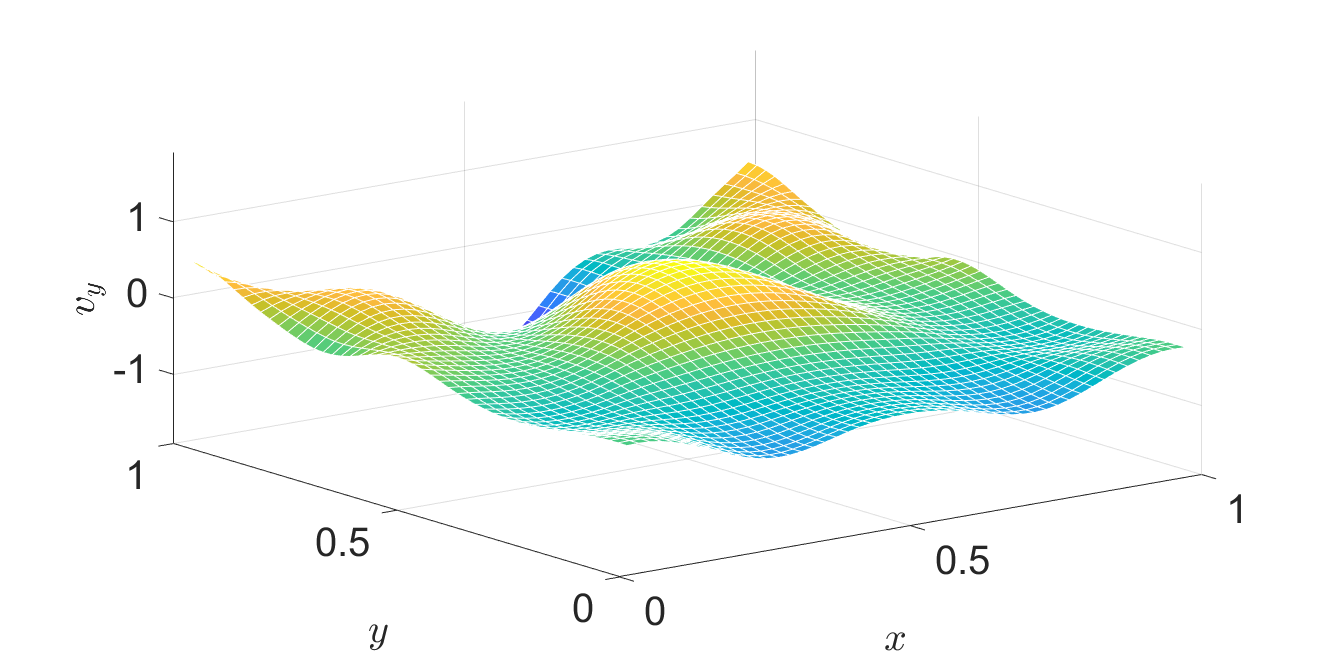}
 \\
\includegraphics[width=\qwidth]{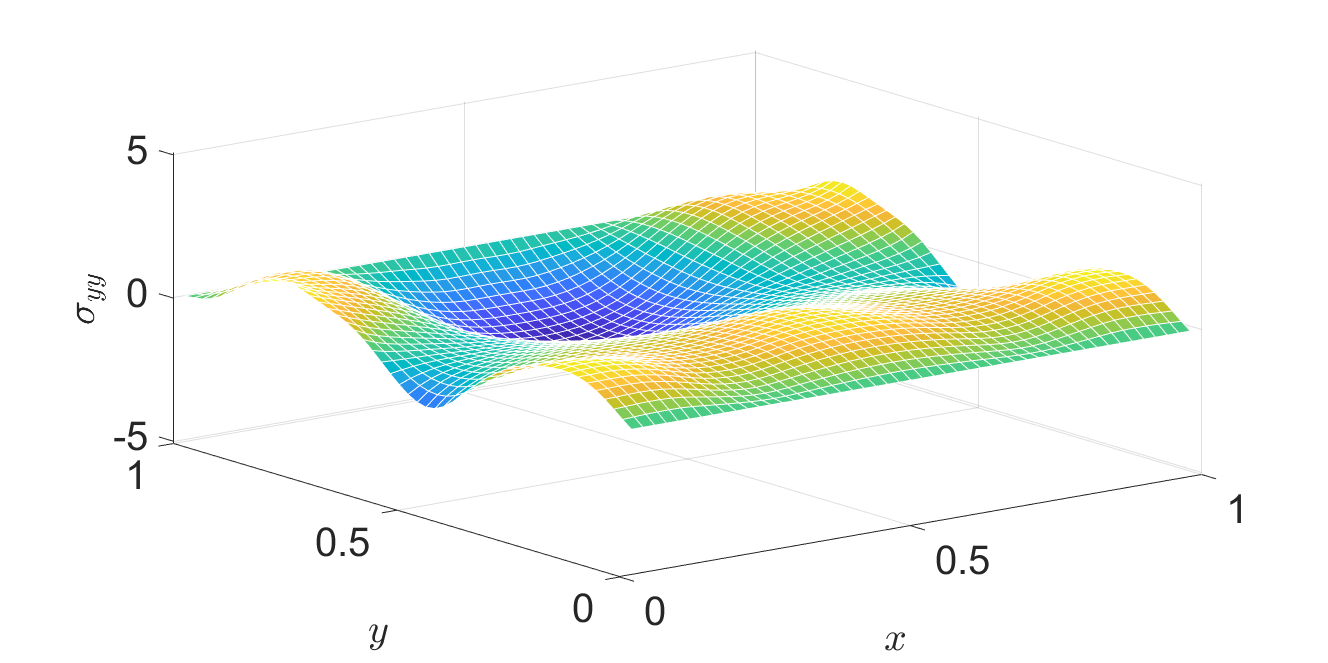}
 \hfill
\includegraphics[width=\qwidth]{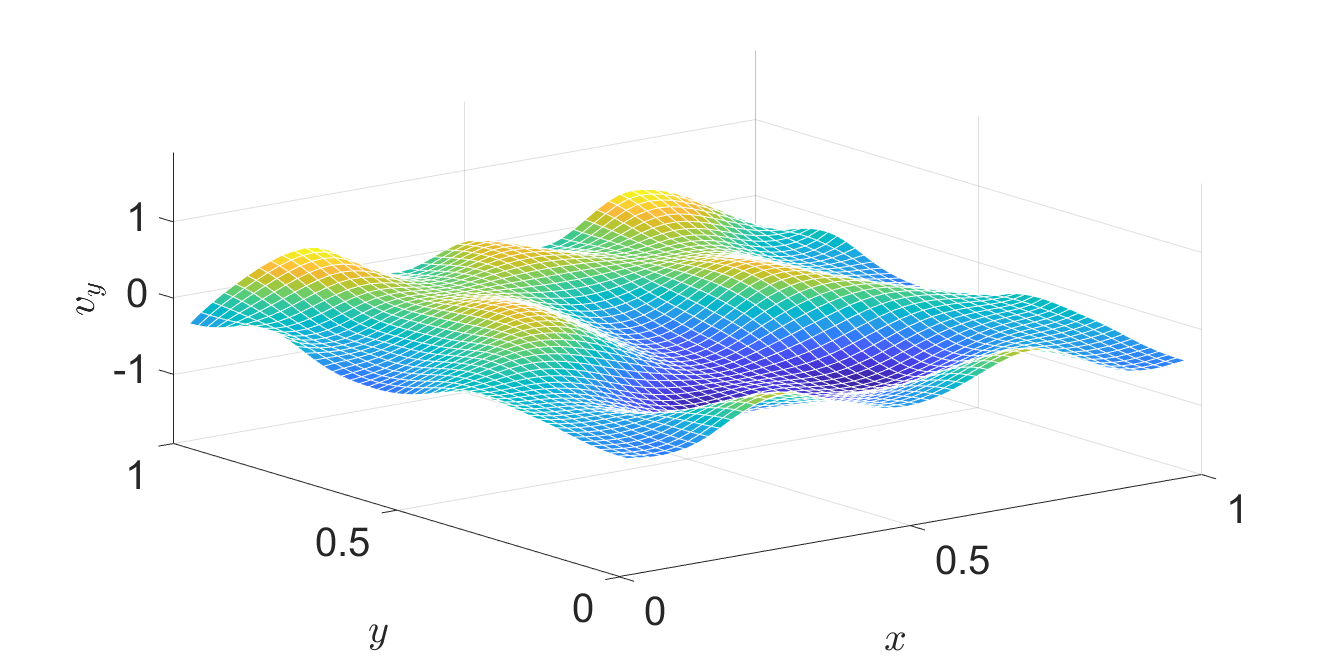}
 \\
\includegraphics[width=\qwidth]{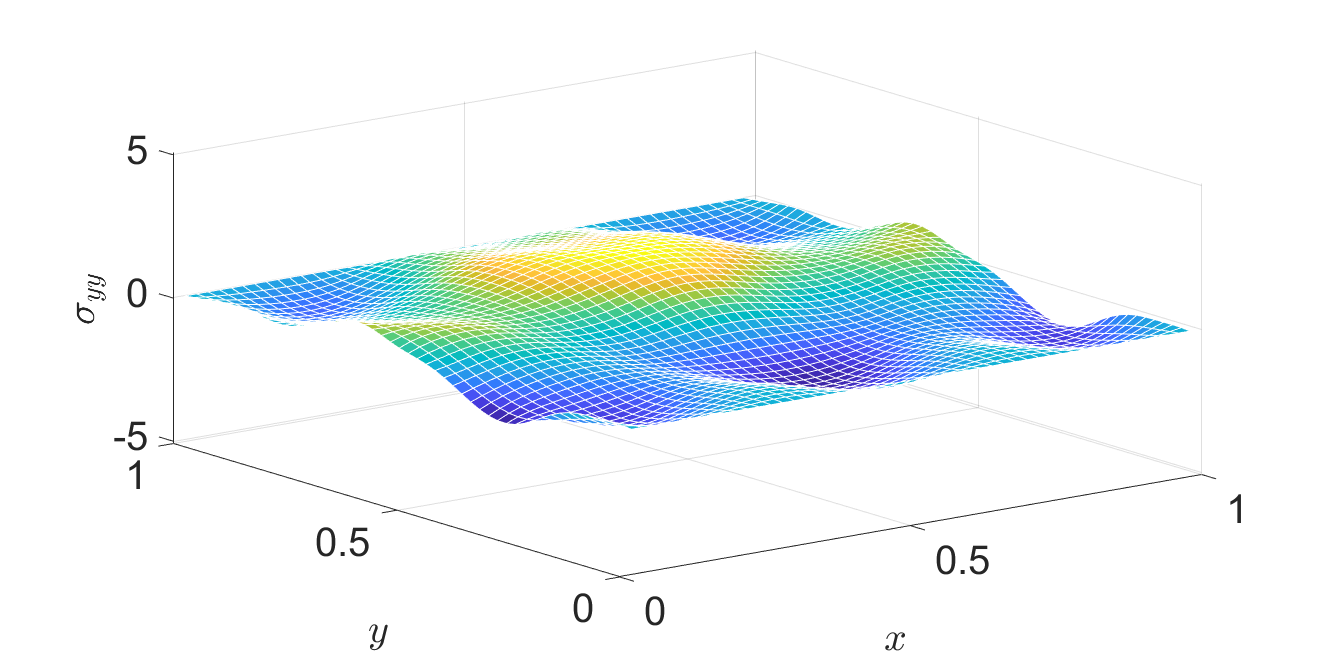}
 \hfill
\includegraphics[width=\qwidth]{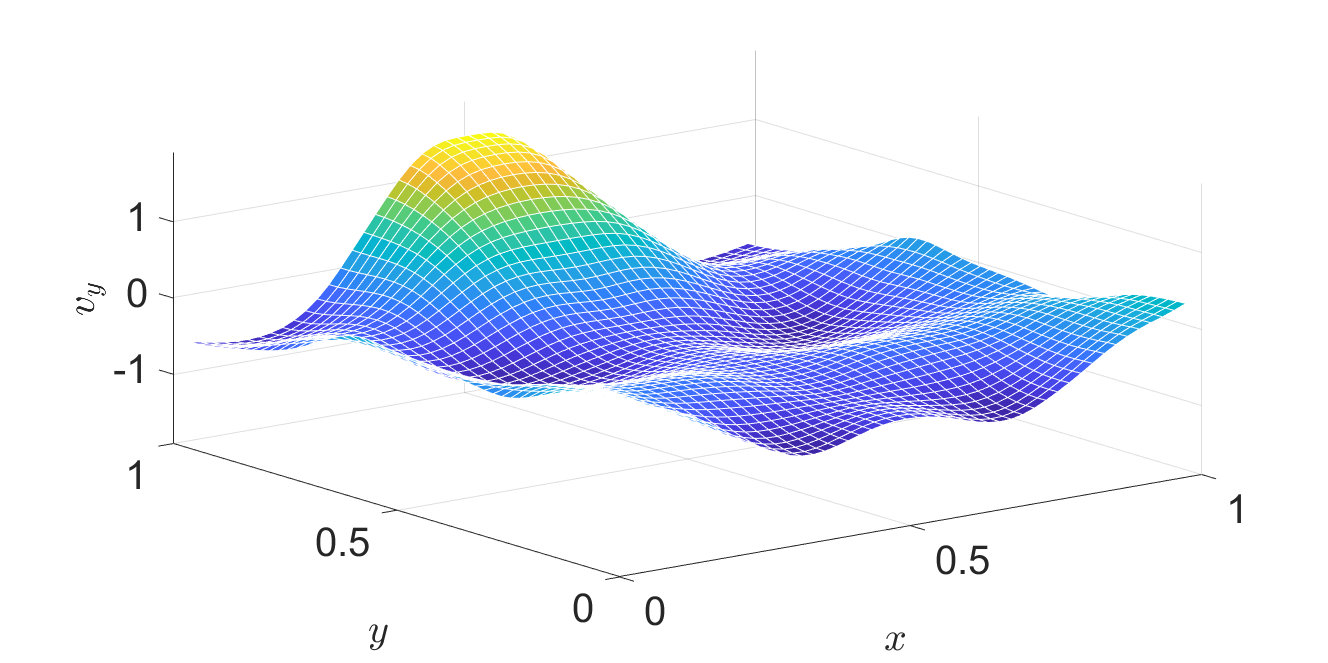}
 \caption{Continuation of Figure \ref{Hs}: distribution of a stress component
\1 1 {left column} and of a velocity component \1 1 {right column} at various
instants, in the Hooke case. From top to bottom: snapshots at instants \m {
\F001{5}{2} \qtaub }, \m{3 \qtaub}, \m { \F001{7}{2} \qtaub }, \m { 4 \qtaub
}, respectively.} 
  \label{Hv}
 \end{figure}
\unskip
\begin{figure}[H]
\centering
\includegraphics[width=0.75\columnwidth]{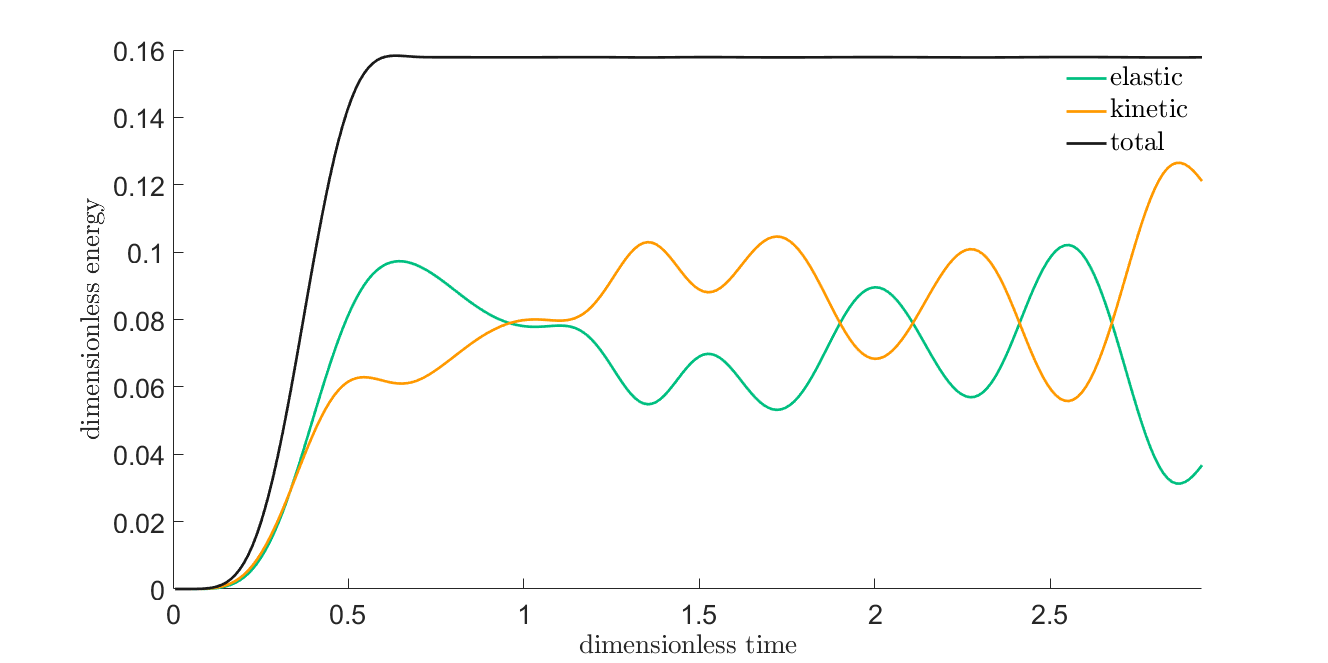}
 \caption{Mechanical energy types as functions of time, for the Hooke case.} 
 \label{He}
 \end{figure}

\subsection{PTZ Case}

In a PTZ medium, the solution of the analogous problem is similarly good.
Figures \ref{PTZs}--\ref{vm2} present snapshots, where Figures
\ref{vm1}--\ref{vm2} display two further quantities: \m { \sqrt{\tr \9 1 {
\Qsighatdevhack^2 }} } \1 1 {essentially the Huber--Mises--Hencky or von
Mises equivalent stress} and temperature. Dissipation is nicely indicated via
temperature.
 Settings \1 1 {in addition to the above ones for the Hooke case} are
\m{\dev\qEE=1.4}, \m{\dev\qtau=0.2}, \m{\qcp=0.001}, \m{T^{0}=0.1},
\m{\alpha=\iOii}.

 \begin{figure}[H]
\centering
\includegraphics[width=\qwidth]{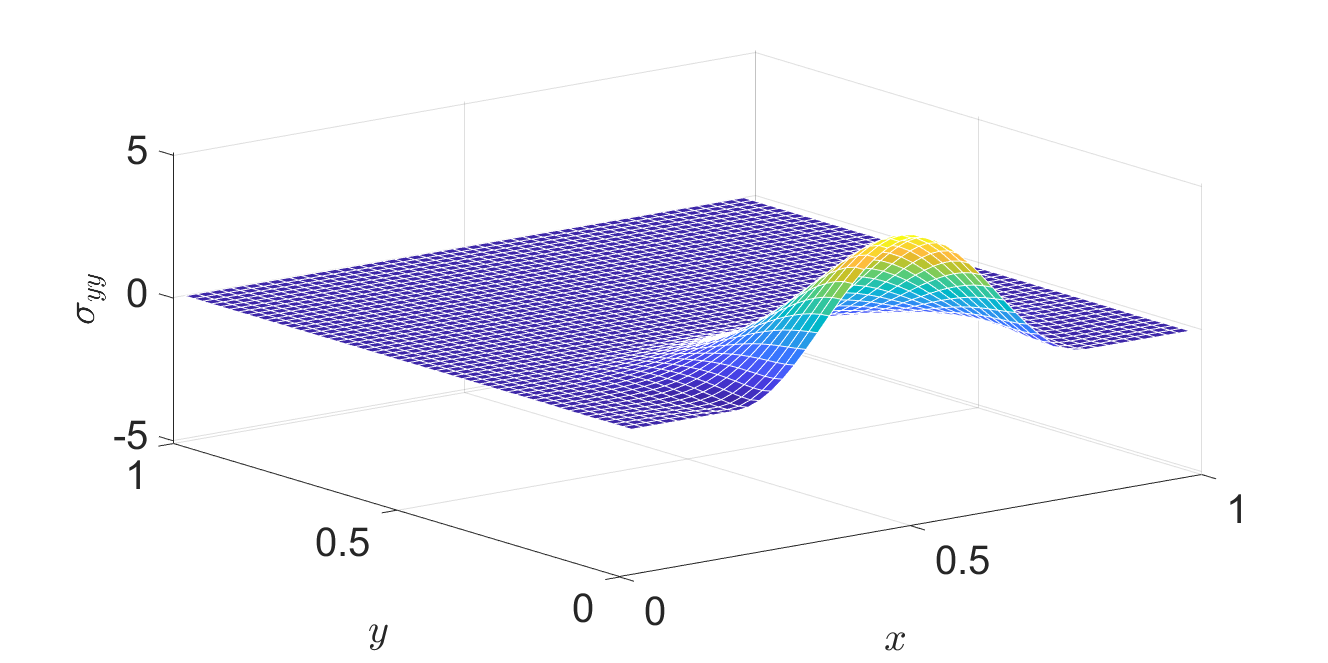}
 \hfill
\includegraphics[width=\qwidth]{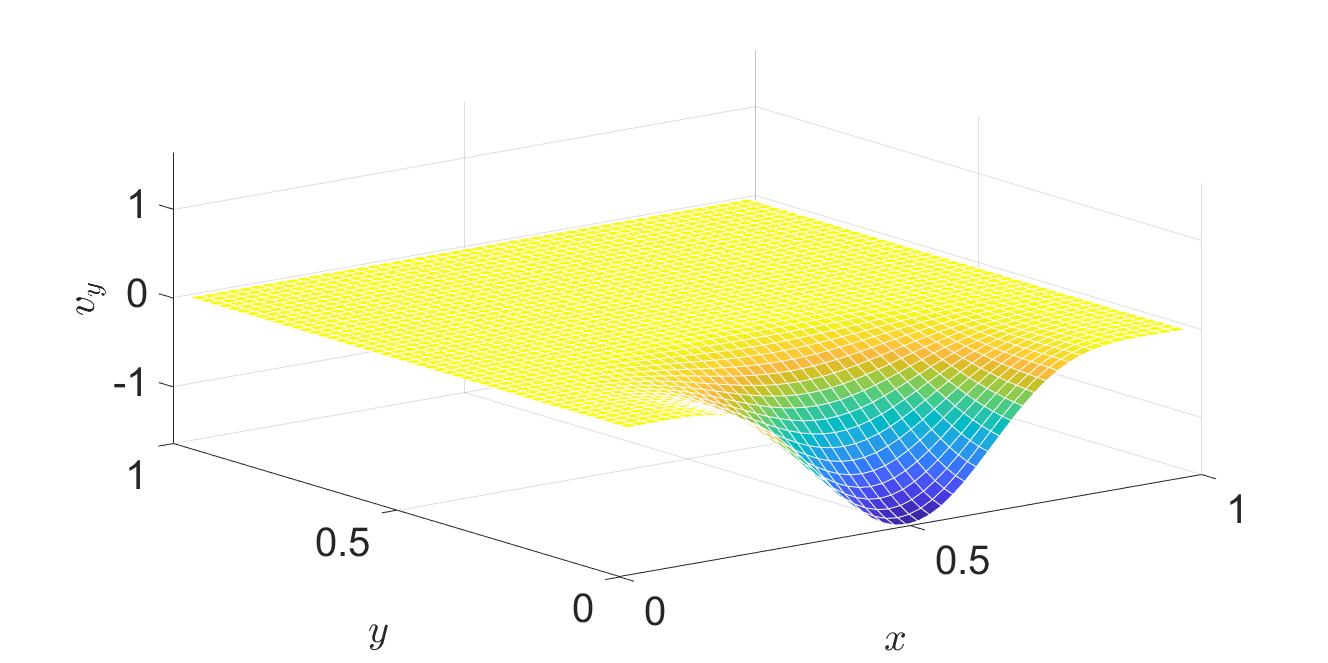}
 \\
\includegraphics[width=\qwidth]{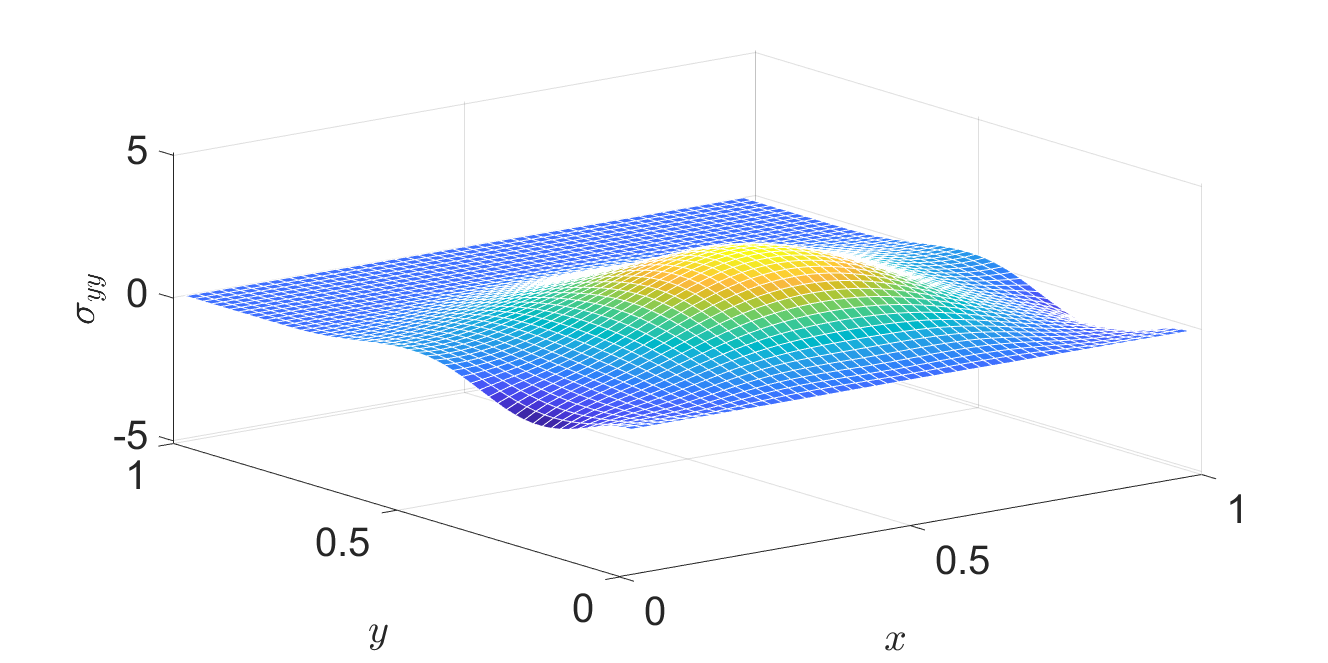}
 \hfill
\includegraphics[width=\qwidth]{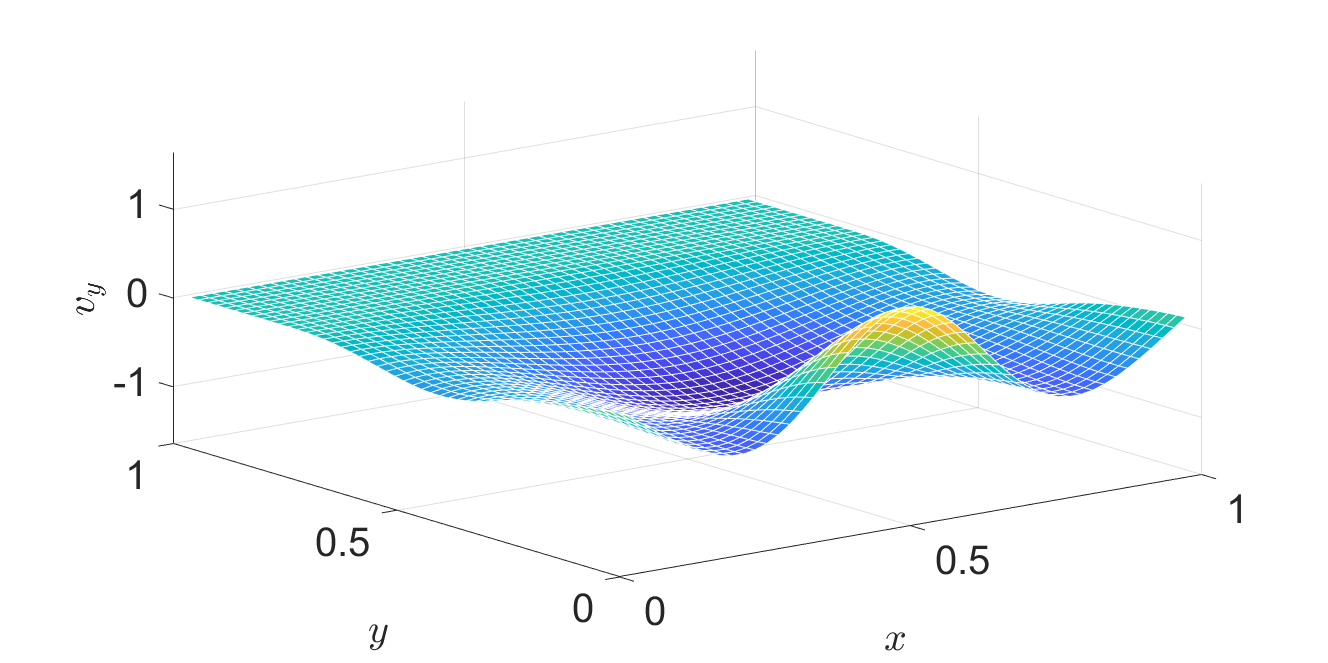}
 \\
\includegraphics[width=\qwidth]{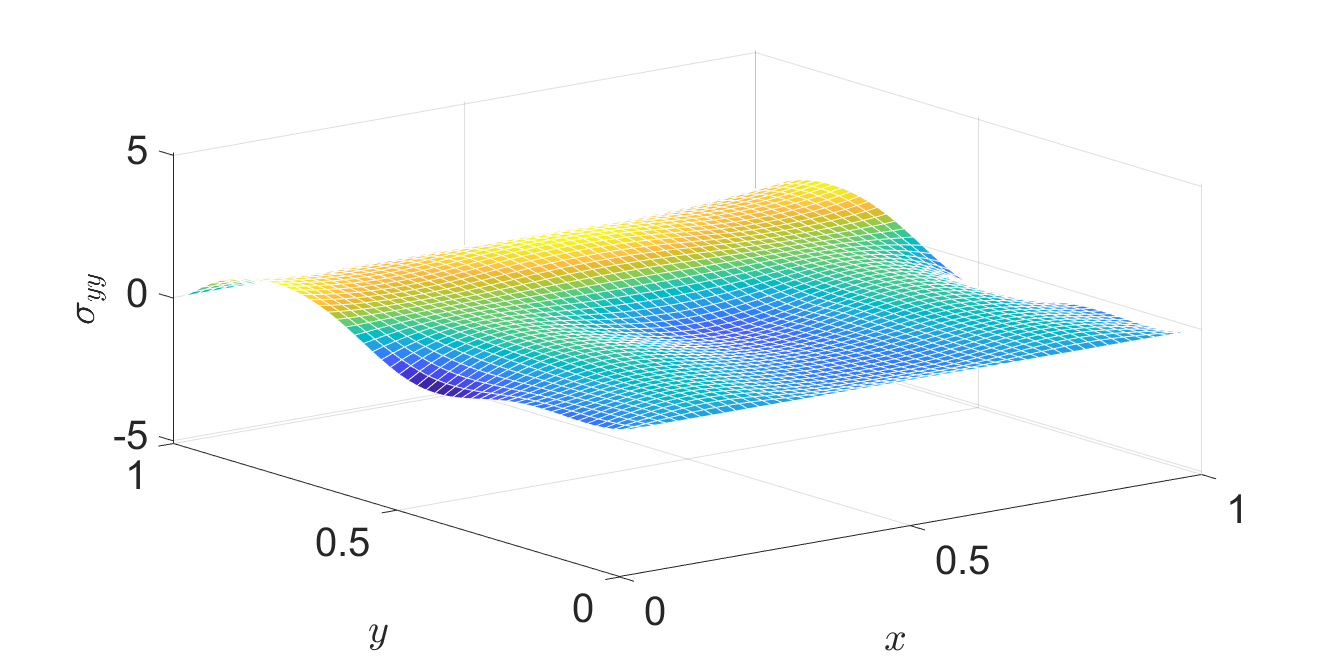}
 \hfill
\includegraphics[width=\qwidth]{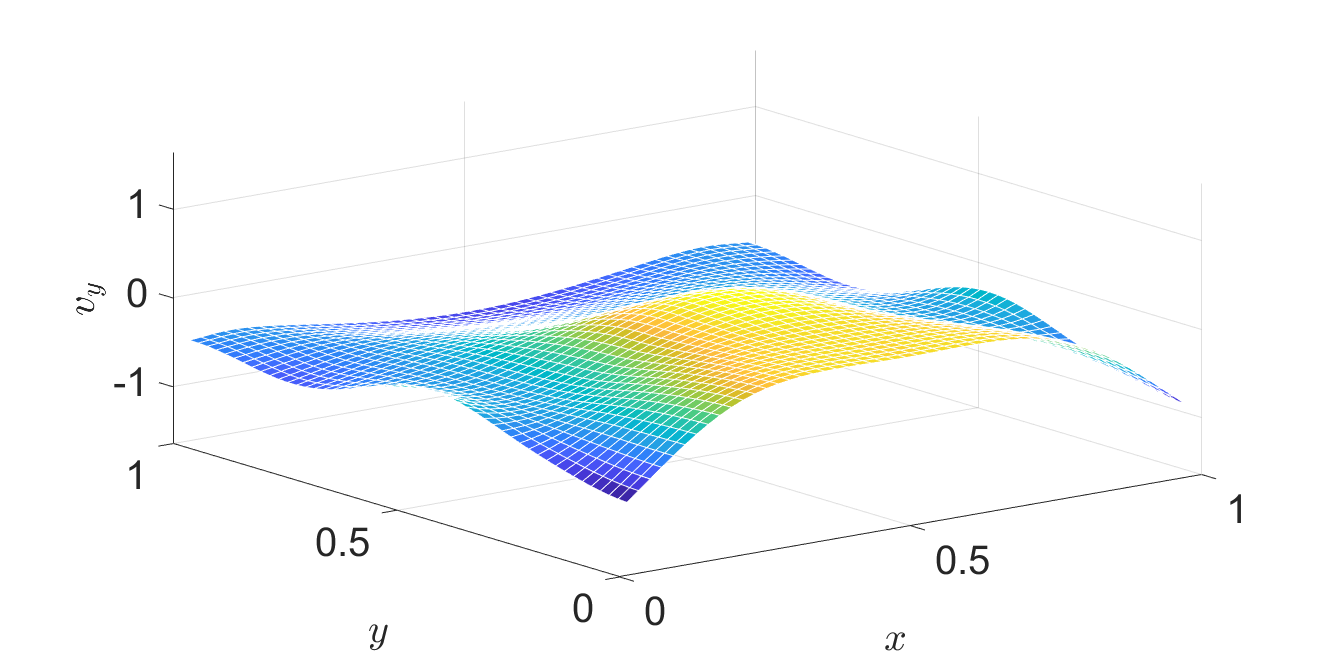}
 \\
\includegraphics[width=\qwidth]{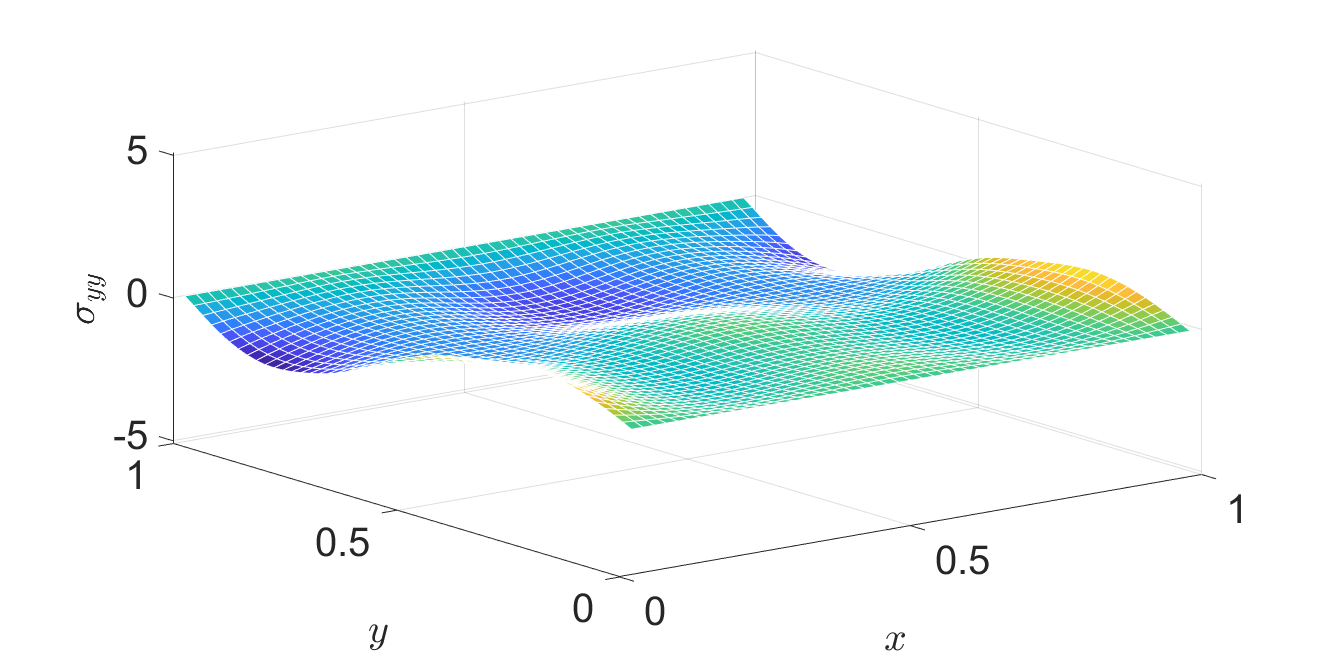}
 \hfill
\includegraphics[width=\qwidth]{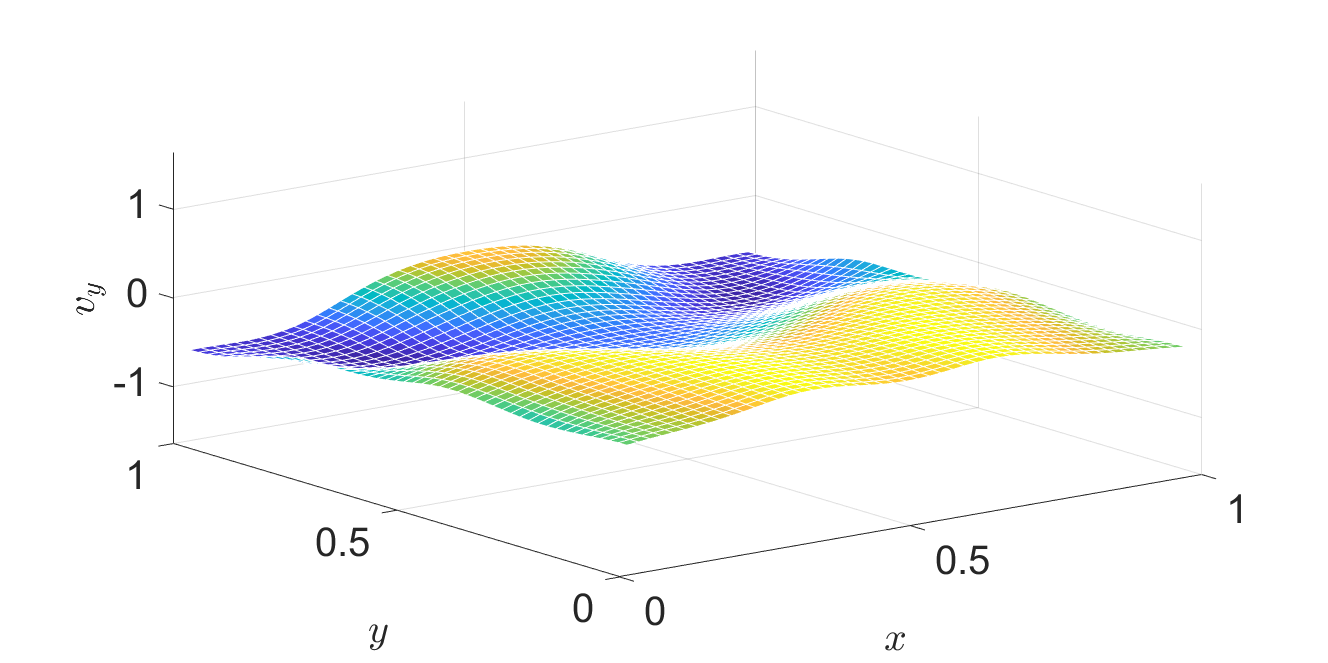}
 \caption{Distribution of a stress component \1 1 {left column} and of a
velocity component \1 1 {right column} at various instants, in the  Poynting--Thomson--Zener (PTZ)
case. From top to bottom: snapshots at instants \m { \F001{1}{2} \qtaub },
\m{\qtaub}, \m { \F001{3}{2} \qtaub }, \m { 2 \qtaub }, respectively.} 
  \label{PTZs}
 \end{figure}

 \begin{figure}[H]
\centering
\includegraphics[width=\qwidth]{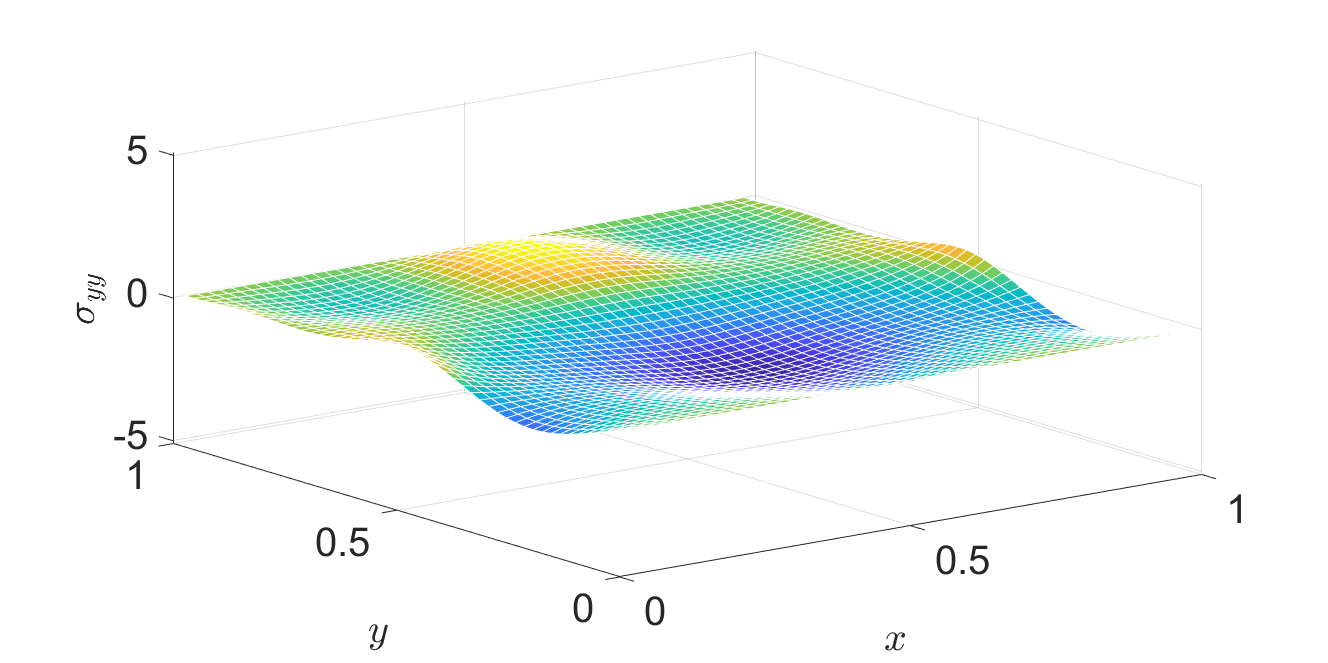}
 \hfill
\includegraphics[width=\qwidth]{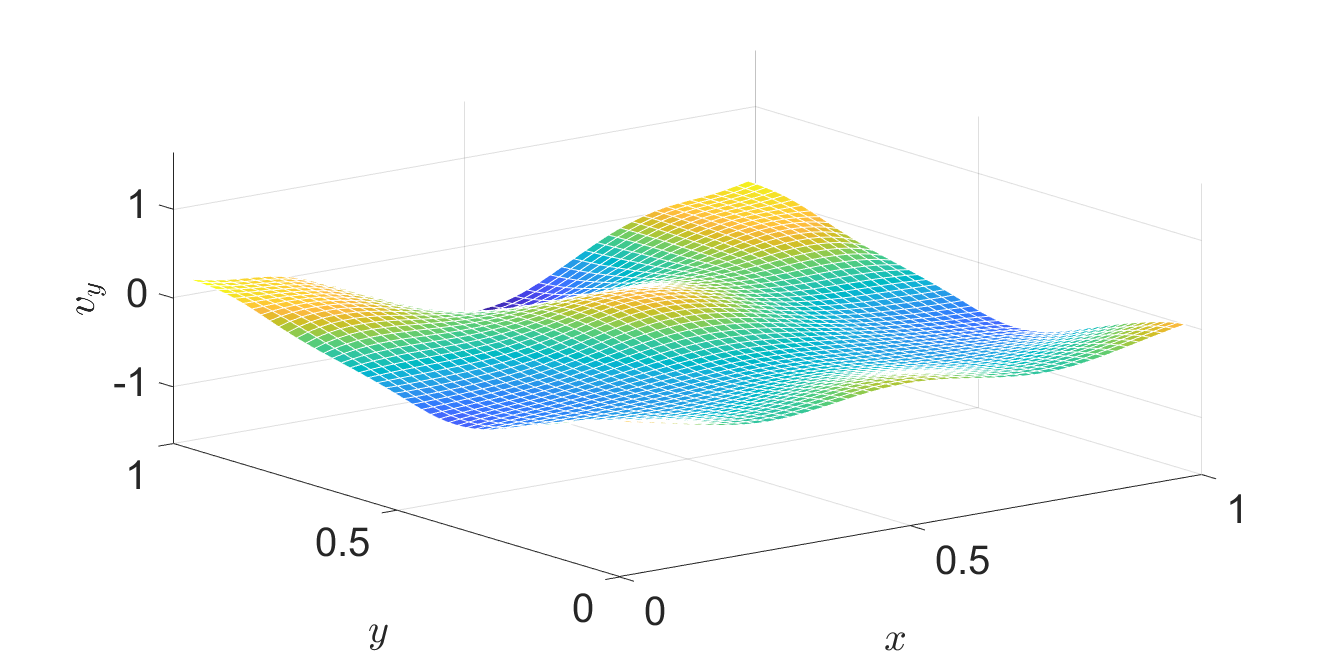}
 \\
\includegraphics[width=\qwidth]{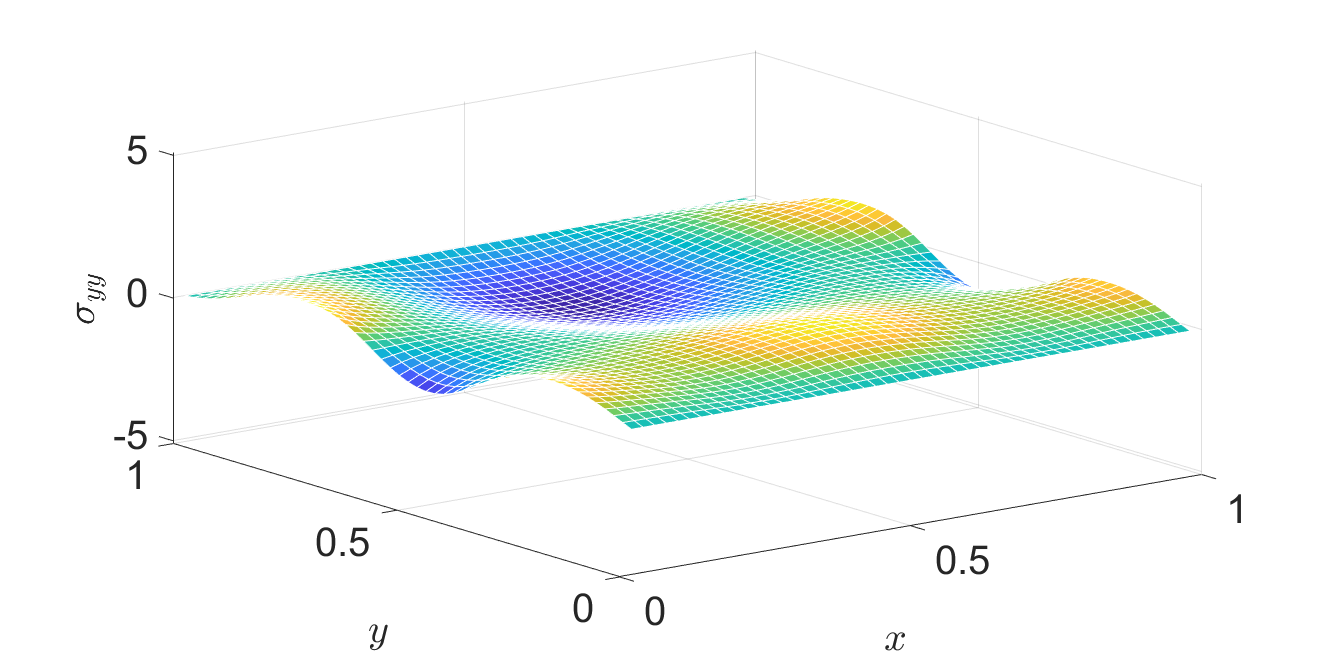}
 \hfill
\includegraphics[width=\qwidth]{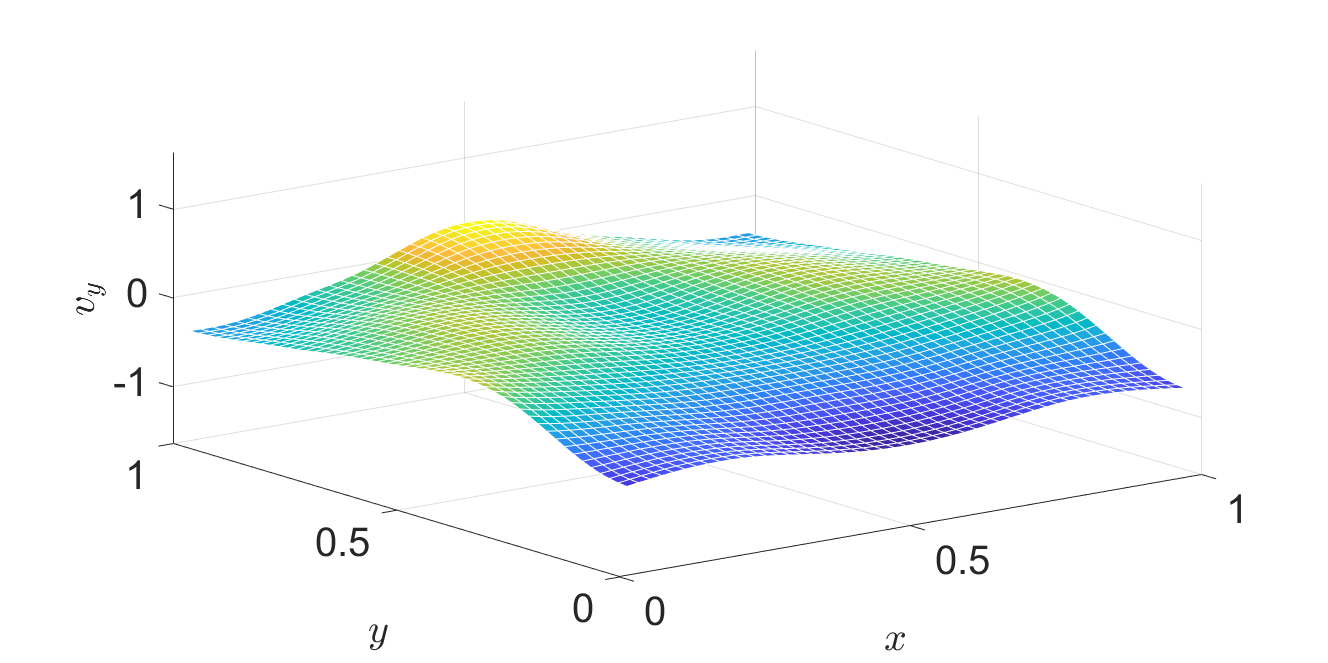}
 \\
\includegraphics[width=\qwidth]{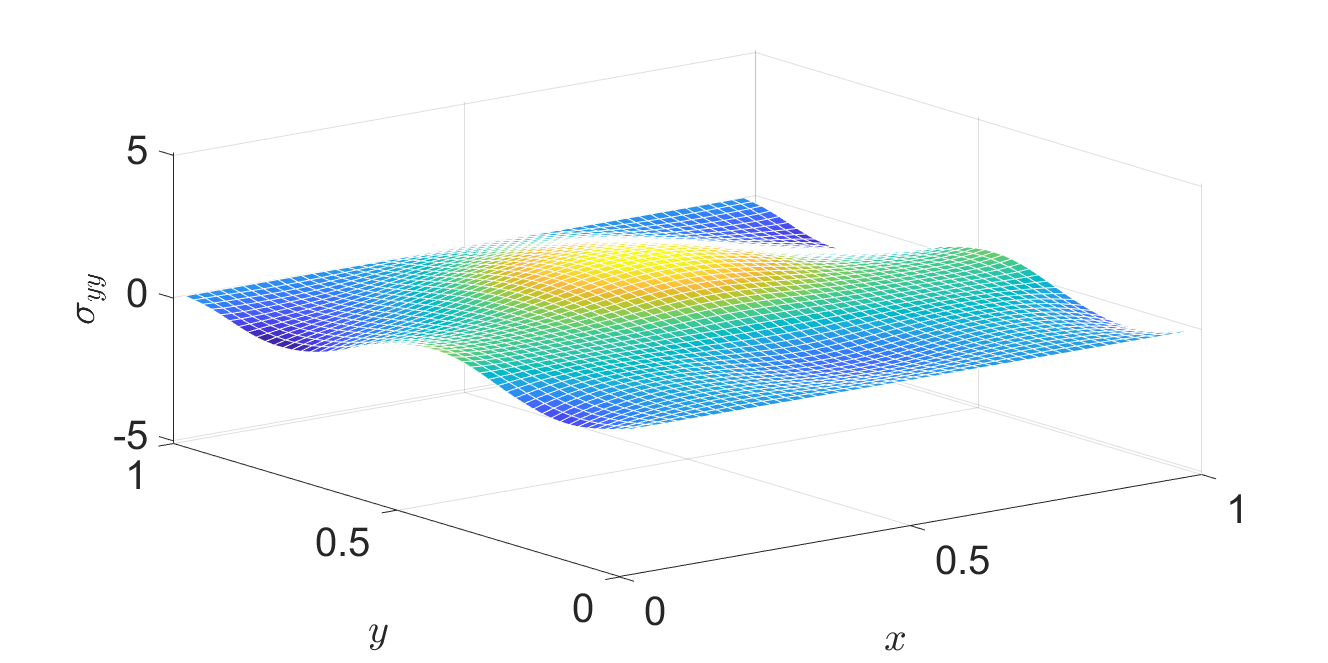}
 \hfill
\includegraphics[width=\qwidth]{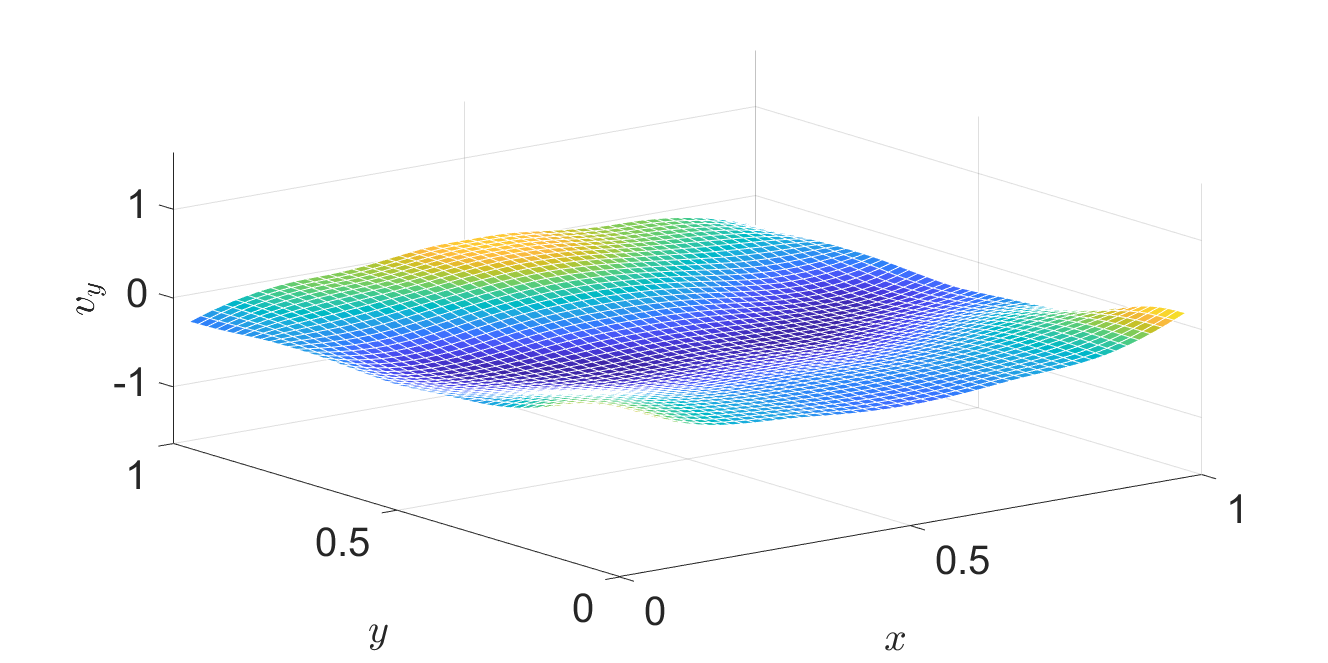}
 \\
\includegraphics[width=\qwidth]{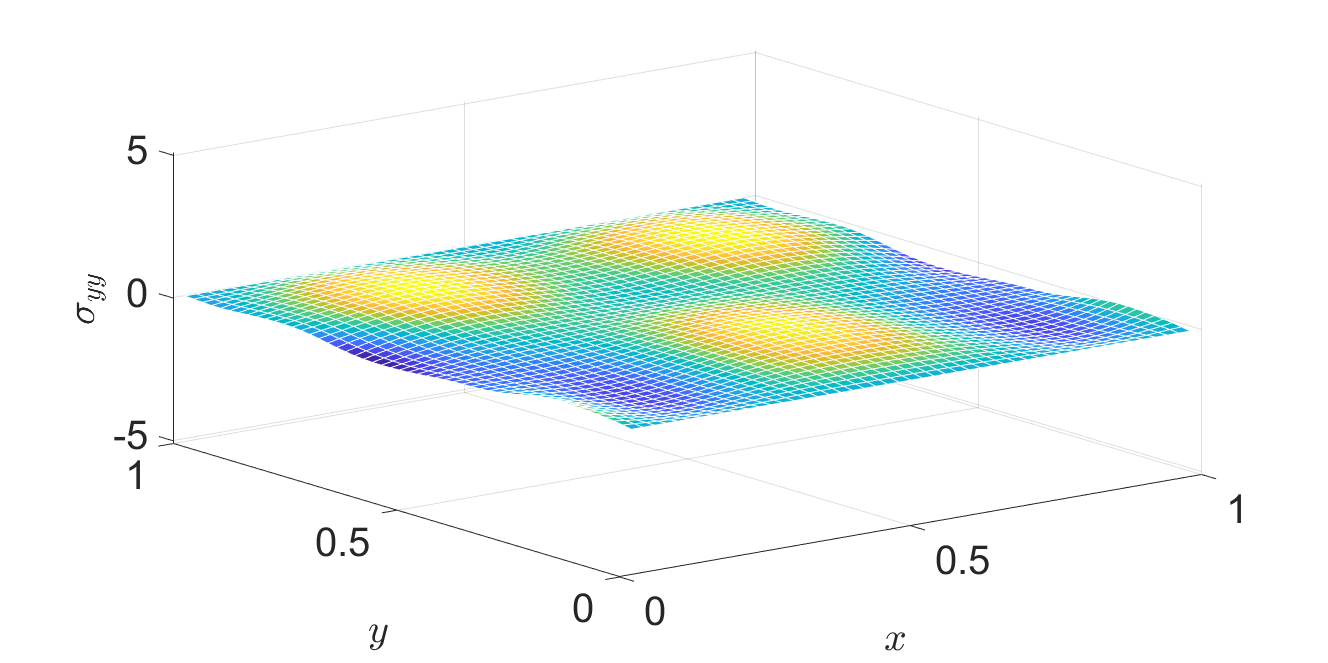}
 \hfill
\includegraphics[width=\qwidth]{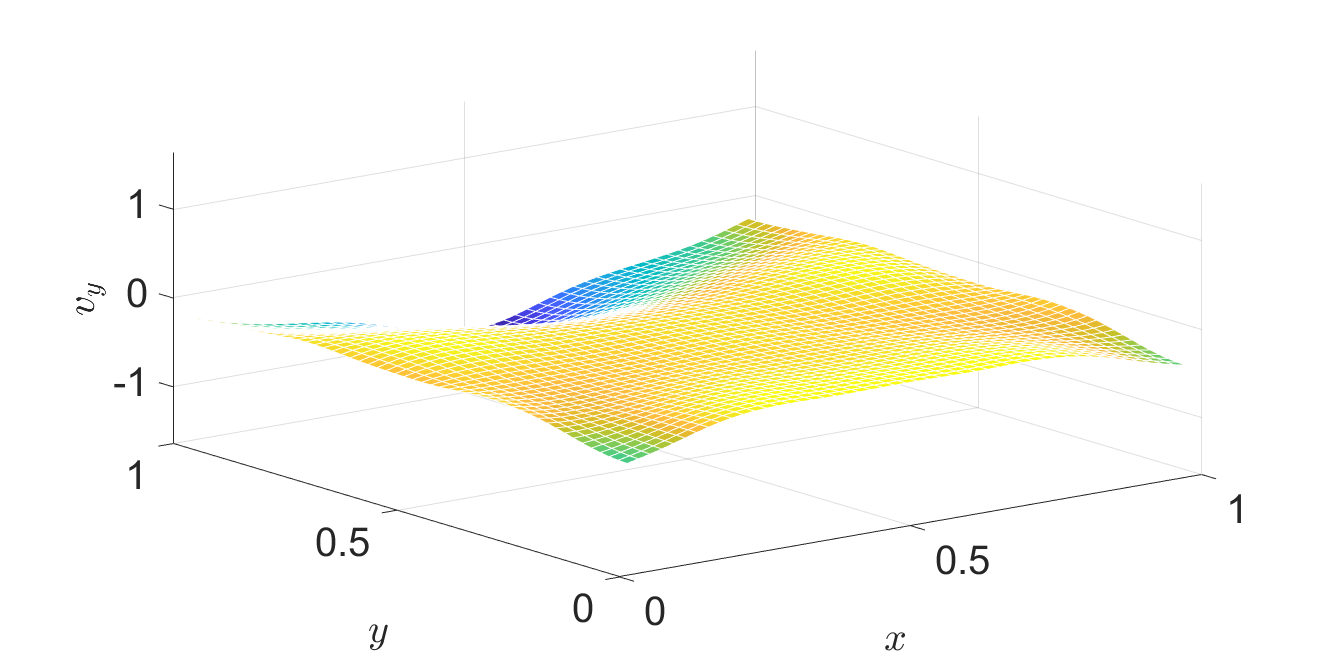}
 \caption{Continuation of Figure \ref{PTZs}: distribution of a stress component
\1 1 {left column} and of a velocity component \1 1 {right column} at various
instants, in the PTZ case. From top to bottom: snapshots at instants \m {
\F001{5}{2} \qtaub }, \m{3 \qtaub}, \m { \F001{7}{2} \qtaub }, \m { 4 \qtaub
}, respectively.} 
  \label{PTZv}
 \end{figure}
\unskip
 \begin{figure}[H]
\centering
\includegraphics[width=\qqwidth]{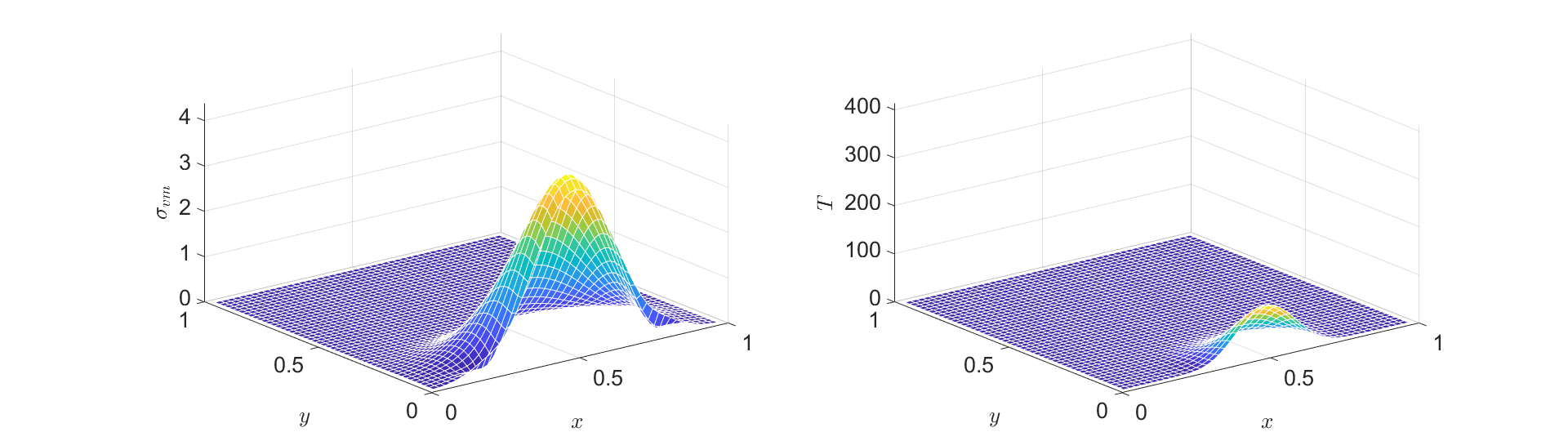}
 \\
\caption{\textit{Cont.}}
\end{figure}

\begin{figure}[H]\ContinuedFloat
\centering
\includegraphics[width=\qqwidth]{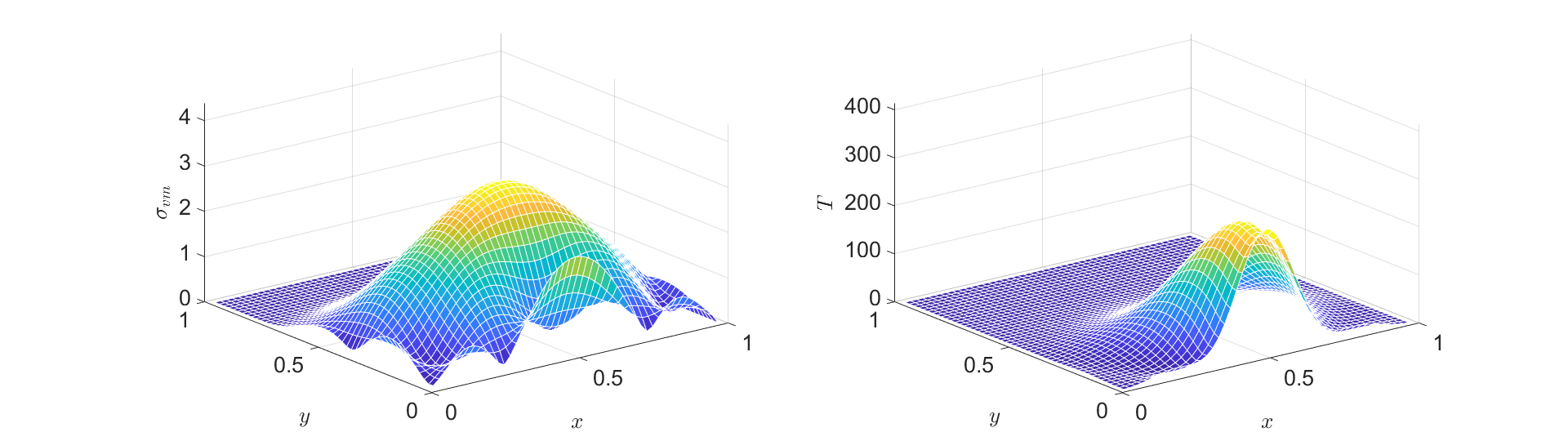}
 \\
\includegraphics[width=\qqwidth]{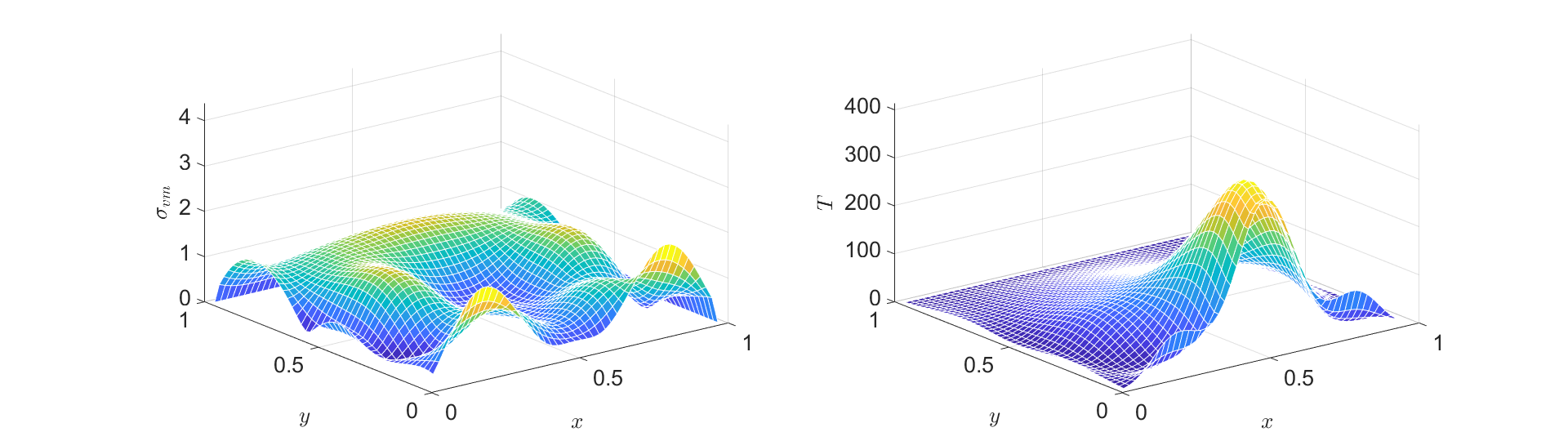}
 \\
\includegraphics[width=\qqwidth]{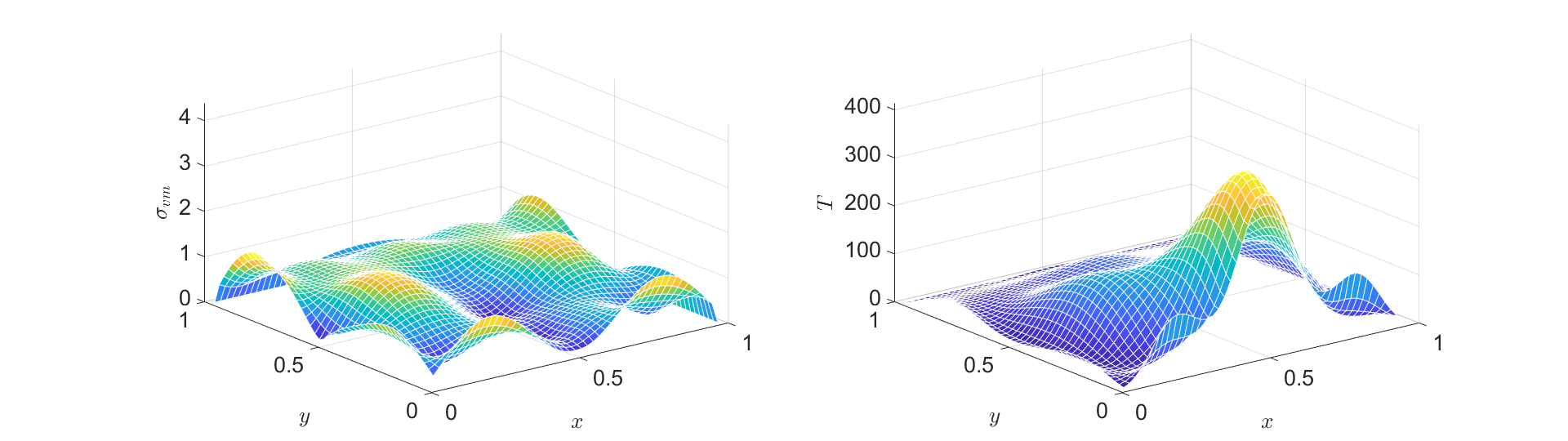}
 \caption{
 Snapshots of the distribution of the stress invariant 
 $\sqrt{\text{tr}(\hat{\bm{\sigma}}^{\text{dev}^{2}})}$  \1 1 {left column} and of temperature
\1 1 {right column} at various instants, in the PTZ case. From top to bottom:
snapshots at instants \m { \F001{1}{2} \qtaub }, \m{\qtaub}, \m { \F001{3}{2}
\qtaub }, \m { 2 \qtaub }, respectively.
}
  \label{vm1}
 \end{figure}
\unskip
 \begin{figure}[H]
\centering
\includegraphics[width=\qqwidth]{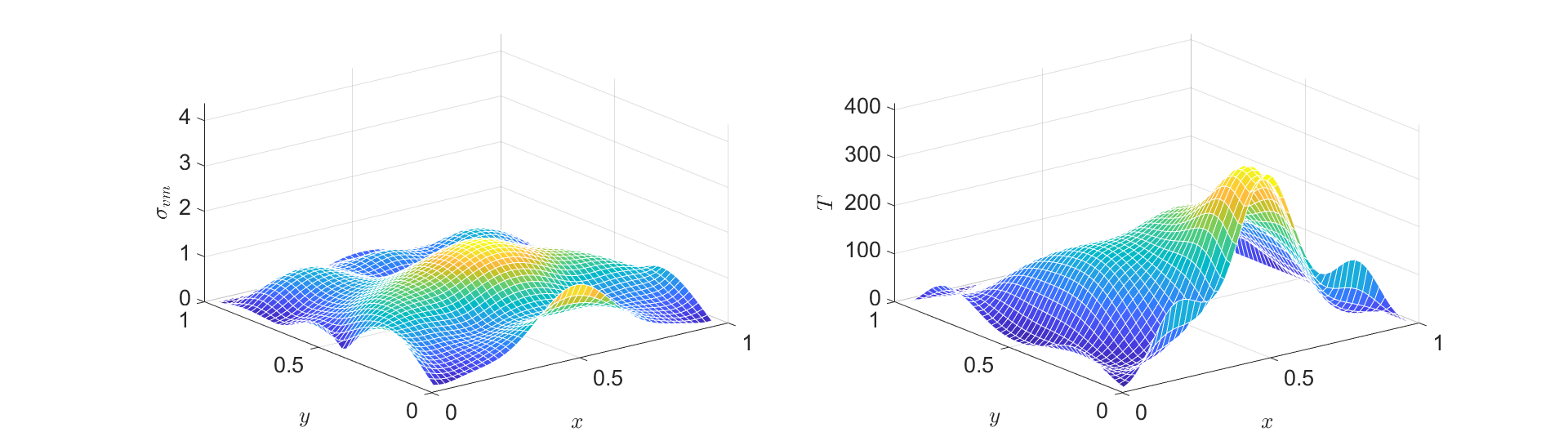}
 \\
\caption{\textit{Cont.}}
\end{figure}

\begin{figure}[H]\ContinuedFloat
\centering
\includegraphics[width=\qqwidth]{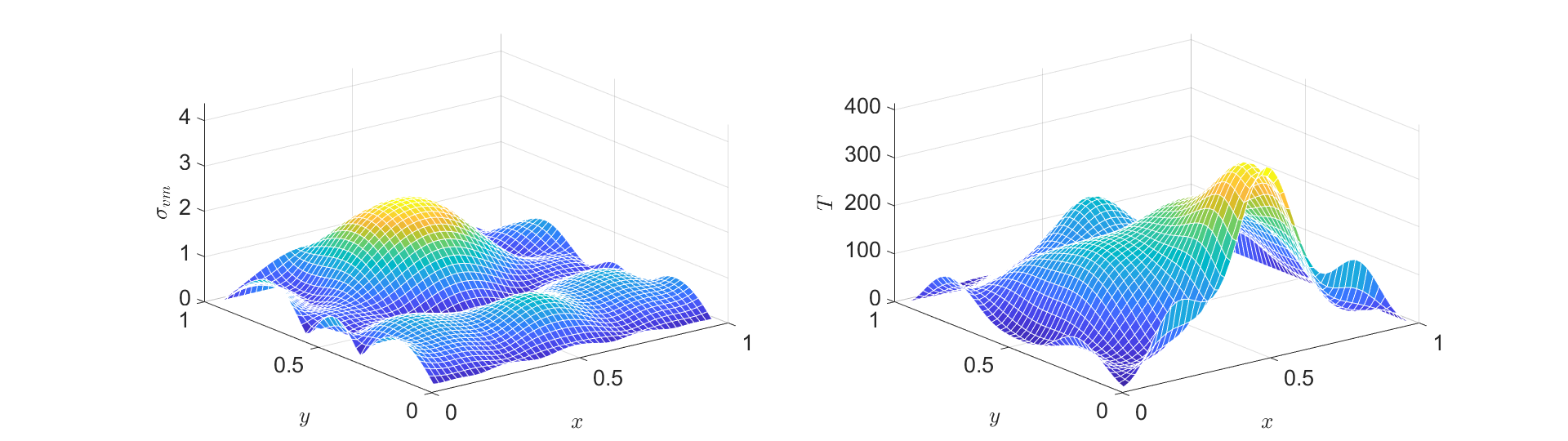}
 \\

\includegraphics[width=\qqwidth]{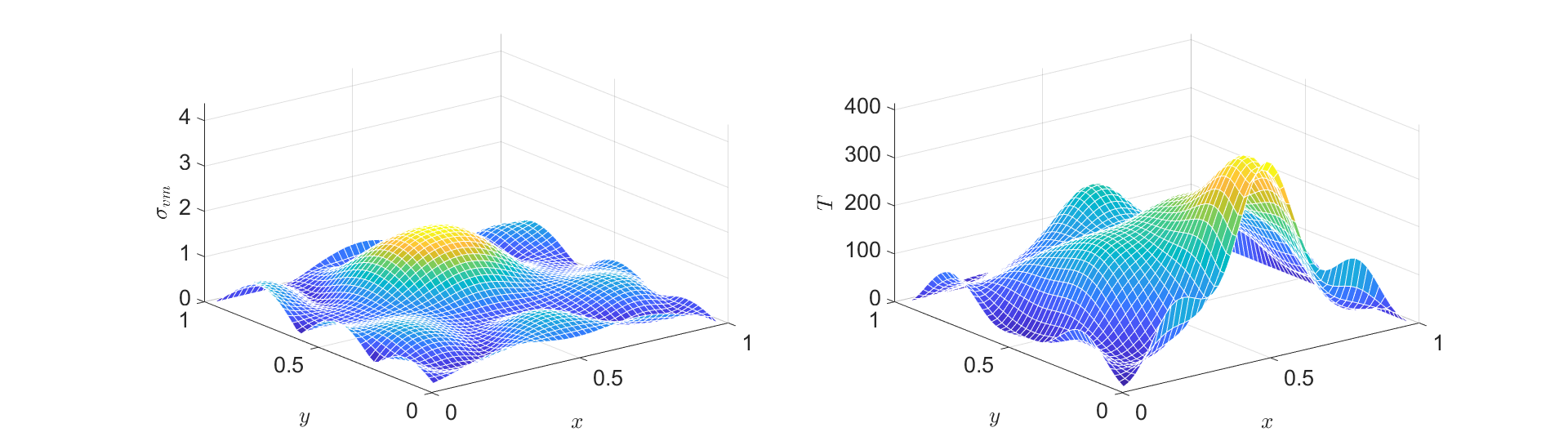}
 \\
\includegraphics[width=\qqwidth]{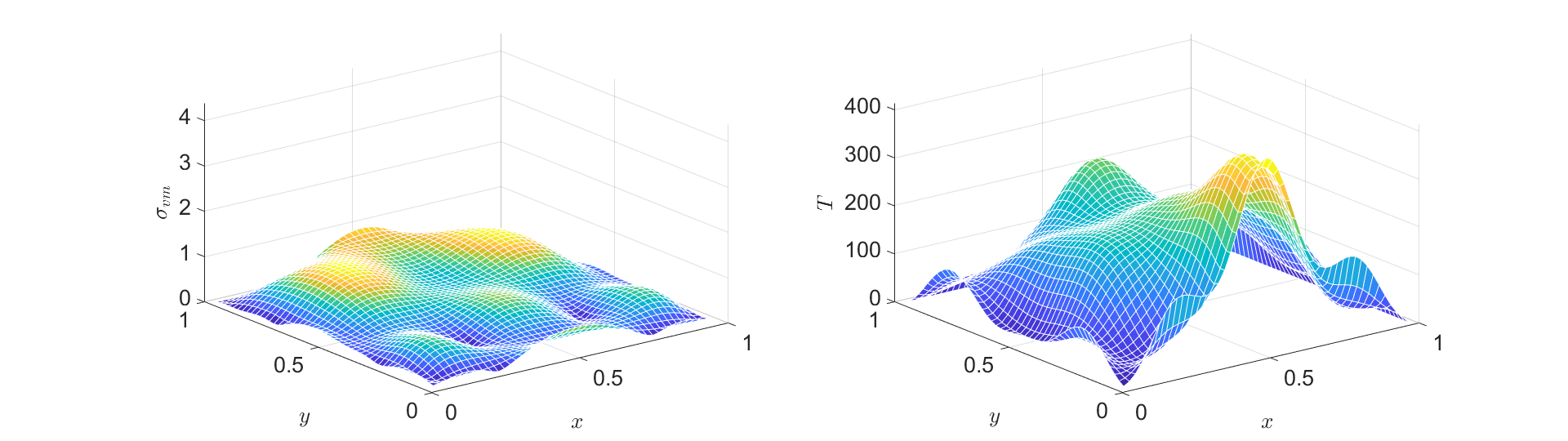}
 \caption{
 Continuation of Figure \ref{vm1}: Snapshots of the distribution of
the stress invariant $\sqrt{\text{tr}(\hat{\bm{\sigma}}^{\text{dev}^{2}})}$ \1 1 {left
column} and of temperature \1 1 {right column} at various instants, in the
PTZ case. From top to bottom: snapshots at instants \m { \F001{5}{2} \qtaub
}, \m{3 \qtaub}, \m { \F001{7}{2} \qtaub }, \m { 4 \qtaub }, respectively.
}
  \label{vm2}
 \end{figure}

\begin{figure}[H]
\centering
\includegraphics[width=0.75\columnwidth]{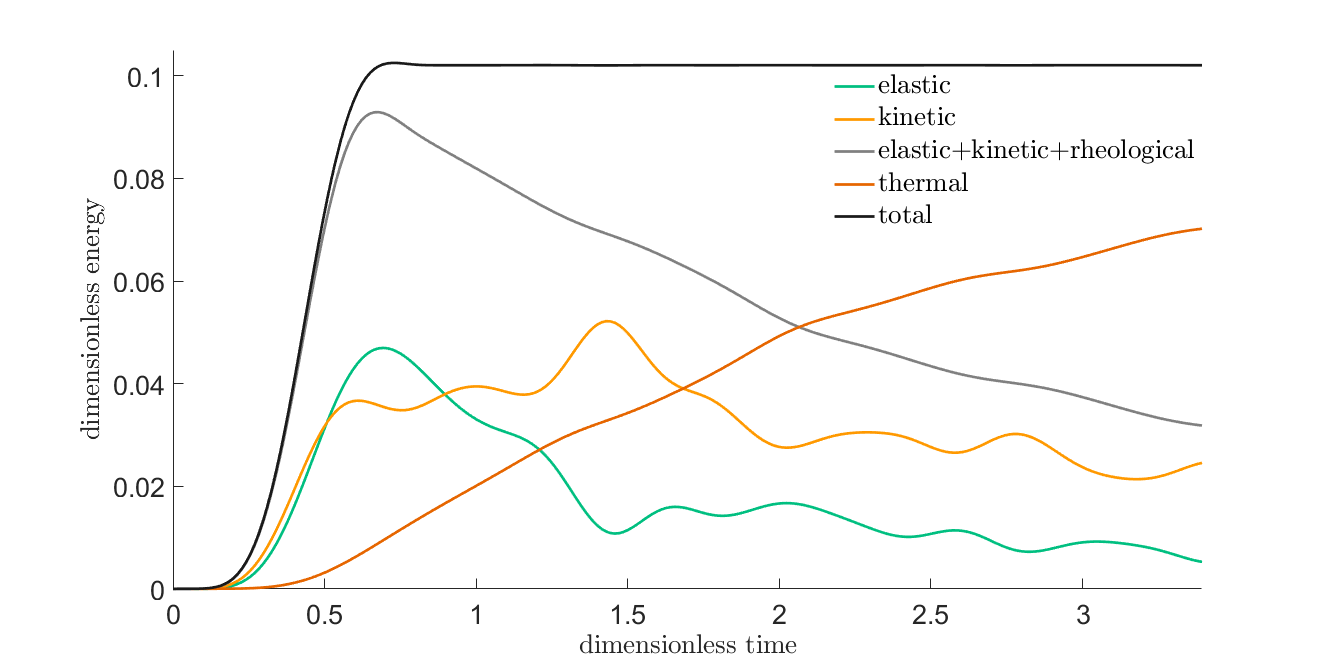}
 \caption{Total energy and the various energy types as functions of time, for
the PTZ case.}
  \label{PTZe}
 \end{figure}

The diagnostic role of the various energies, and especially their sum, is a
great help again for checking whether the simulation performs acceptably.
Figure~\ref{PTZe} illustrates how the scheme that is introduced above behaves in this
respect.

\section{Solution for a Cube, and the Role of Entropy Production}

In the second example treated, a cube is considered, initially relaxed, as in
the previous example, and then one of its sides being pressed by a cosine
\quot{bump} in time, as well as in both spatial directions:
  \Par
 \begin{align}  \label{coscos}
\qsig_{\qyy} \0 1 { \qt, \qx, 0, \qz } =
 \begin{cases}
\qsigb \Bigl\{ \f {1}{2} \9 2 { 1 - \cos \0 1 { 2 \pi \f {\qt}{\qtaub} } }
 \cdot
\f {1}{2} \9 2 { 1 - \cos \0 1 {2 \pi \f {\qx - \qX/2 }{\qWb} } }
 & \\
\phantom{\qsigb \Bigl\{ \f {1}{2} \9 2 { 1 - \cos \0 1 { 2 \pi \f
{\qt}{\qtaub} } }}  
\cdot \f {1}{2} \9 2 { 1 - \cos \0 1 {2 \pi \f {\qz - \qX/2 }{\qWb} } }
\Bigr\}  \rule{0em}{3.5ex}  
& \text{if } \:
\1 0 { 0 \le \qt \le \qtaub \: \text{ and}
 \\ & \f{\qX - \qWb}{2} \le \qx, \qz \le \f{\qX + \qWb}{2} } ,
 \\
0
& \1 0 {\text{otherwise}}
 \end{cases}
 \end{align}
  \Par
\1 1 {see Figure~\ref{3dbump}}, where the notations are analogous to the ones
of the previous case: \m { \qtaub } is the duration of the pulse, \m { \qsigb
} is its amplitude, \m { \qWb } is its spatial width, and \m { \qX } is the
length of the edges of the cube. On the other sides, normal stress is
constantly zero \1 1 {free surfaces}, like in the previous example.

 \begin{figure}[H]
\centering
 \includegraphics[width=.5\columnwidth]{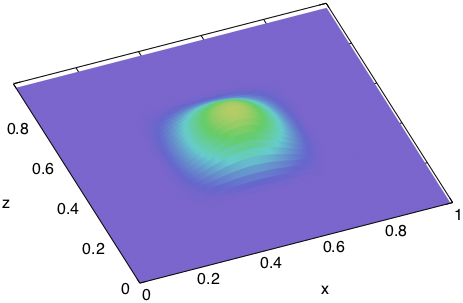}
\caption{The spatial distribution of normal stress as the boundary condition
on one side of a cube. The excitation is also a single cosine-shaped \quot{bump} in time.}  \label{3dbump}
 \end{figure}

Here, we only present the PTZ model, on a \m { 25 \times 25 \times 25 }
grid, with all other settings being the same as in the previous example.

The solution proves similarly satisfatory, as in the former case.
Figures~\ref{3dvm1} and \ref{3dvm2} show snapshots of the distribution of the
von Mises stress invariant on two mid-planes of the cubes.

 \begin{figure}[H]
\centering
\includegraphics[width=\qwidth]{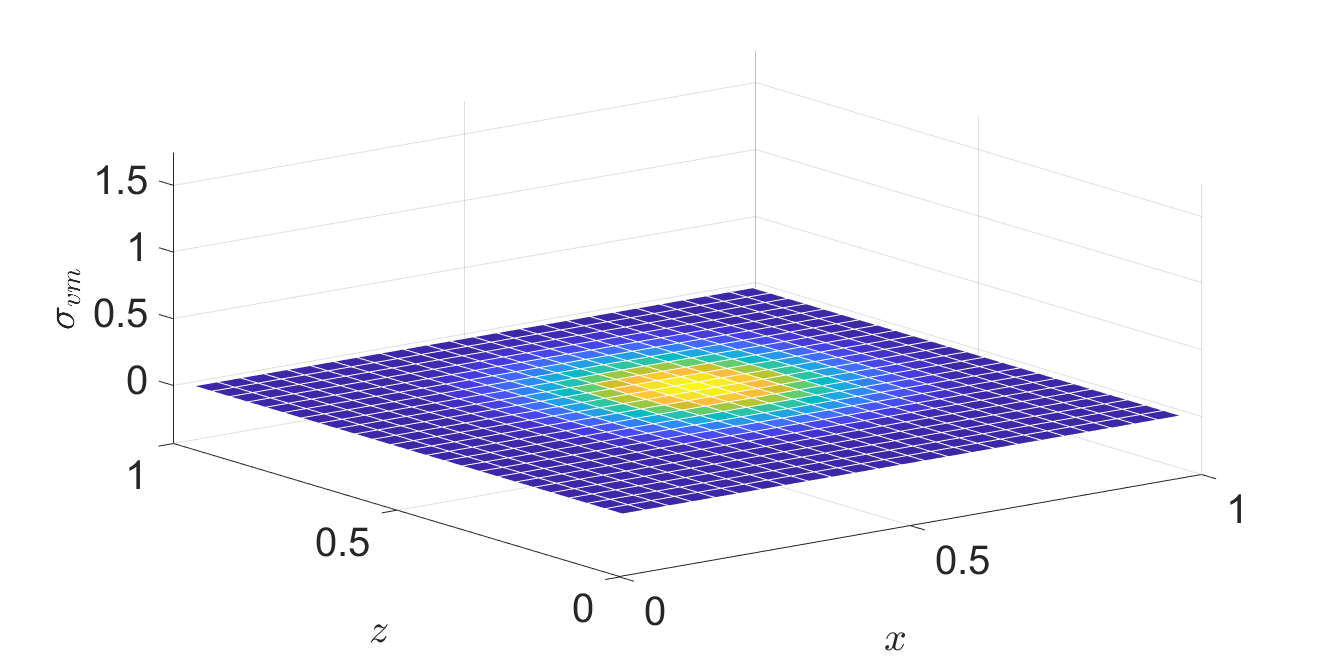}
\includegraphics[width=\qwidth]{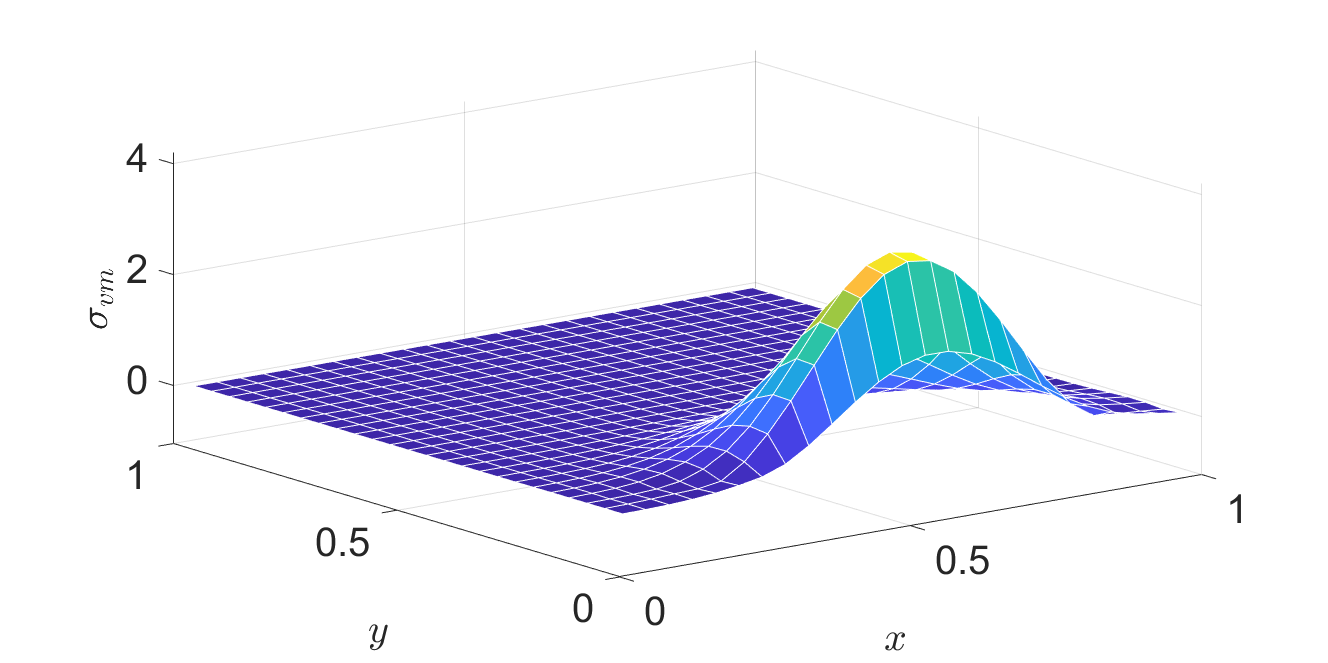}
 \\
\includegraphics[width=\qwidth]{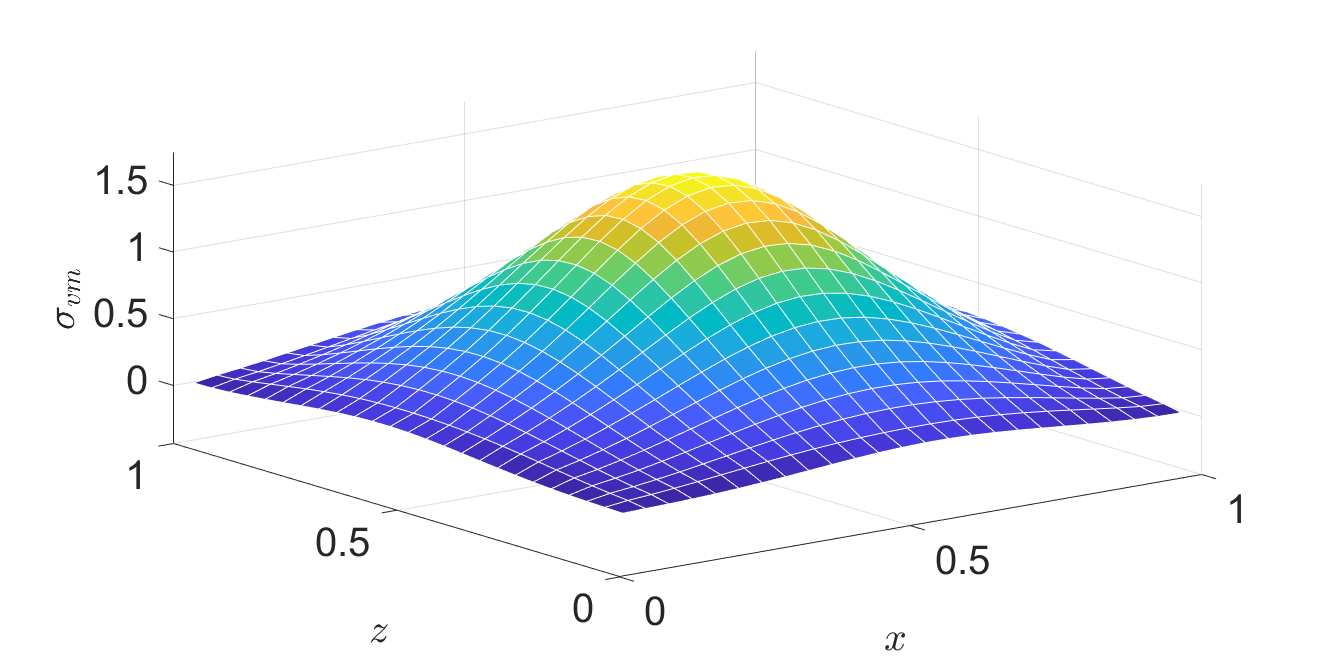}
\includegraphics[width=\qwidth]{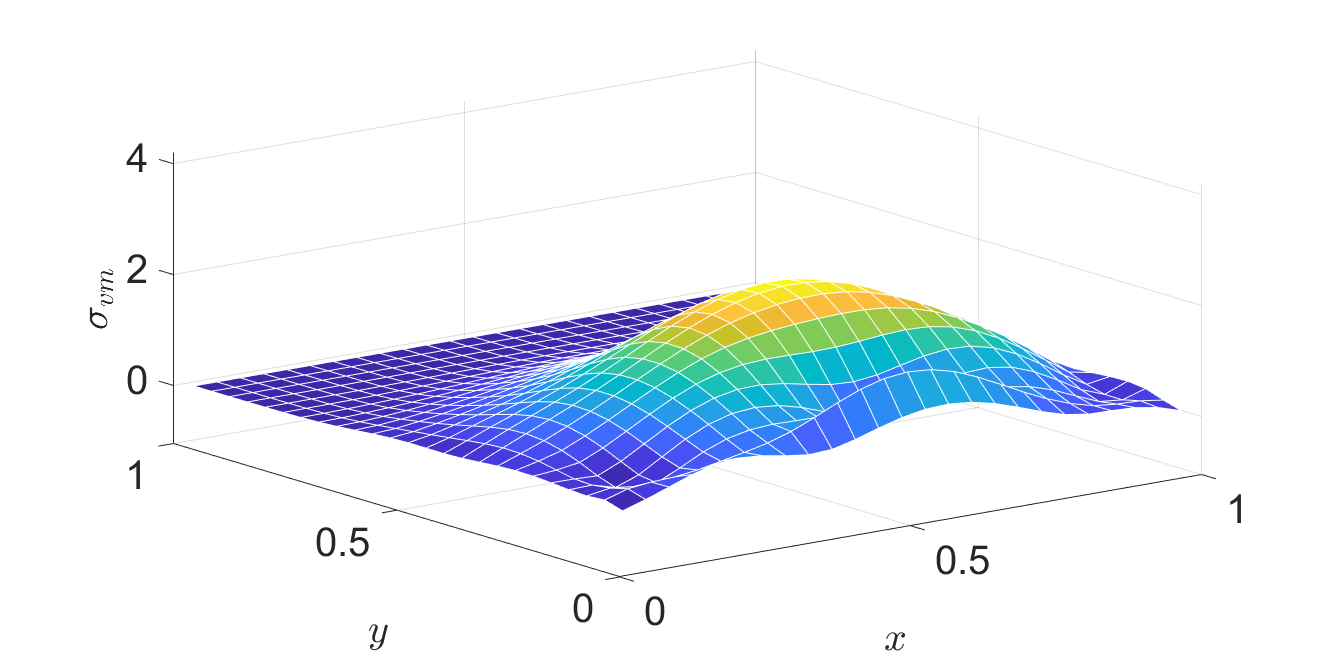}
 \\
\caption{\textit{Cont.}}
\end{figure}

\begin{figure}[H]\ContinuedFloat
\centering
\includegraphics[width=\qwidth]{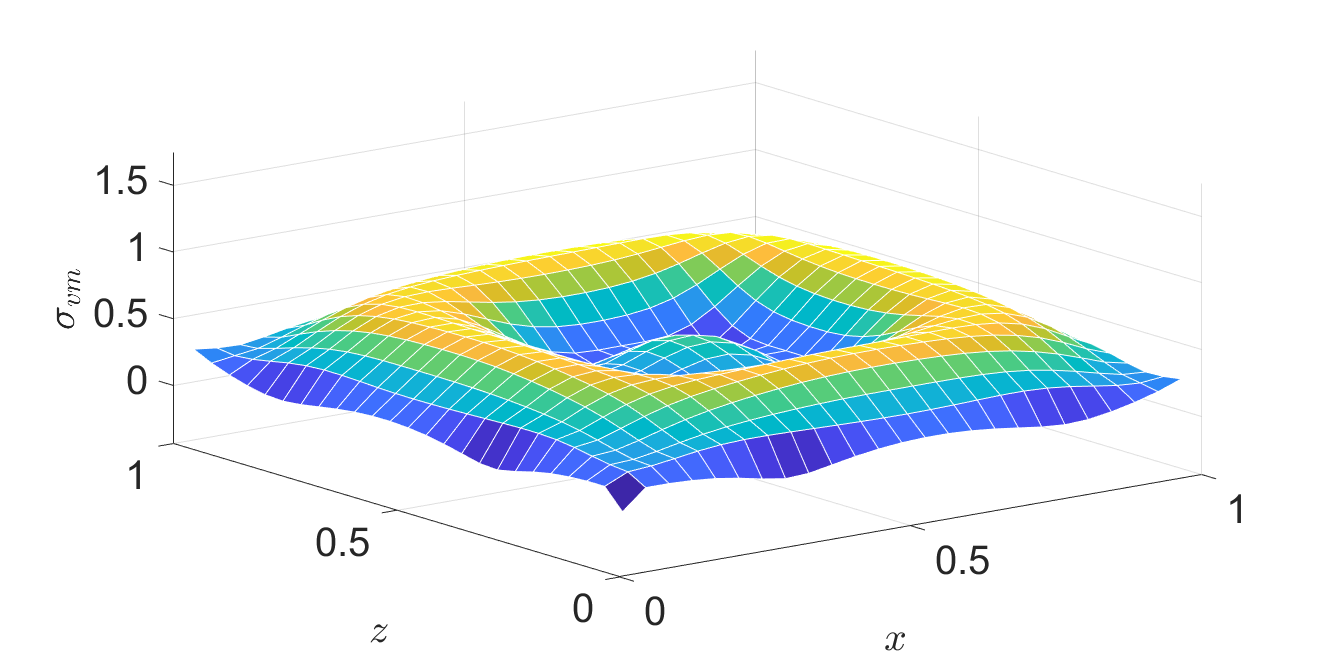}
\includegraphics[width=\qwidth]{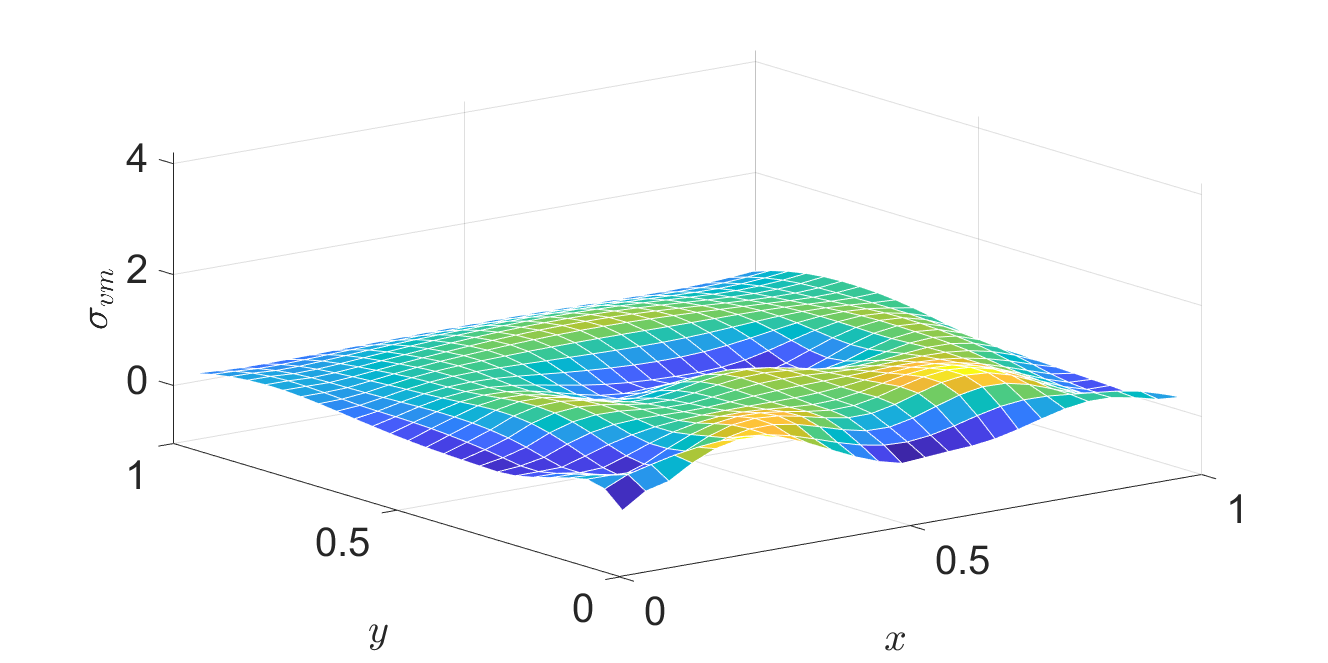}
 \\
\includegraphics[width=\qwidth]{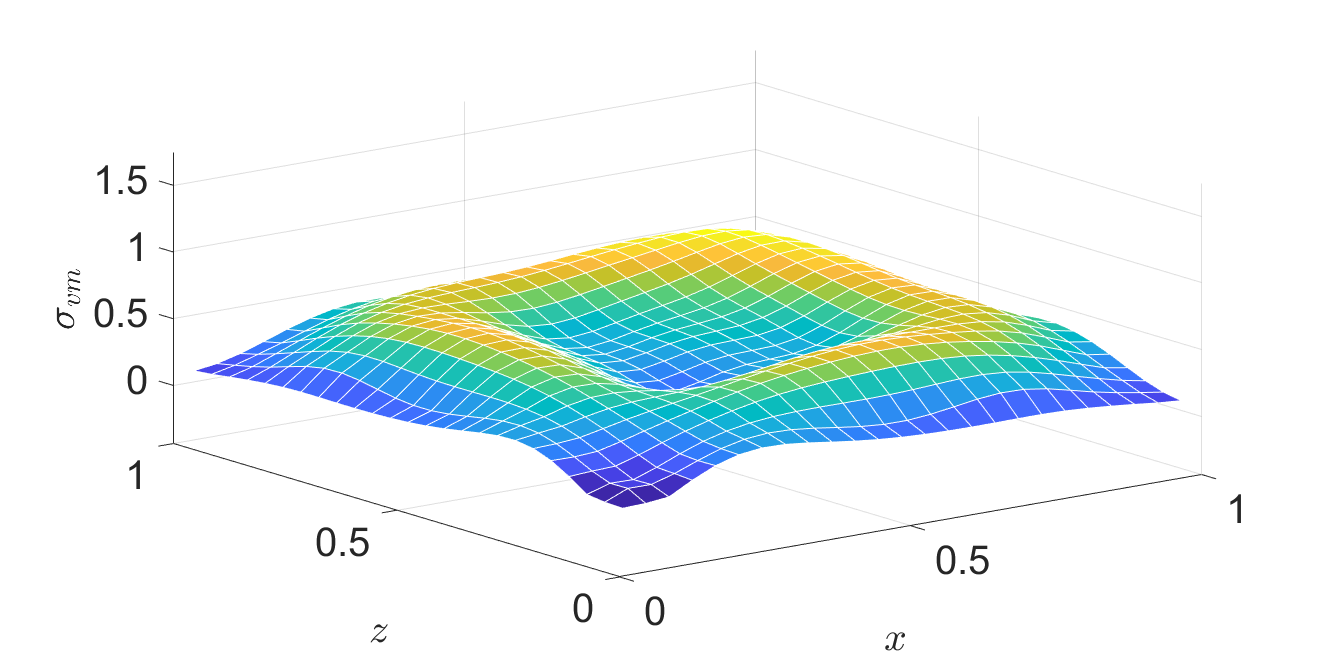}
\includegraphics[width=\qwidth]{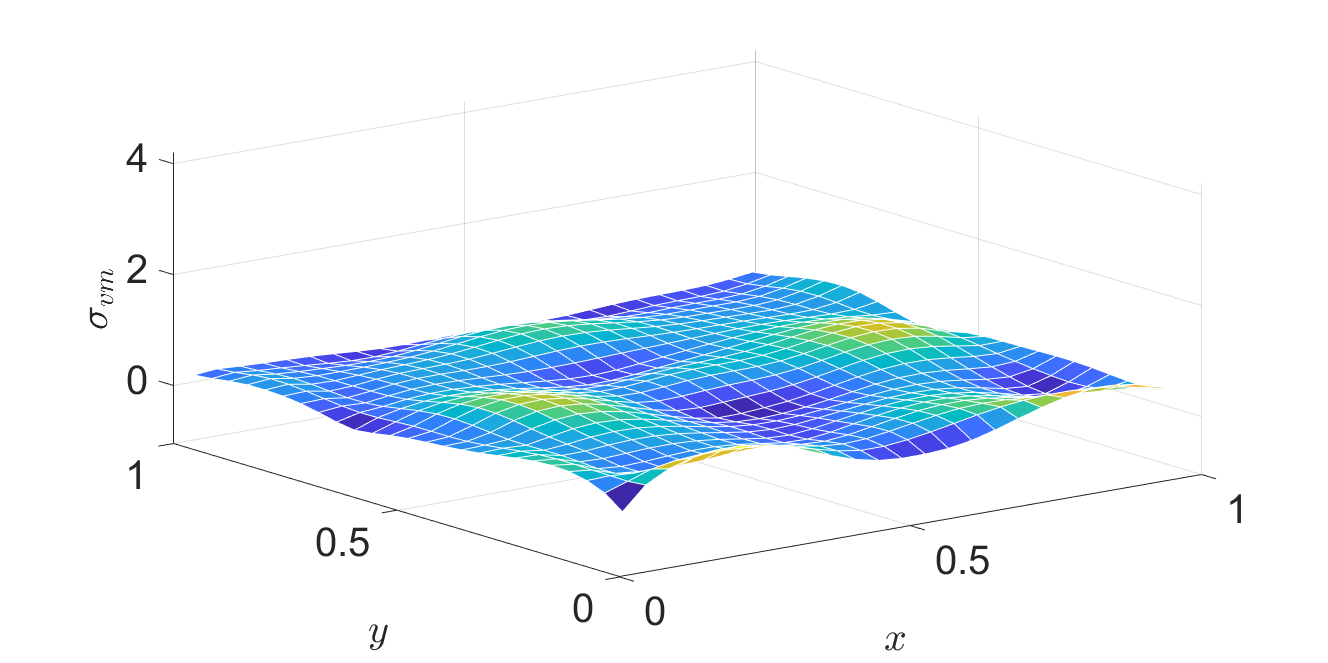}
 \caption{
 Snapshots of the distribution of the stress invariant $\sqrt{\text{tr}(\hat{\bm{\sigma}}^{\text{dev}^{2}})}$ at the mid-plane parallel to the
excited side \1 1 {left column} and at a mid-plane orthogonal to it
\1 1 {right column} at various instants, in~the PTZ case. From top to bottom:
snapshots at instants \m { \F001{1}{2} \qtaub }, \m{ \qtaub}, \m {
\F001{3}{2} \qtaub }, \m { 2 \qtaub }, respectively.
}
  \label{3dvm1}
 \end{figure}
\unskip
 \begin{figure}[H]
\centering
\includegraphics[width=\qwidth]{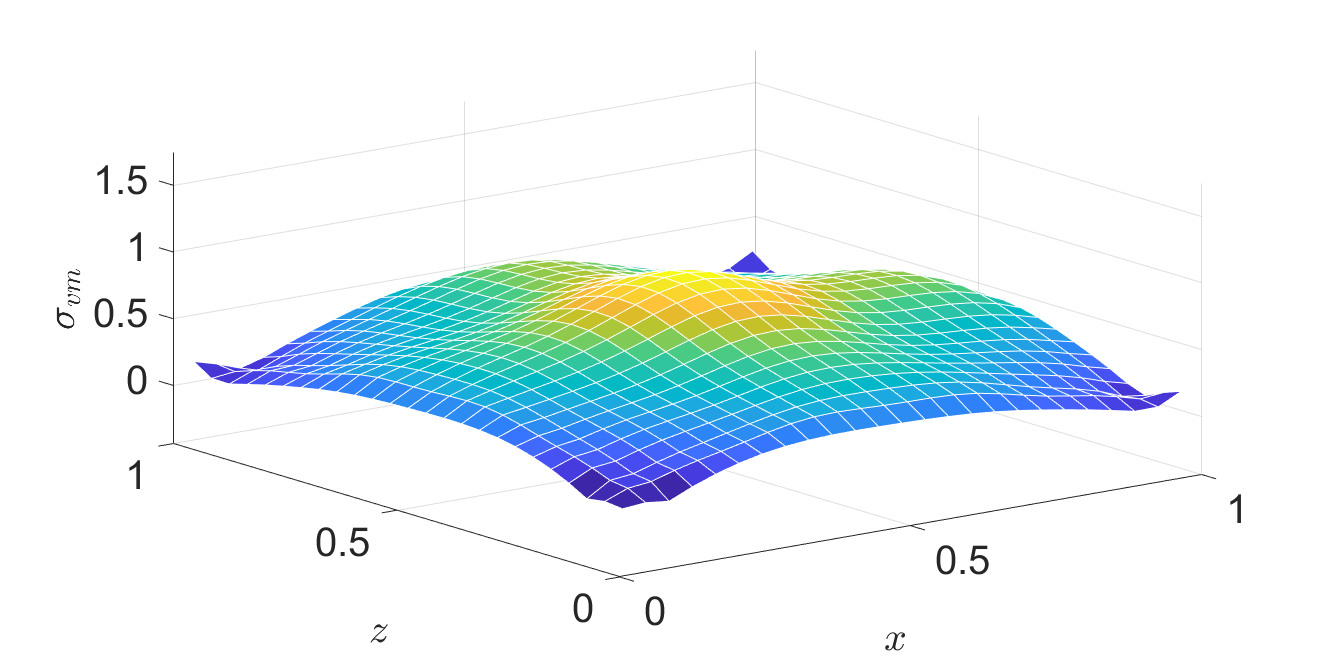}
\includegraphics[width=\qwidth]{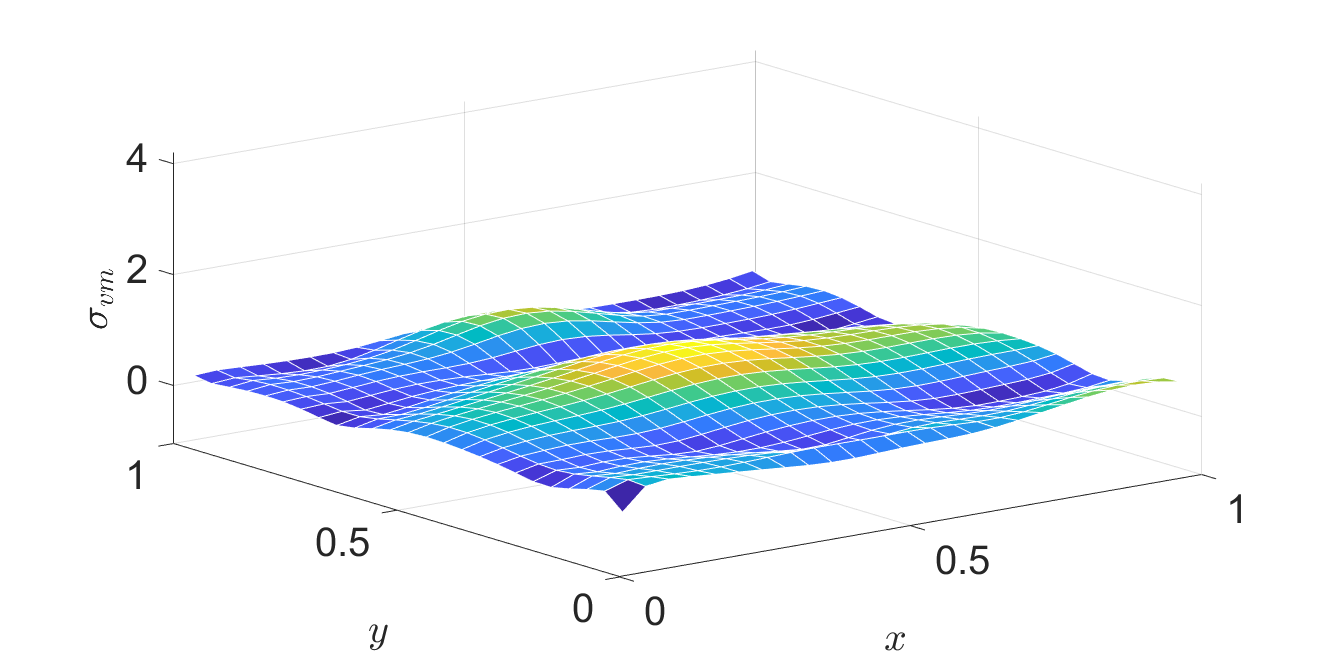}
 \\
\includegraphics[width=\qwidth]{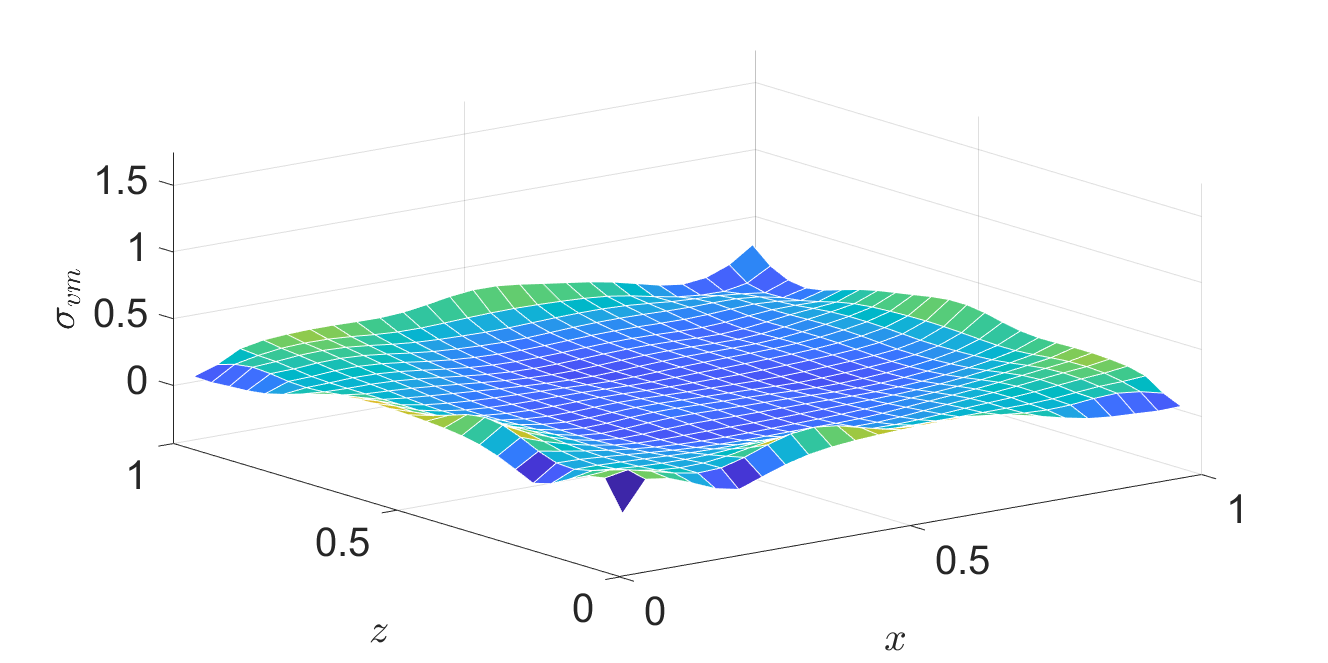}
\includegraphics[width=\qwidth]{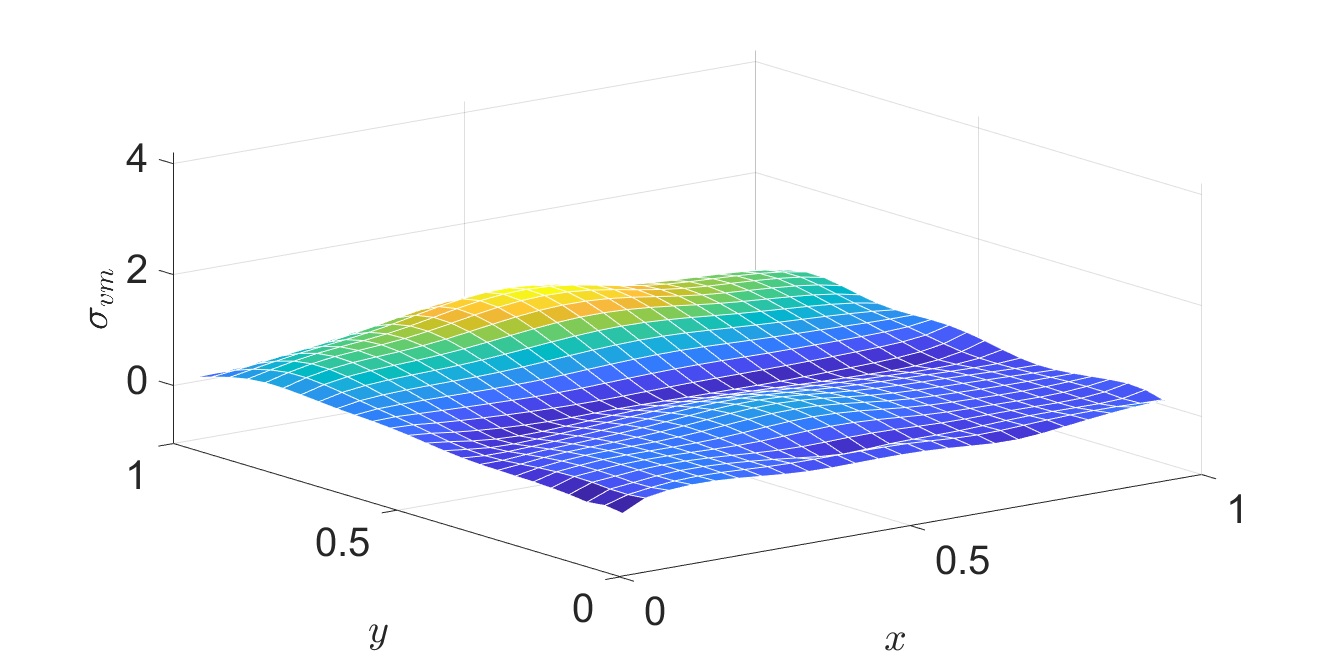}
 \\
\includegraphics[width=\qwidth]{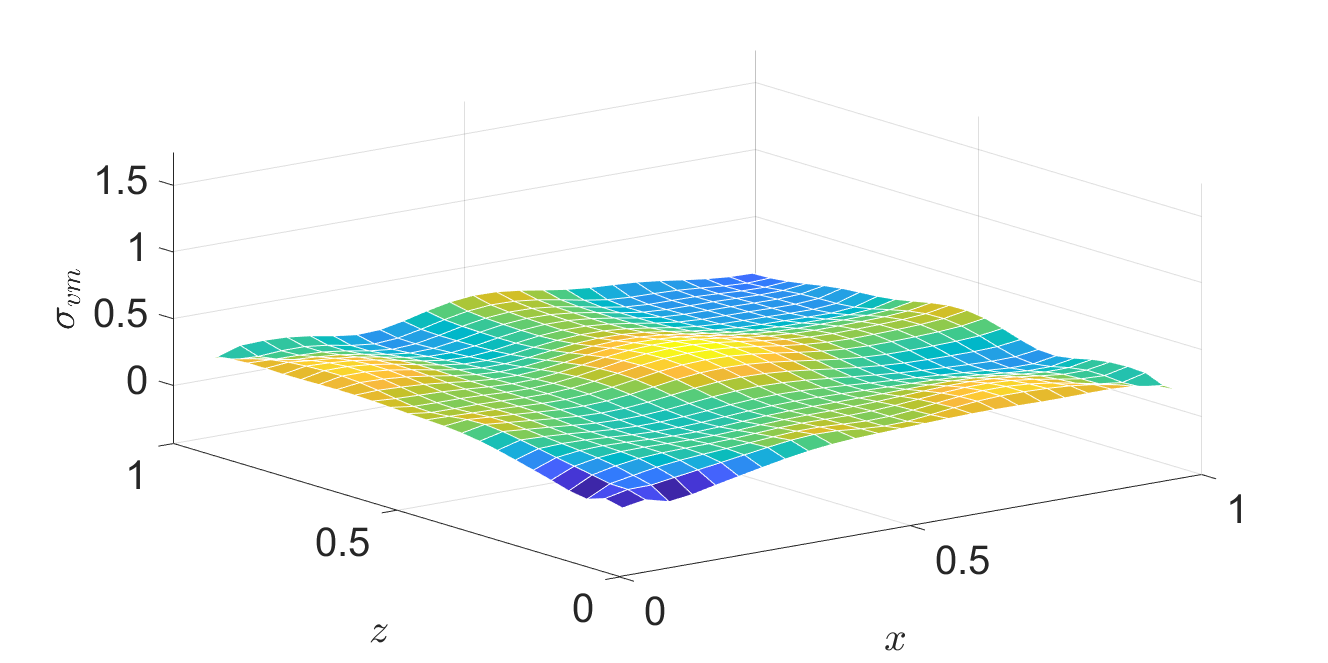}
\includegraphics[width=\qwidth]{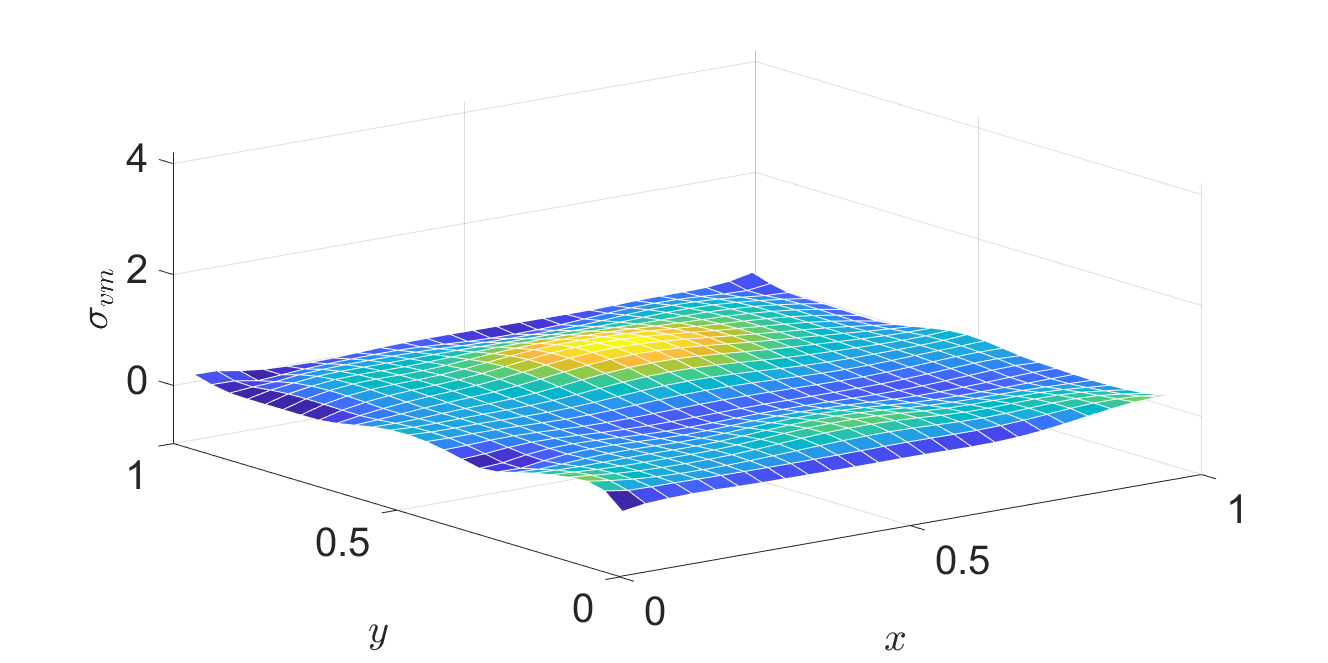}
 \\
\caption{\textit{Cont.}}
\end{figure}

\begin{figure}[H]\ContinuedFloat
\centering
\includegraphics[width=\qwidth]{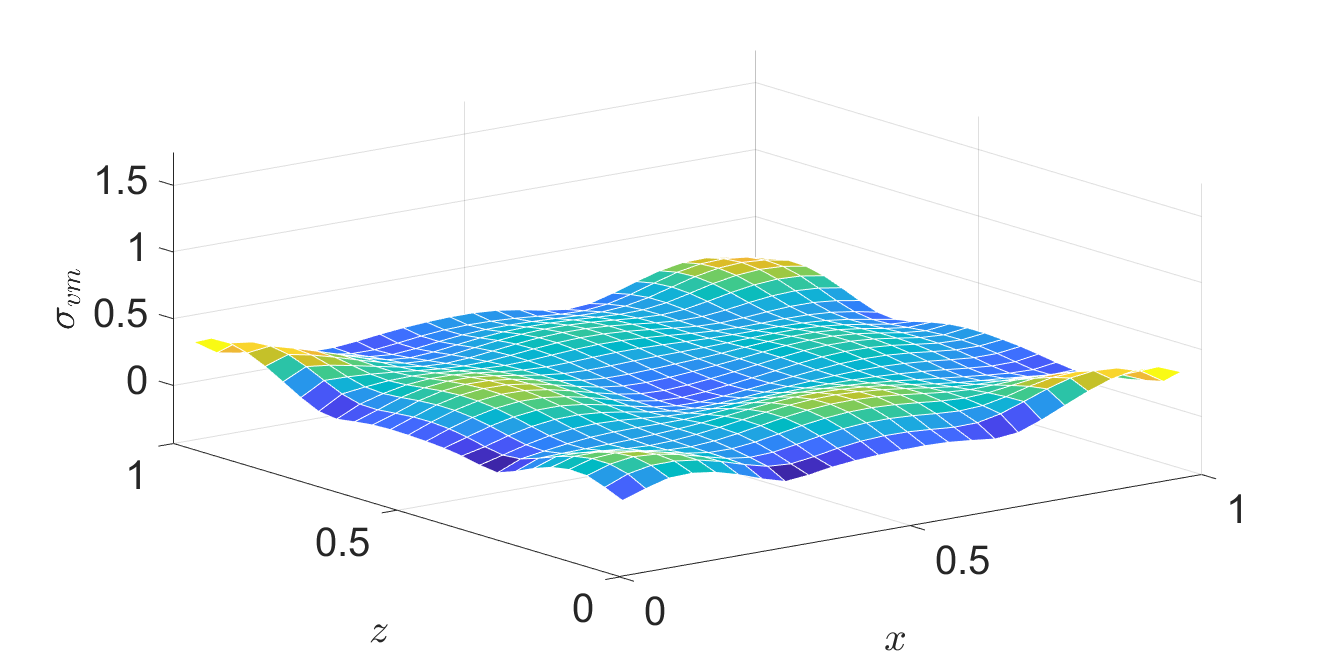}
\includegraphics[width=\qwidth]{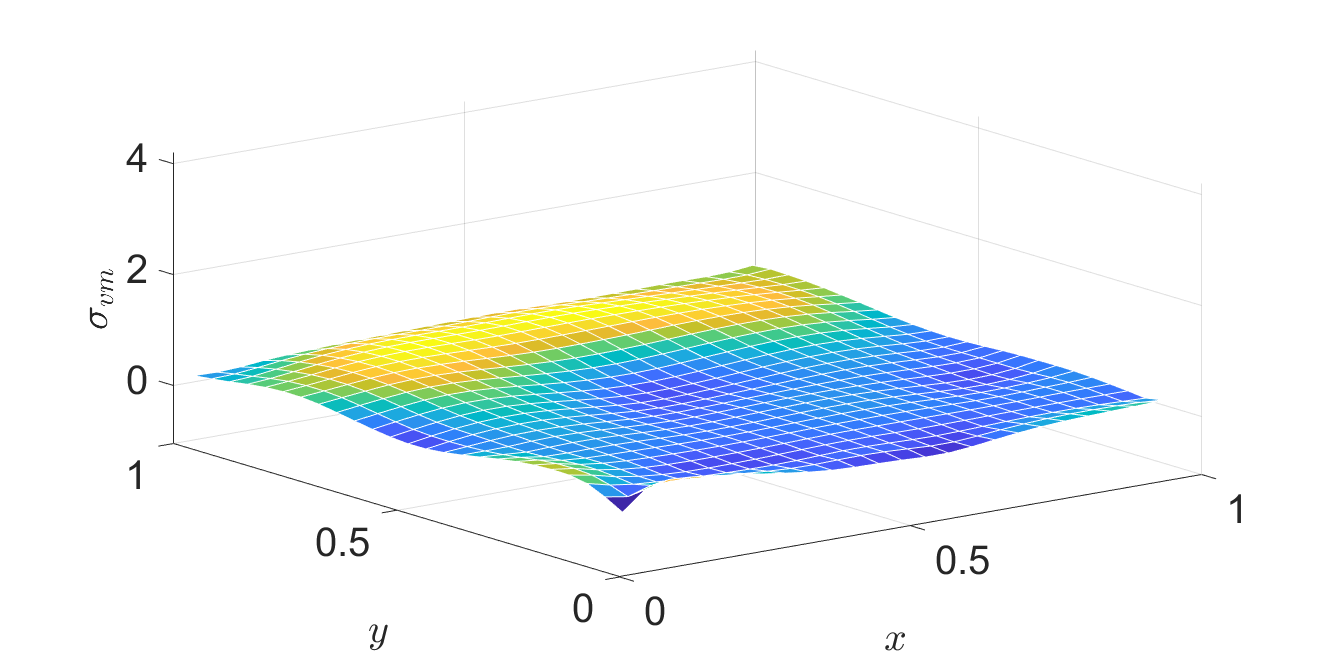}
 \caption{
 Continuation of Figure \ref{3dvm1}: snapshots of the distribution of
the stress invariant $\sqrt{\text{tr}(\hat{\bm{\sigma}}^{\text{dev}^{2}})}$ at the
mid-plane parallel to the excited side \1 1 {left column} and at a mid-plane
orthogonal to it \1 1 {right column} at various instants, in the PTZ case.
From top to bottom: snapshots at instants \m { \F001{5}{2} \qtaub }, \m{ 3
\qtaub}, \m { \F001{7}{2} \qtaub }, \m { 8 \qtaub }, respectively.
}
  \label{3dvm2}
 \end{figure}

Next, in this example, we demonstrate the usefulness of another
thermodynamical quantity, entropy production rate density.

It is instructive to start with showing how the idea works in the simpler,
one space dimensional, setting \1 1 {the rod discussed in \cite{1Dcikk}}. The
one space dimensional analogue of \re{entprodA} is
  \Par
 \begin{align}  \label{entprodA1D}
\qpi_{\qs} & = \f {1}{\qT} \f {1}{\qII} \, \Qsighat \cdot \9 1 { \qEE
{\qdot\Qeps} - \qtau {\qdot\Qsig} }
 \end{align}
  \Par
Let us introduce four different discretizations of this product, embodying
the patterns
 \begin{itemize}[leftmargin=*,labelsep=5.8mm]
\item \m{ \text{old} \cdot (\text{new} - \text{old}) },
\item \m{ \text{old} \cdot (\text{new} - \text{older}) },
\item \m{ \text{new} \cdot (\text{new} - \text{old}) }, and
\item \m{ \text{new} \cdot (\text{new} - \text{older}) }:
 \end{itemize}
  \Par
 \begin{align}  \label{entprodA1Da}
& \f {1}{\qT^{\qj}_{\qn}} \f {1}{\qII} \, \Qsighat^{\qj}_{\qn} \cdot \9 1 {
\qEE \f{\Qeps^{\qj+1}_{\qn} - \Qeps^{\qj}_{\qn}}{\Delta t} - \qtau
\f{\Qsig^{\qj+1}_{\qn} - \Qsig^{\qj}_{\qn}}{\Delta t} } ,
 \\ \label{entprodA1Db}
& \f {1}{\qT^{\qj}_{\qn}} \f {1}{\qII} \, \Qsighat^{\qj}_{\qn} \cdot \9 1 {
\qEE \f{\Qeps^{\qj+1}_{\qn} - \Qeps^{\qj-1}_{\qn}}{2\Delta t} - \qtau
\f{\Qsig^{\qj+1}_{\qn} - \Qsig^{\qj-1}_{\qn}}{2\Delta t} } ,
 \\ \label{entprodA1Dc}
& \f {1}{\qT^{\qj}_{\qn}} \f {1}{\qII} \, \Qsighat^{\qj+1}_{\qn} \cdot \9 1 {
\qEE \f{\Qeps^{\qj+1}_{\qn} - \Qeps^{\qj}_{\qn}}{\Delta t} - \qtau
\f{\Qsig^{\qj+1}_{\qn} - \Qsig^{\qj}_{\qn}}{\Delta t} } ,
 \\ \label{entprodA1Dd}
& \f {1}{\qT^{\qj}_{\qn}} \f {1}{\qII} \, \Qsighat^{\qj+1}_{\qn} \cdot \9 1 {
\qEE \f{\Qeps^{\qj+1}_{\qn} - \Qeps^{\qj-1}_{\qn}}{2\Delta t} - \qtau
\f{\Qsig^{\qj+1}_{\qn} - \Qsig^{\qj-1}_{\qn}}{2\Delta t} } ,
 \end{align}
  \Par
where \m { \qT^{\qj}_{\qn} } denotes the time average \m { \9 1 {
\qT^{\qjm}_{\qn} + \qT^{\qjp}_{\qn}} / 2}.

\begin{figure}[H]
\centering
\includegraphics[width=\qwidth]{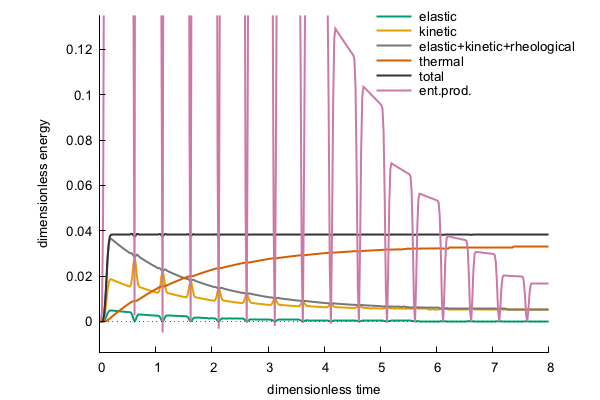}
\includegraphics[width=\qwidth]{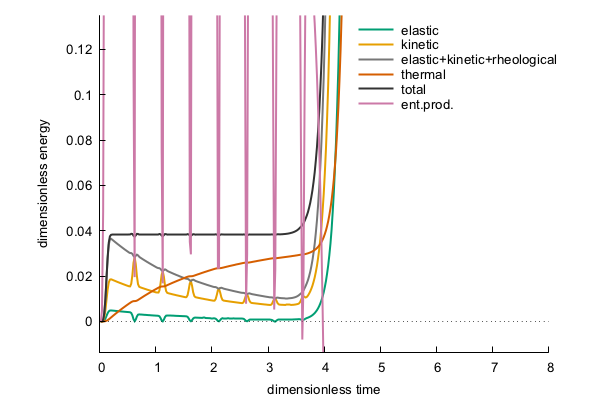}
 \\
\includegraphics[width=\qwidth]{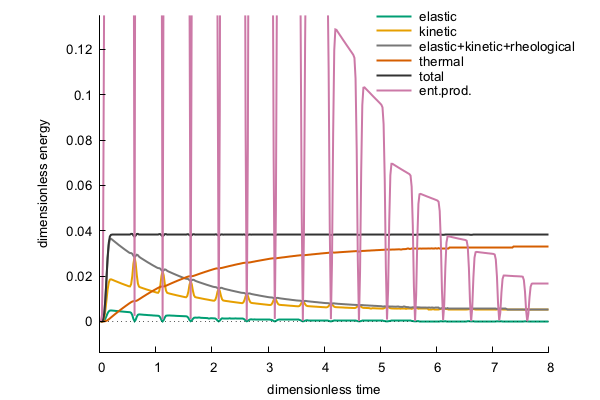}
\includegraphics[width=\qwidth]{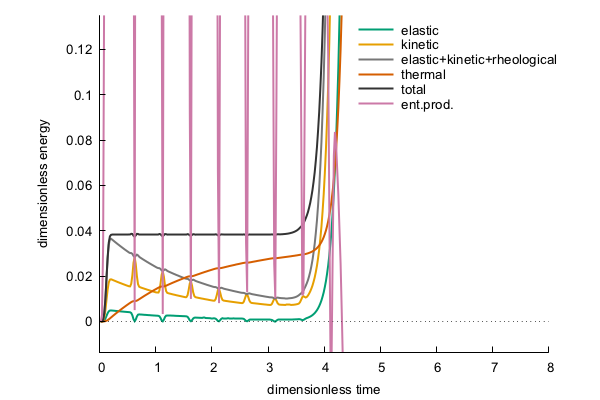}
 \\
\includegraphics[width=\qwidth]{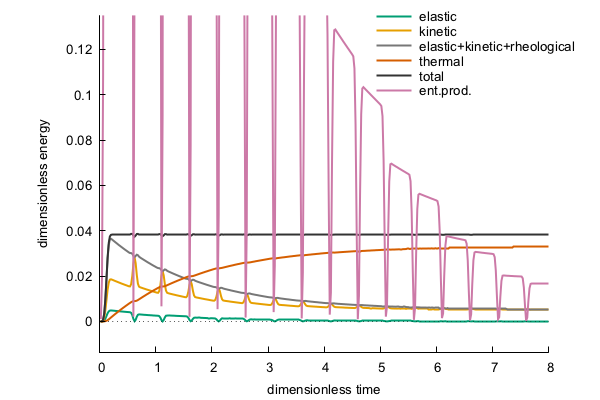}
\includegraphics[width=\qwidth]{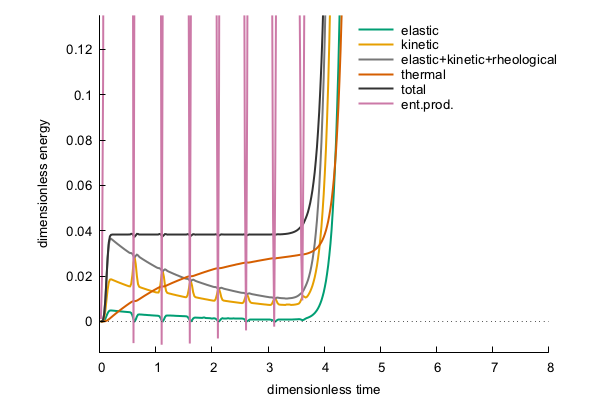}
 \\
\includegraphics[width=\qwidth]{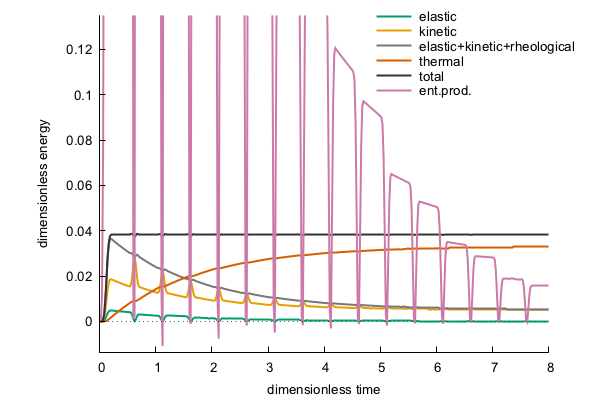}
\includegraphics[width=\qwidth]{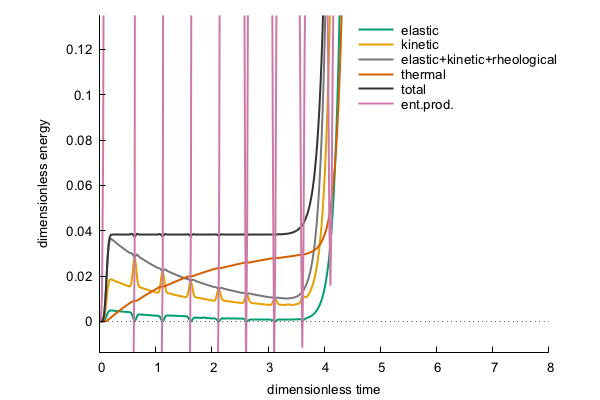}
 \caption{Entropy production rate according to the four discretized versions
\re{entprodA1Da}--\re{entprodA1Dd}, respectively, for~a time step resulting
in a stable run \1 1 {left column} \1 3 {namely, when the rheological Courant
number \m { \qcc \Delta t/\Delta x } is 1, where \m{ \qcc =
\sqrt{\qEE/\11{\qtau\qrho}} } is the high-frequency rheological wave
propagation speed%
} and for a choice
leading to an unstable outcome \1 1 {right column} \1 3 {rheological Courant
number \m { \qcc \Delta t/\Delta x = 1.003 }}.
PTZ~model with \m { \qtau \qEY / \qEE = 0.25 };
the sample length \m { \qX } is such that \m { \qEE / \qEY = 5 \qX/\qc } with
the low-frequency/elastic wave propagation speed \m { \qc = \sqrt{\qEY/\qrho}
}. See \cite{1Dcikk} for further details.
}  \label{1ds}
 \end{figure}

These four versions are integrated in space and plotted in Figure \ref{1ds},
the left column for a stable setting and the right column for an unstable
one. The energies are also displayed. Only 25 space
cells have been chosen to enhance artefacts.

Visibly, certain versions become negative when instability gets exposed.
Moreover, some become negative even before that, showing, at an early stage,
that there is a problem to come.

Next, let us see how the three space dimensional generalizations behave for
the problem of the pressed cube: the outcomes can be seen in
Figure~\ref{3ds}. An intentionally rude \m { 10 \times 10 \times 10 } grid is
taken, the time unit is divided to 125 time steps in order to lead to a stable
solution and to 100 time steps producing an unstable one. Further settings
are \m{\qrho=1}, \m{\dev\qEY=3}, \m{\sph\qEY=5}, \m{\dev\qEE=20},
\m{\dev\qtau=0.391}, \m{\qcp=0.001}, \m{T^{0}=0.1}, \m{\qtaub=0.25}, and
\m{\qsigb=3}, \m { \qWb/\qX=0.6}, \m{\alpha=\iOii}.

 \begin{figure}[H]
\centering
\includegraphics[width=\qwidth]{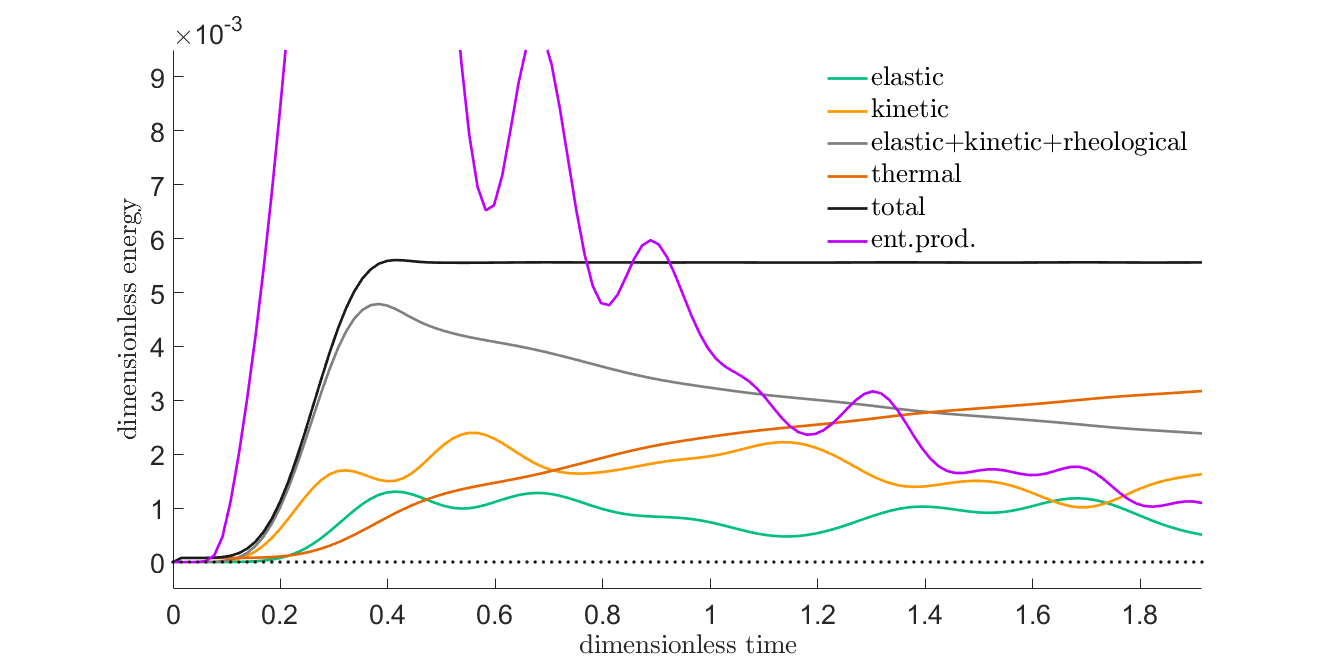}
\includegraphics[width=\qwidth]{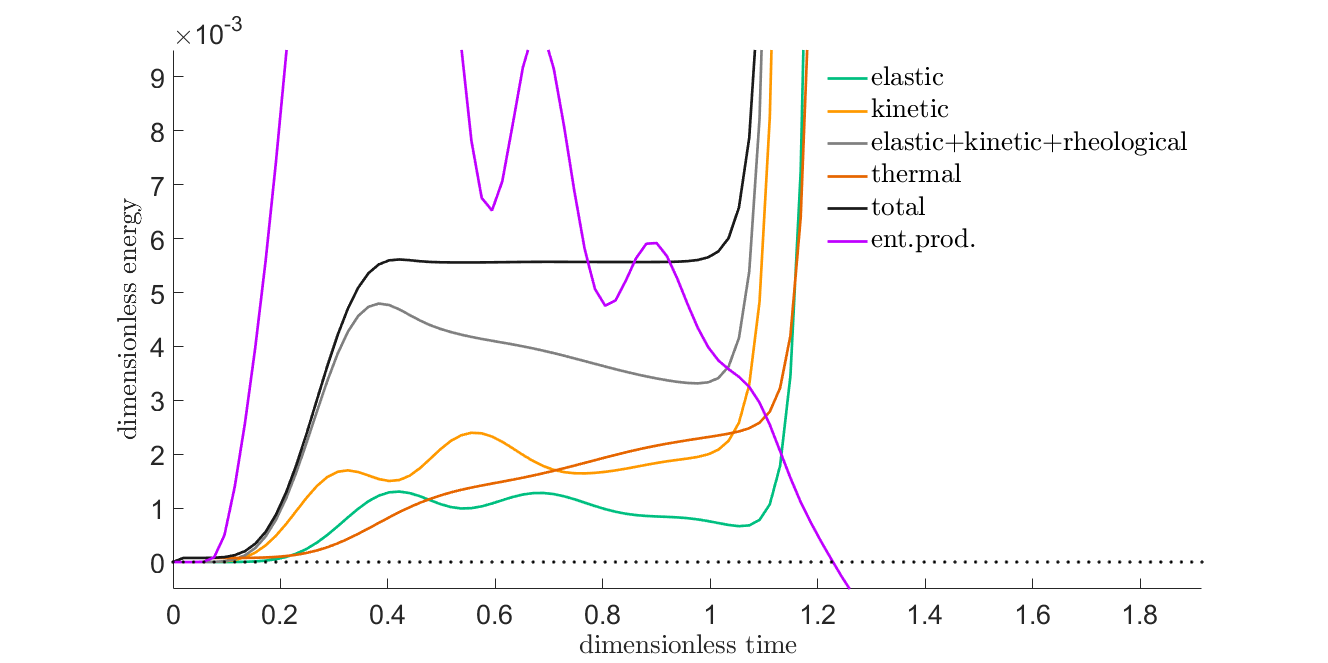}
 \\
\includegraphics[width=\qwidth]{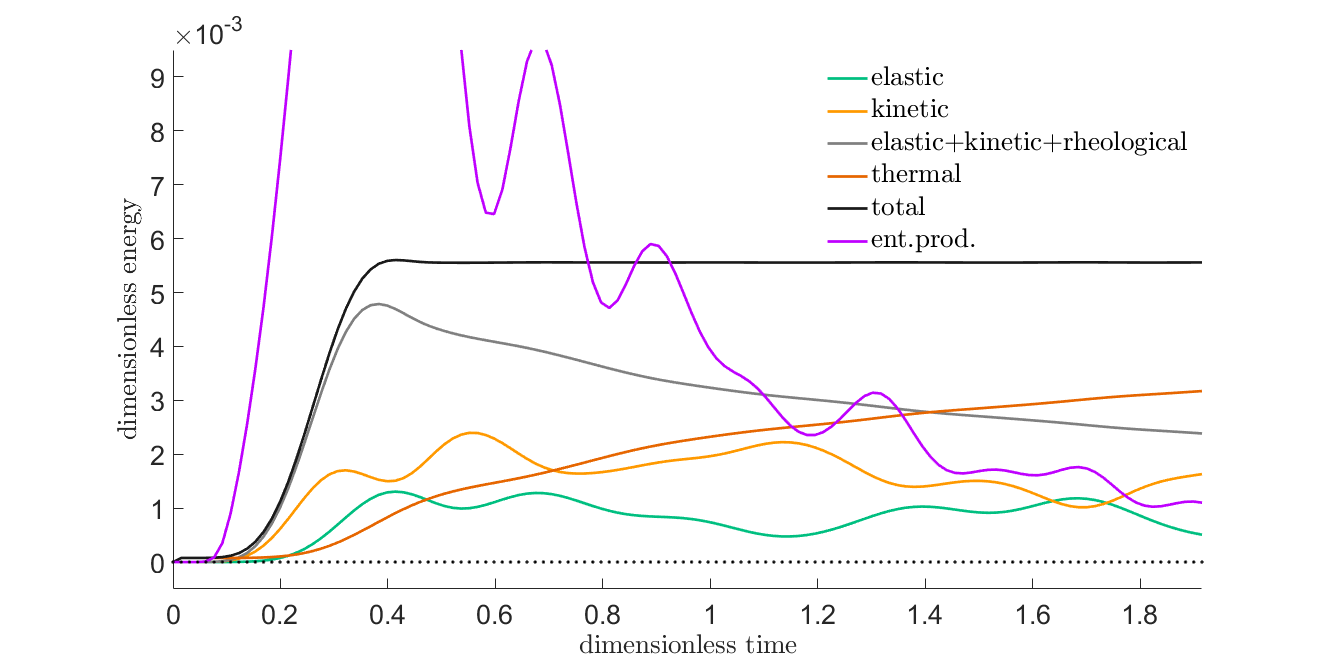}
\includegraphics[width=\qwidth]{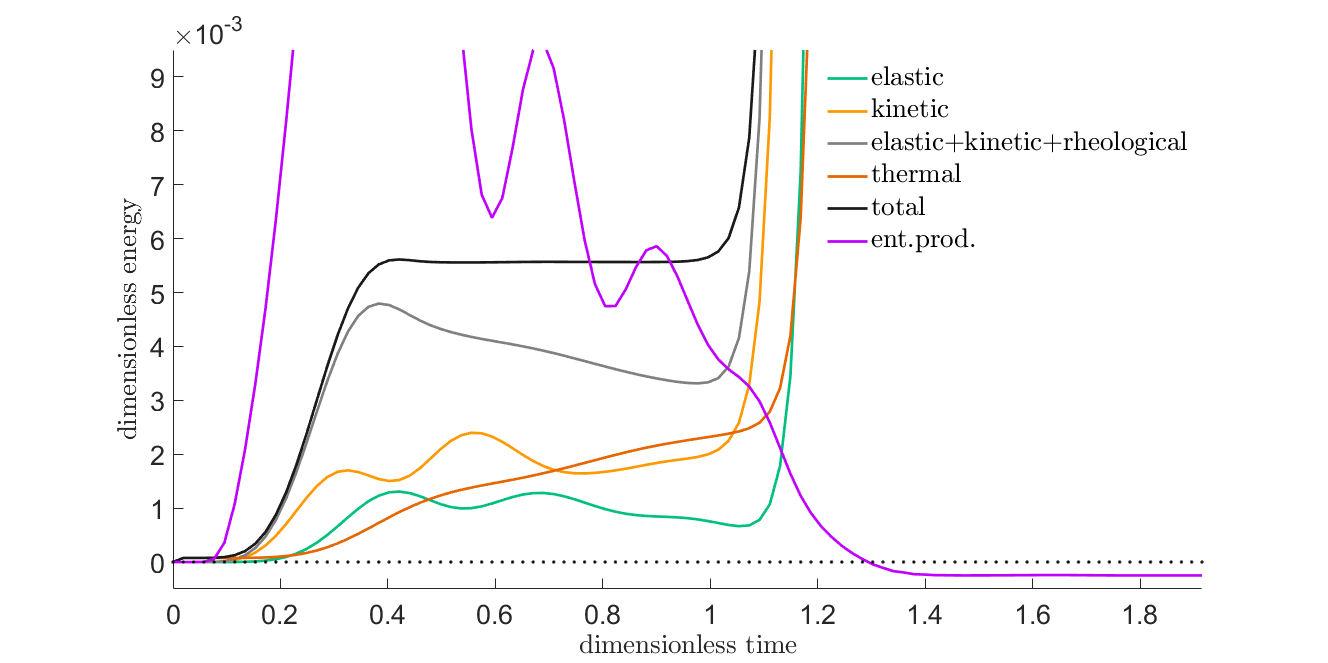}
 \\
\includegraphics[width=\qwidth]{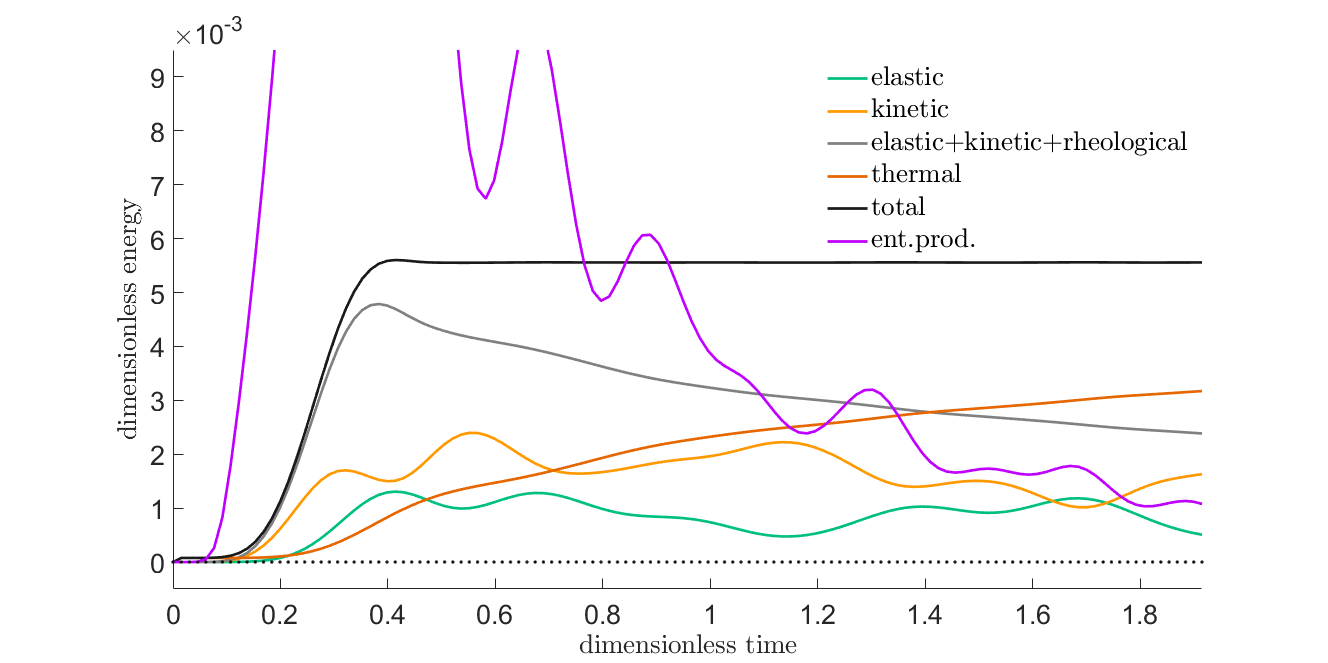}
\includegraphics[width=\qwidth]{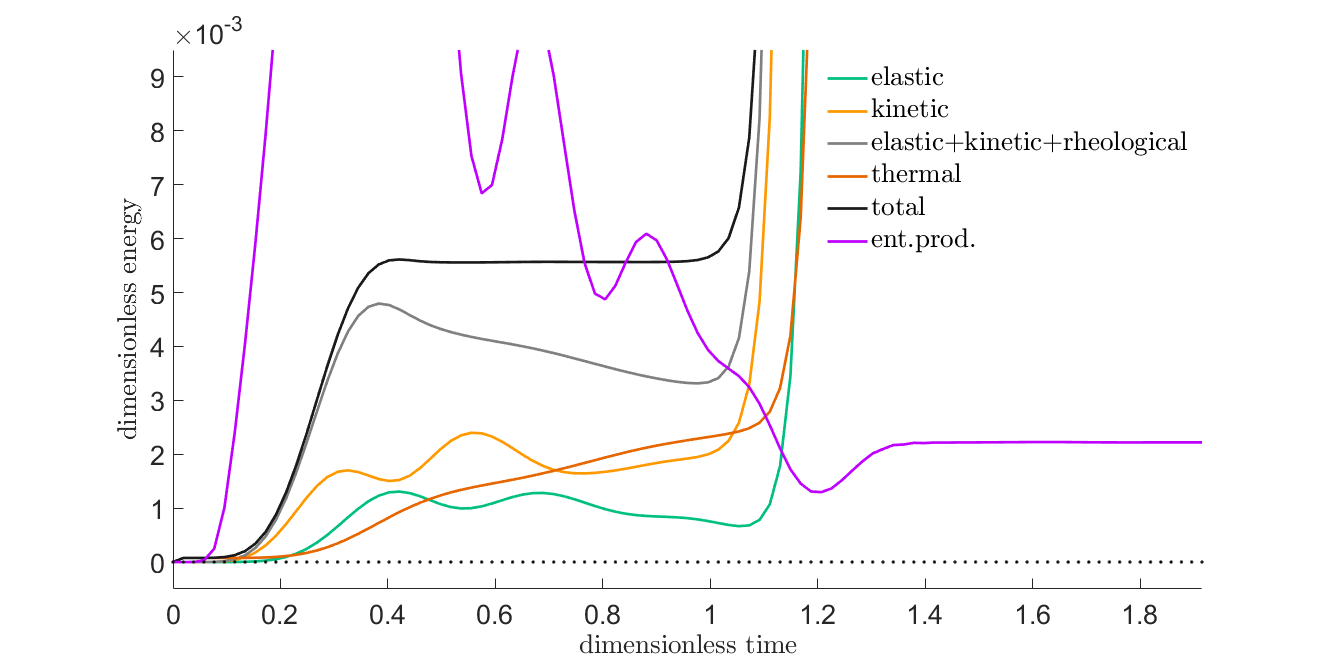}
 \\
\includegraphics[width=\qwidth]{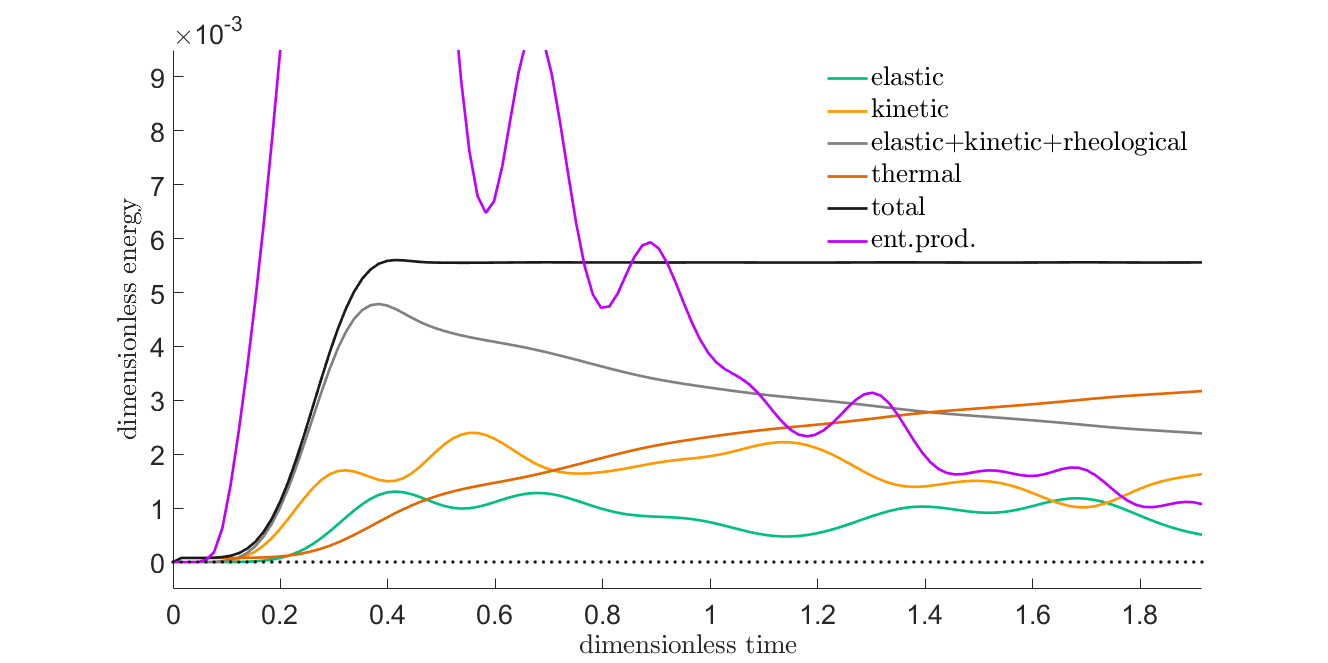}
\includegraphics[width=\qwidth]{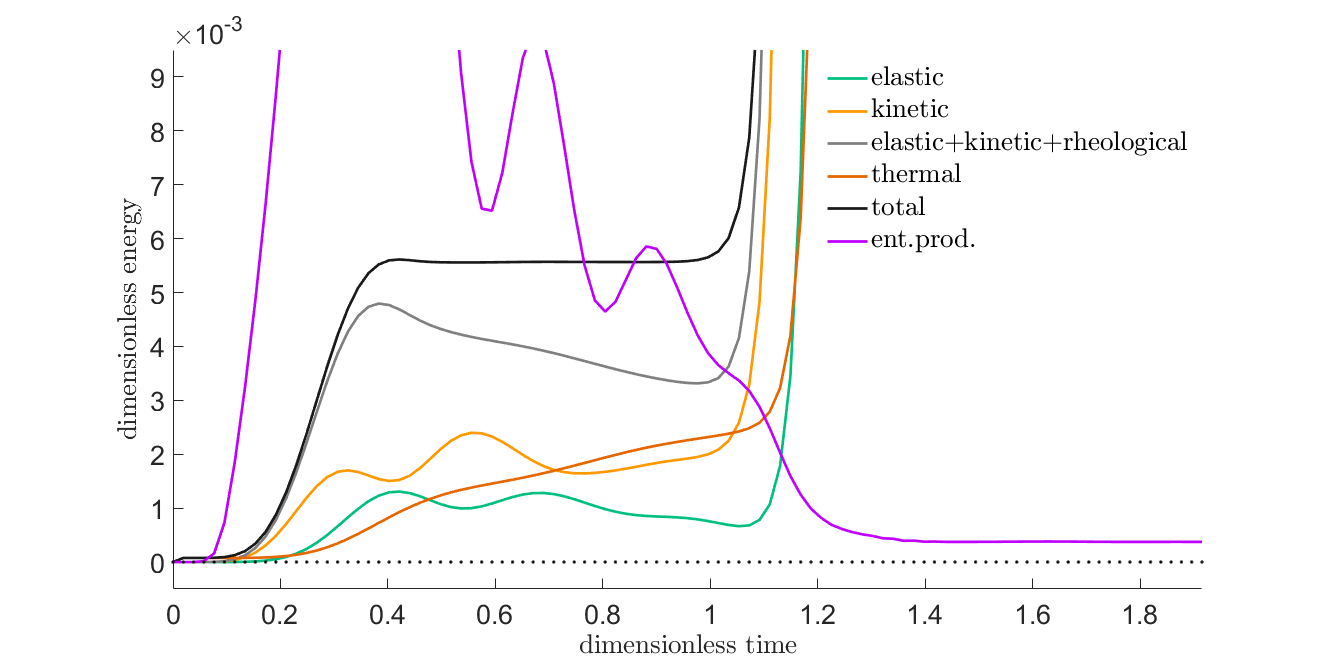}
 \caption{Entropy production rate according to the four discretized versions
\re{entprodA1Da}--\re{entprodA1Dd} generalized to three space dimensions,
respectively, computed for a PTZ cube, with a time step resulting in a stable
run \1 1 {left column} and for a choice leading to an unstable outcome \1 1
{right column}.}  \label{3ds}
 \end{figure}

One can see that an appropriately chosen discretization diagnoses instability.

Such numerical comparisons between stable and unstable parameter domains,
which are probably enhanced by some analytical considerations, can help in the future in deciding which discretized entropy production rate is useful for
diagnosing what.

\section{Two-Dimensional Wave Propagation According to the Finite Element
Software COMSOL}

In \cite{1Dcikk}, a comparison of solutions via the one space dimensional
scheme with corresponding finite element solutions was presented. Repeating
it for the present three space dimensional scheme would be informative. We found it advisable to start investigating the higher dimensional wave propagation describing the possibilities of finite element softwares in a much simpler setting than the one treated above, based on the \1 1 {discouraging} experience that is gained with the one-dimensional case, and since a rheological model like PTZ in the deviatoric--spherical formulation is difficult to translate to the capabilities of classic finite element softwares.

Namely, we
 consider
the wave equation in two spatial dimensions, for a single scalar degree of
freedom \m { \qu = \qu \1 1 {\qt, \qx, \qy} }, with constant \1 1 {unit}
coefficients and no source term in the equation:

 \begin{align}
\f {\partial^2 \qu}{\partial \qt^2} = 
\f {\partial^2 \qu}{\partial \qx^2} +
\f {\partial^2 \qu}{\partial \qy^2} . 
 \end{align}

A rectangular sample
 is
taken with \quot{flux} \1 1 {Neumann} boundary
condition \1 1 {normal spatial derivative component is prescribed}: on one of
the edges, one cosine-type pulse of the kind \re{cos}
is
applied, while the other edges
 are
kept \quot{free} \1 1 {zero normal derivative}. Initially, both \m { \qu }
and \m { \partial \qu / \partial \qt }
 are
set to zero \1 1 {\quot{relaxed initial state}}.

To stay similar to the numerical calculations that are presented here, a 50 \m {
\times } 50 spatial grid
 is
taken. The software that we
 use is
COMSOL v5.3a, where this wave equation problem is a built-in possibility. The
pulse duration
 is
0.3 time unit. Figure~\ref{COM} displays the resulting spatial distribution
of the scalar degree of freedom after two units of time, obtained with various
available settings of COMSOL v5.3a.

 \begin{figure}[H]
\centering
\includegraphics[width=\Qwidth]{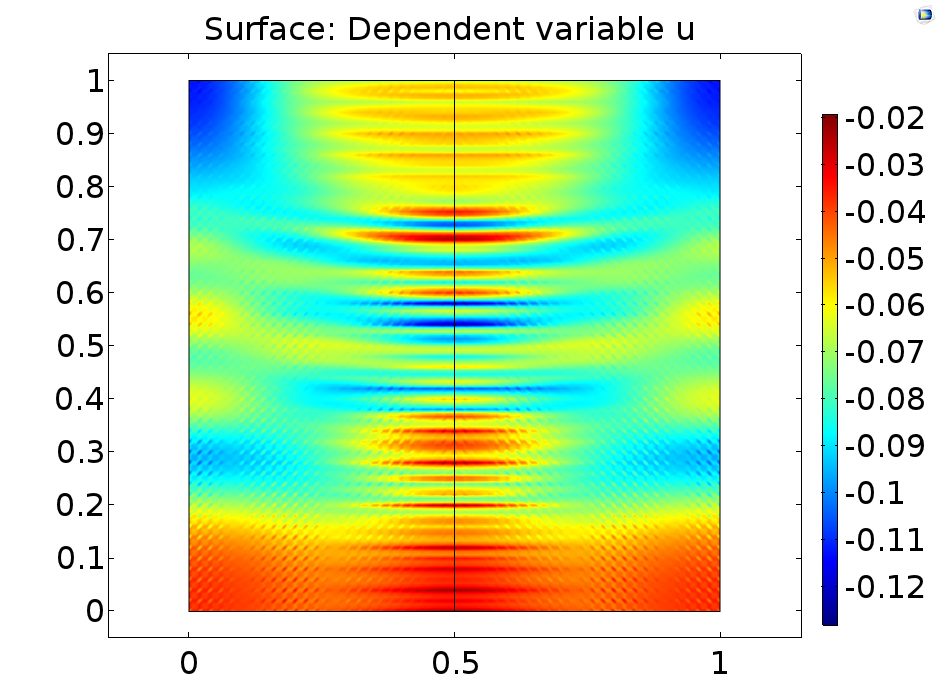}
\includegraphics[width=\QWidth]{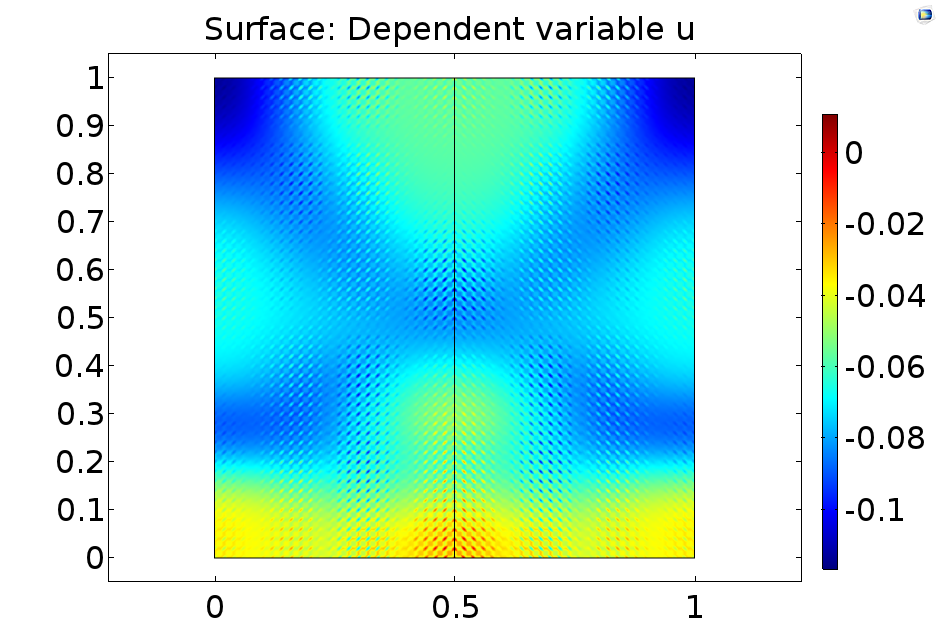}
 \\
\includegraphics[width=\Qwidth]{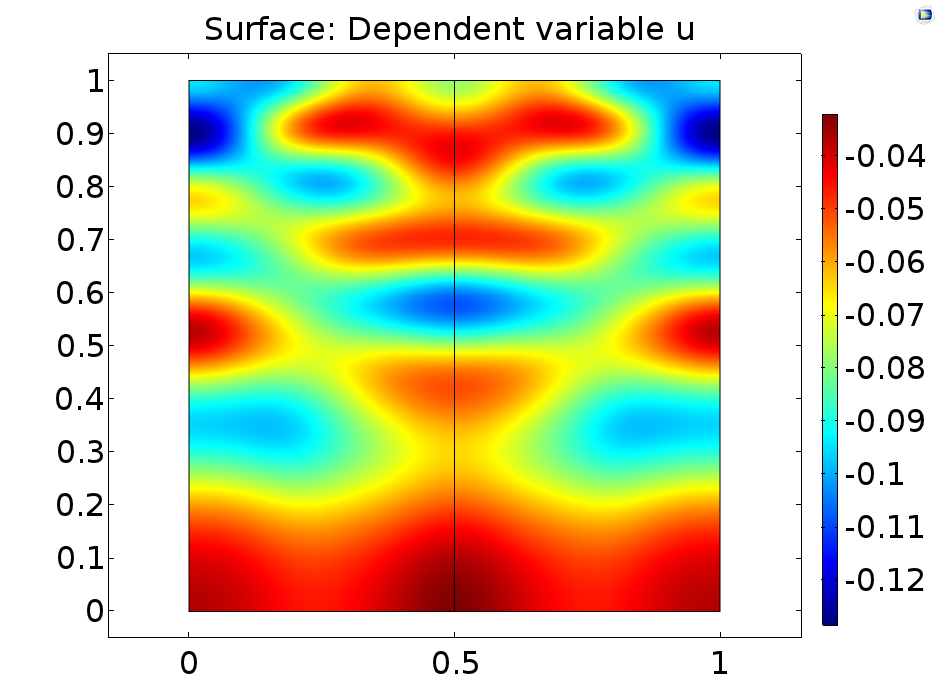}
\includegraphics[width=\QWidth]{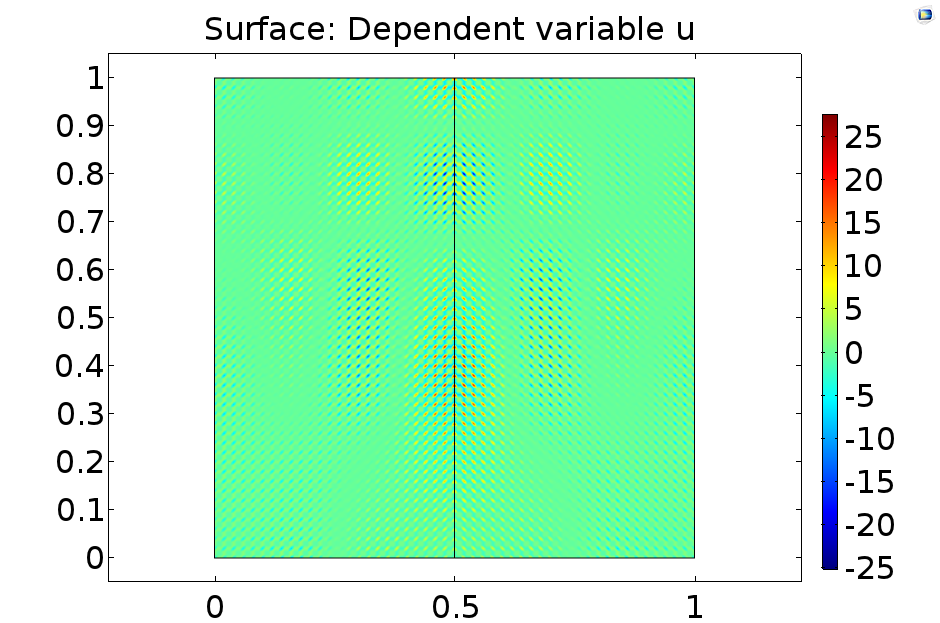}
 \\
\caption{\textit{Cont.}}
\end{figure}

\begin{figure}[H]\ContinuedFloat
\centering
\includegraphics[width=\Qwidth]{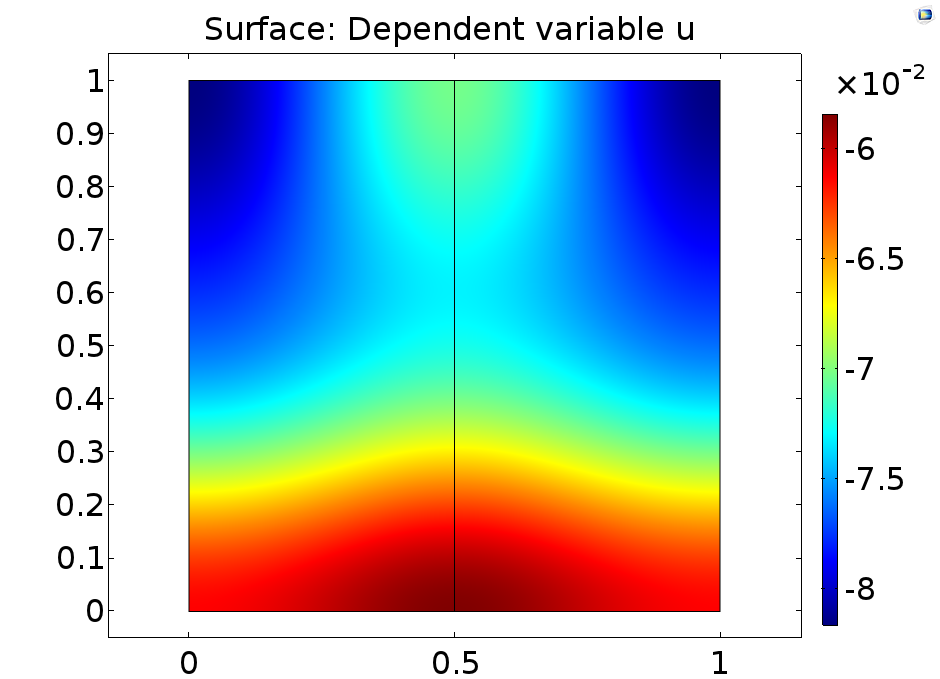}
\includegraphics[width=\QWidth]{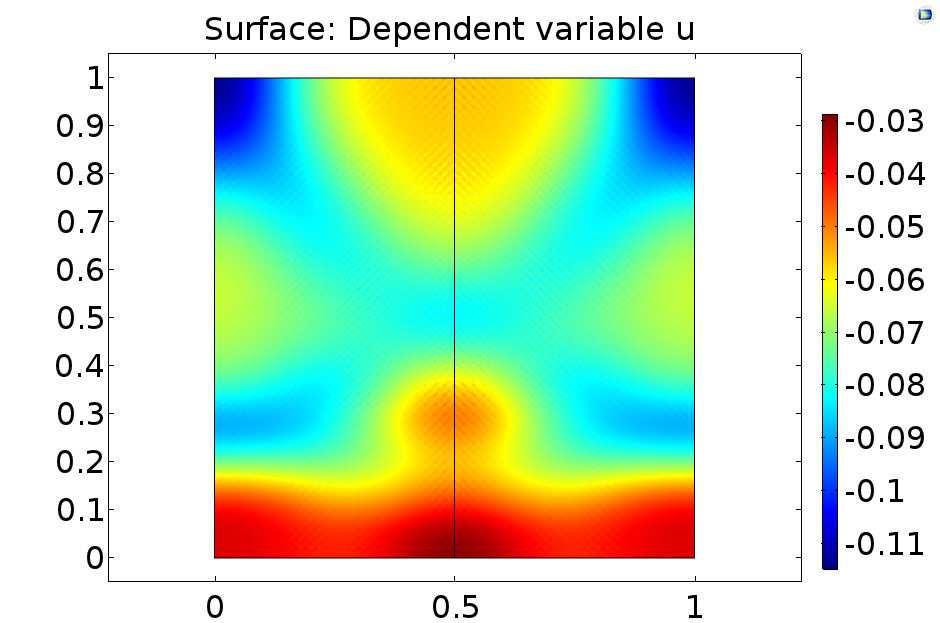}
 \\
\includegraphics[width=\Qwidth]{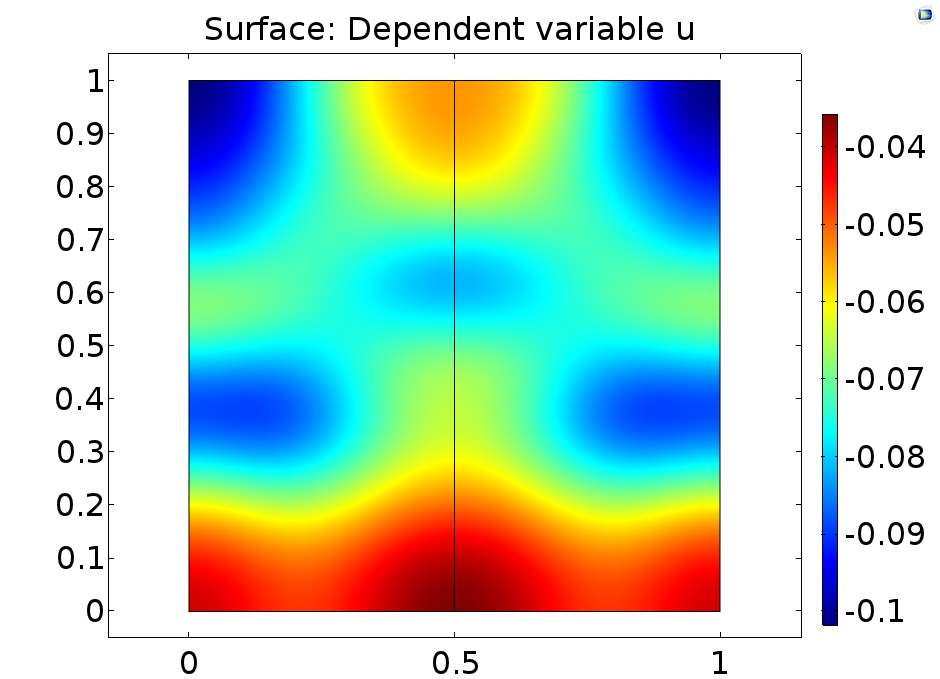}
\includegraphics[width=\QWidth]{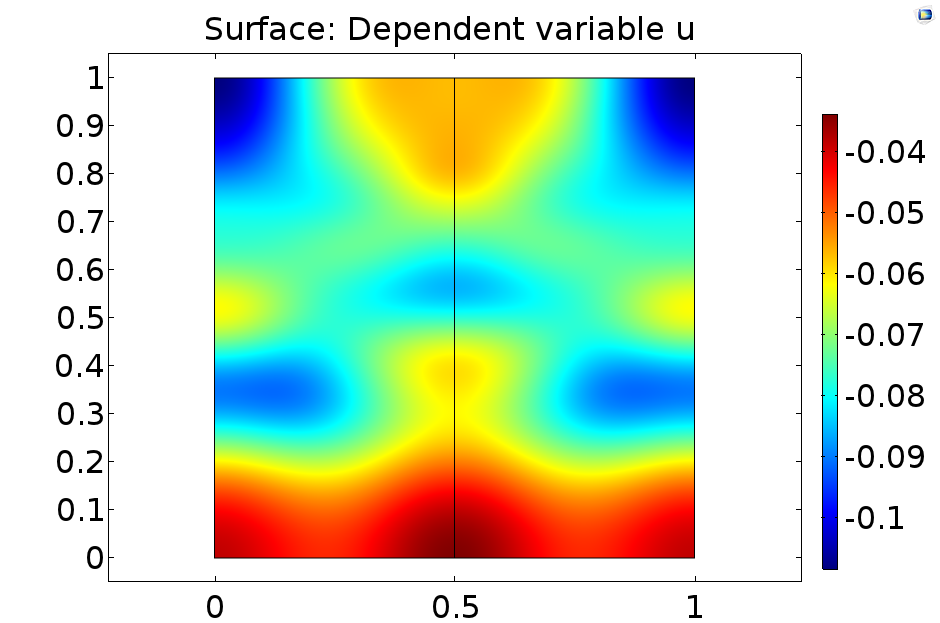}
 \\
\includegraphics[width=\Qwidth]{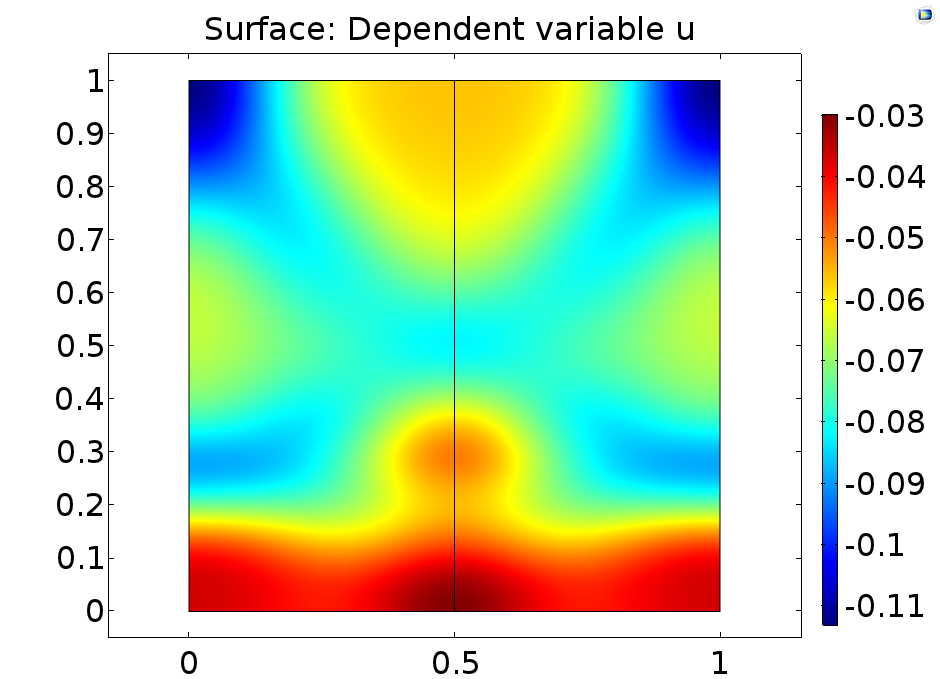}
\includegraphics[width=\QWidth]{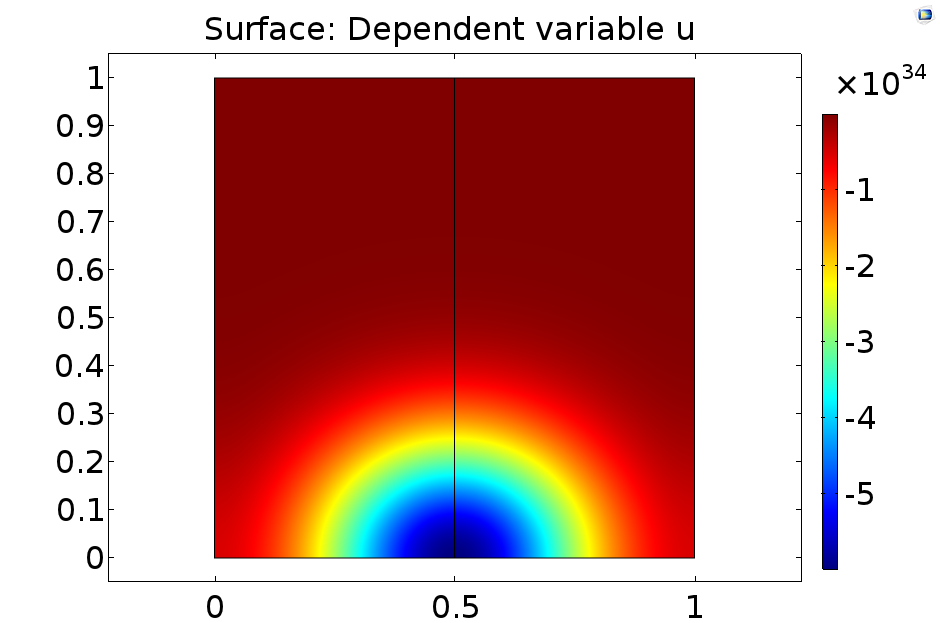}
 \caption{%
\m { \qu \1 1 {2, \qx, \qy} } with various COMSOL settings \1 1 {cf.\
Figure~\ref{com}}. Left column, from top to bottom:
BDF Max. order 5 \& min. order 2,                       
BDF Max. order 3 \& min. order 2,                       
BDF Max. order 1 \& min. order 1,                       
BDF Max. order 5 \& min. order 3 Strict time stepping,  
RK34 Default.                                           
Right column, from top to bottom:
DP5 PI-Smooth,                                          
DP5 PI-Standard,                                        
DP5 PI-Quick,                                           
Generalized-\m { \alpha } \1 1 {GA},                    
GA Predictor Constant.                                  
 Not~shown (failed like DP5 PI-Standard or GA Predictor Constant):
DP5 Pi-Off, RK34 Manual, Cash-Karp 5 Free or Strict \1 1 {Manual with a small
time step gives a result similar to that of RK~Default}.
 }  \label{COM}
 \end{figure}

One can see that the solution depends on the settings very seriously. There
are large-scale differences and fine-structured irregularities.
\textcolor{\qred}{Moreover, one has no means of validation which outcome is correct to what
extent.}

When taking into account that our full problem has 16 coupled degrees of freedom
in three spatial dimensions and with further time derivatives \1 1 {in the
PTZ model}, it does not seem to be reasonable to try to represent it via any
commercial finite element software as long as such a much simpler problem
cannot be confidently treated.

\section{Discussion}

The numerical scheme presented here, due to its symplectic root, second-order
accuracy, and the equation-friendly and spacetime geometry friendly
arrangement of discretized quantities, has been found to provide reliable
results in a fast and resource-friendly way. Being a finite difference
scheme, it is not very flexible to simulate arbitrary shaped samples, but,
already, the extension of the Cartesian formulae to cylindrical and spherical
geometries promises useful applications, including the various
wave-based measurement methods that are used in rock mechanics (see, e.g., \cite{Mal}),
many of which rely on simple and easily treatable sample shapes. Fitting a
rheological model on experimental data may require many runs so good finite
difference schemes find their applicability.

The investigation of stability and dissipative and dispersion error is
expected to be much more involved than in the corresponding one space
dimensional situation, where the analysis was done in \cite{1Dcikk}.
Nevertheless, it is an important task for the future, because the outcomes
support efficient applications of the scheme.

It is an interesting challenge to apply the presented scheme for other
dissipative situations \1 1 {like~\cite{SzKo}, just to mention one example}.

A systematic and general framework could be
obtained, which is also beneficial for other purposes, if the spacetime background is strengthened further, by using
four-quantities, four-equations on them, and formulating discretization in a
fully four-geometrical way. It would, for example, help
in building connection to a finite element---spacetime finite element---approach, along which way objects of general shape could also be treated.
Notably, the current finite element paradigm has deficiencies and probably needs to be renewed, as indicated by results found here and earlier works \cite{Rieth18,1Dcikk}.

 In addition, the use  the thermodynamical full description of a system for
monitoring and controlling numerical artefacts during a computer simulation
is a promising perspective. The~steps made here: recognizing the usefulness
of total energy and its various parts, and of entropy production rate
density, are hoped to contribute to a future routine in numerical
environments, where thermodynamics could be utilized.

\vspace{6pt} 

\authorcontributions{
\'A.P.: numerical simulations, figures, analysis.
M.S.: analysis, numerical simulations, figures.
R.K.: numerical simulations, figures, manuscript text, funding.
T.F.: conceptual idea, numerical simulations, figures, manuscript text.
All authors have read and agreed to the published version of the manuscript.
}

\funding{%
The research reported in this paper and carried out at BME has been supported
by the grants National Research, Development and Innovation Office-NKFIH KH
130378, and K124366(124508), and by the NRDI Fund (TKP2020 NC,Grant No.
BME-NC) based on the charter of bolster issued by the NRDI Office under the
auspices of the Ministry for Innovation and Technology.
This paper was supported by the J\'anos Bolyai Research Scholarship of the
Hungarian Academy of Sciences.}
\acknowledgments{The authors are thankful to Tam\'as Kov\'acs for the discussions.}

\conflictsofinterest{The authors declare no conflict of interest.}

\reftitle{References}





\end{document}